\documentclass[twocolumn]{aastex701}
\usepackage{graphicx}

\usepackage{amssymb}
\usepackage{amsmath}
\usepackage{bbm}
\usepackage{url}
\usepackage[version=4]{mhchem}

\usepackage{hyperref}

\usepackage{color, colortbl}
\definecolor{Gray}{gray}{0.75}

\newcommand\hp{HAT-P-11}
\newcommand{\caii}{Ca\,\textsc{ii}}
\defcitealias{Narrett2024}{N24}

\newcommand\HST{HST}
\newcommand\hubble{\HST{}}
\newcommand\JWST{JWST}
\newcommand\Kepler{Kepler}

\newcommand\TESS{TESS}
\newcommand\Gaia{Gaia}

\newcommand\Spitzer{\textit{Spitzer}}

\newcommand\phoenix{\texttt{PHOENIX}}
\newcommand\spex{\texttt{SpeX}}

\newcommand\stis{\texttt{STIS}}
\newcommand\wfcthree{\texttt{WFC3}}

\begin{document}

\title{Unlocking HST's Stellar Treasure Trove: Stellar Activity Minima for HAT-P-11 Offer Prime Windows for Transmission Spectroscopy}

\author[0000-0002-8052-3893]{Prajwal Niraula}
\affiliation{Department of Earth, Atmospheric and Planetary Sciences, Massachusetts Institute of Technology, Cambridge, MA 02139, USA}
\affiliation{Kavli Institute for Astrophysics and Space Research, Massachusetts Institute of Technology, Cambridge, MA 02139, USA}
\email{pniraula@mit.edu}

\author[0000-0002-3627-1676]{Benjamin V.\ Rackham}
\affiliation{Department of Earth, Atmospheric and Planetary Sciences, Massachusetts Institute of Technology, Cambridge, MA 02139, USA}
\affiliation{Kavli Institute for Astrophysics and Space Research, Massachusetts Institute of Technology, Cambridge, MA 02139, USA}
\email{brackham@mit.edu}

\author[0000-0003-2415-2191]{Julien de Wit}
\affiliation{Department of Earth, Atmospheric and Planetary Sciences, Massachusetts Institute of Technology, Cambridge, MA 02139, USA}
\email{jdewit@mit.edu}

\author[0000-0003-3714-5855]{Daniel Apai}
\affiliation{Department of Astronomy and Steward Observatory, The University of Arizona, 933 North Cherry Avenue, Tucson, AZ 85721, USA}
\affiliation{Lunar and Planetary Laboratory, The University of Arizona, Tucson, AZ, USA}
\affiliation{James C. Wyant College of Optical Sciences, The University of Arizona, AZ 85721, USA}
\email{apai@arizona.edu}

\author[0000-0002-2132-5264]{Mark S. Giampapa}
\affiliation{Department of Astronomy and Steward Observatory, The University of Arizona, 933 North Cherry Avenue, Tucson, AZ 85721, USA}
\affiliation{Lunar and Planetary Laboratory, The University of Arizona, Tucson, AZ, USA}
\email{giampapa@arizona.edu}

\author[0000-0001-6298-412X]{David Berardo}
\affiliation{Department of Earth, Atmospheric and Planetary Sciences, Massachusetts Institute of Technology, Cambridge, MA 02139, USA}
\affiliation{Kavli Institute for Astrophysics and Space Research, Massachusetts Institute of Technology, Cambridge, MA 02139, USA}
\email{dberardo@mit.edu}

\author[0000-0001-5989-7594]{Chia-lung Lin}
\affiliation{Department of Astronomy and Steward Observatory, The University of Arizona, 933 North Cherry Avenue, Tucson, AZ 85721, USA}
\email{chialunglin@arizona.edu}

%

\begin{abstract}
 HAT-P-11 is a well-studied, active K dwarf hosting an eccentric, misaligned transiting sub-Neptune. As part of the HST Stellar Treasure Trove program (HST-AR-17551), we analyze absolutely calibrated out-of-transit \HST{} spectra from \texttt{STIS} and \texttt{WFC3} across the \textsc{G430L}, \textsc{G750L}, \textsc{G102}, and \textsc{G141} bandpasses to constrain the surface heterogeneity of HAT-P-11 and its impact on transmission spectroscopy. Grid-based spectral retrievals using NewEra \texttt{PHOENIX} models robustly favor two-component photospheres in the \texttt{WFC3} G102 and G141 data, with a ${\sim}4950$\,K photospheric component and a cooler ($\sim$3400\,K) component covering 26{--}33\% of the stellar disk. By contrast, retrievals on the \texttt{STIS} optical spectra do not yield a satisfactory fit, reflecting current limitations of stellar atmosphere models in the optical regime compared to the \HST{} observational precision.  We contextualize these results using long-term photometric monitoring and chromospheric activity indices. The inferred high spot covering fractions are broadly consistent with the elevated photometric variability observed during the \textit{Kepler} era ($f_{\rm spot}$$\sim$10--20\%) but are in tension with the much lower rotational amplitudes observed from TESS in the mid 2020s ($f_{\rm spot}$$\sim$1--10\%). This secular decline in variability is mirrored by a $\sim$20\% decrease in the Ca\,\textsc{ii} H\&K index. These results imply that HAT-P-11 undergoes comparatively quiescent phases that offer more favorable windows for atmospheric characterization, which serendipitously coincided with some of the recent JWST observations. More generally, our study demonstrates that multi-epoch, space-based stellar spectra provides a physically grounded pathway for mitigating stellar contamination in high-precision transmission spectra in the JWST era. 
\end{abstract}


\section{Introduction} 
\label{sec:intro}

Stellar photospheric heterogeneity is a major challenge in the interpretation of exoplanet transmission spectra. Unocculted starspots and faculae alter the depth of planetary transits, imprinting spectroscopic signals in transmission spectra that bias inferences of planetary atmospheric properties \citep{Pont2008, Sing2011, McCullough2014, rackham2017, Rackham2023, Pinhas2018, Espinoza2019, Iyer2020}. These effects are significant for late-type and/or active stars, where typical levels of heterogeneity can mimic or mask planetary signals \citep{rackham2018, rackham2019}. With the increasing precision of modern observations, correcting for stellar contamination has become the key bottleneck in atmospheric characterization \citep[e.g.,][]{Lim2023, Moran2023, Rackham2023, Rackham2024, Radica2025, Rathcke2025}. However, despite growing awareness of this issue, systematic efforts to  characterize the surface heterogeneity of exoplanet host stars remain limited. 

\begin{figure*}[t!]
\centering
\includegraphics[trim={0.0cm 0.0cm 0.0cm 0.0cm},clip, angle=0, width=0.995\textwidth]{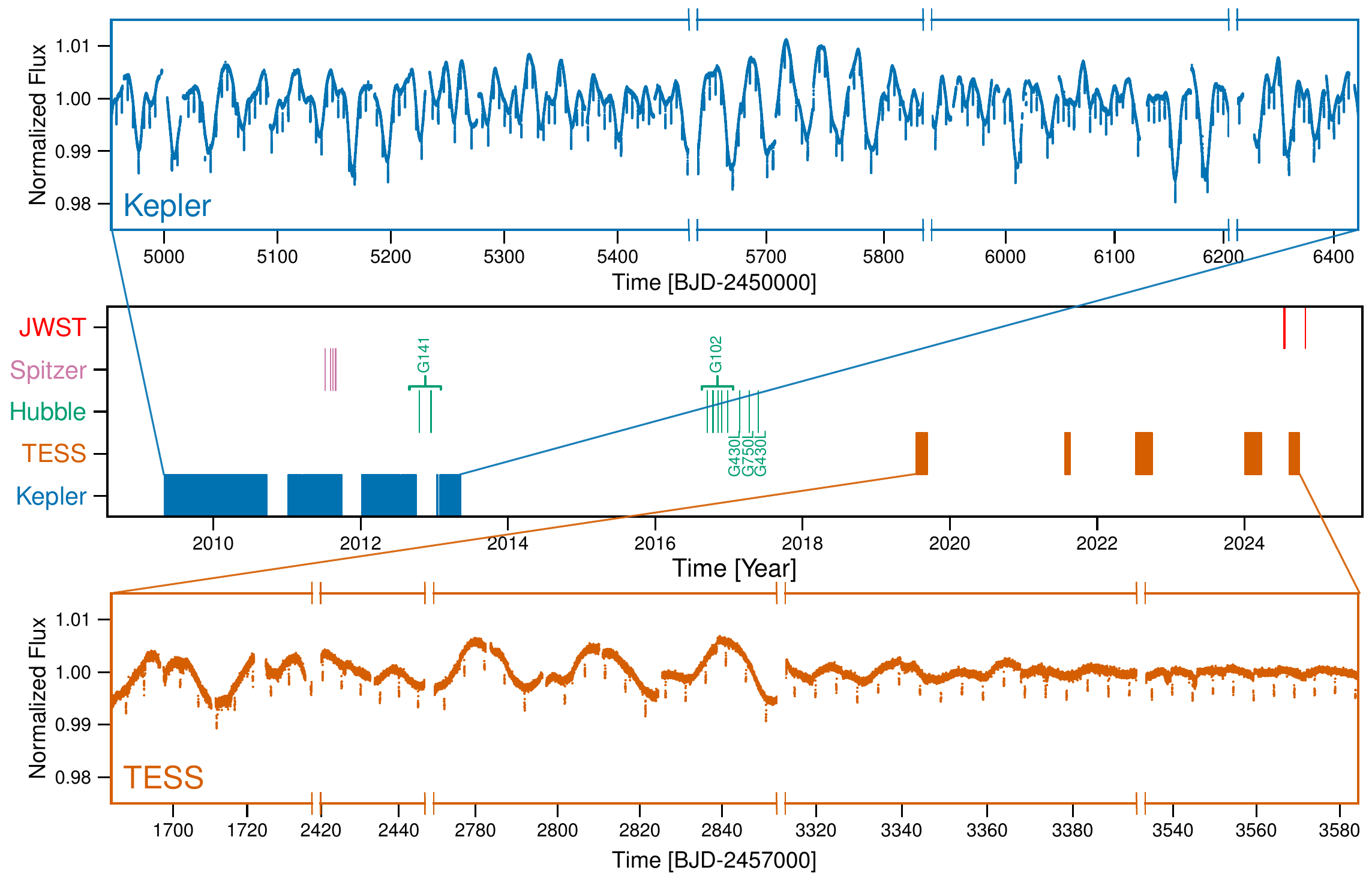}
\caption{
Summary of selected space-based observations of HAT-P-11.
\textbf{Top:} Four years of \Kepler{} short-cadence photometry showing pronounced rotational modulation and numerous spot-crossing events.
\textbf{Middle:} Observation timeline for five facilities---\Kepler{}, \TESS{}, \hubble{}, \Spitzer{}, and \JWST{}---spanning optical to infrared wavelengths.
\textbf{Bottom:} The \TESS{} light curve, displayed on the same normalized-flux scale as the \Kepler{} data for direct comparison.
The rotational modulation amplitude in \textit{TESS} is smaller than in \Kepler. 
The high-cadence data are binned to highlight the modulation signal. 
The rotational periods inferred from both datasets are consistent with the previously reported value of 29.2$\pm$0.5\,days. \label{fig:TESSKeplerPhotometry} }
\end{figure*}

The Hubble Space Telescope (\HST{}) Stellar Treasure Trove program\footnote{\url{https://hst-stt.github.io/}} (HST AR 17551; PI: Rackham) is an initiative to systematically analyze archival \HST{} transit observations to characterize the surface heterogeneities of exoplanet host stars and quantify their impact on transmission spectra. This program is targeting a sample of exoplanet host stars spanning spectral types from F to late M, covering a broad range of stellar temperatures, metallicities, and activity levels. A pilot analysis of HD\,189733 \citep{Narrett2024} (henceforth \citetalias{Narrett2024}) demonstrated that stellar component temperatures and covering fractions can be robustly extracted from \HST{} \wfcthree/G141 spectra.
These spectral inferences point to larger spot filling factors than previously estimated for this active K dwarf, motivating the extension of this approach to other active planet-hosting stars of high interest for transmission spectroscopy.

Among such compelling targets is the \object{HAT-P-11} system. With a favorable brightness ($V = 9.5$), \hp{} is a chromospherically active K4 dwarf (log $R^{'}_{HK}$ $\approx$ -4.35) hosting a misaligned transiting Neptune-sized planet, \hp{}\,b, on a short-period (4.88\,d), eccentric ($e \approx 0.26$) orbit \citep{bakos2010, ojeda2011, morris2017}. Early transmission spectroscopy efforts reported water absorption in the atmosphere of \hp{}\,b \citep{fraine2014}, later confirmed with additional \HST{} observations \citep{chachan2019, cubillos2022}. More recent high-resolution ground-based spectroscopy independently supports these detections, reporting additional molecular signatures of \ce{NH3} and \ce{CO2} \citep{basilicata2024}. Helium absorption from the planet’s extended exosphere has also been detected \citep{mansfield2018}.

Interpreting these signals, however, is complicated by the star's active and variable surface. 
\hp{}'s spots are unusually well characterized, given the fortuitous combination of the planet's high obliquity and four years of high-precision \Kepler{} photometry showing multiple clear spot crossings.
While previous analyses have focused on in-transit data---including spot-crossing anomalies \citep{morris2017} and the planetary transmission spectrum \citep{fraine2014, chachan2019}---to understand the spottedness of \hp{} and its impact on observations of the planetary atmosphere, none has systematically probed what can be gleaned from the out-of-transit spectra themselves.

Here we leverage the extensive database of \HST{} stellar spectra to directly constrain the disk-integrated photospheric heterogeneity of \hp{} and evaluate its implications for transmission spectroscopy. 
We develop a robust framework for characterizing stellar heterogeneity in multi-epoch, multi-instrument \HST{} datasets. 
We place our results in context by comparing them to independent estimates of spot properties derived from photometric monitoring \citep{rackham2019}, spot-crossing analyses \citep{morris2017}, molecular band modeling \citep{morris2019} and chromospheric activity estimates \citep{morris2017CaHK, morris2025, yee2018} spanning more than a decade of observations.

This paper is organized as follows. In \autoref{sec:data}, we describe the multi-epoch and multi-instrument \HST{} observations, including our reduction framework to extract these spectra, and complementary datasets used in our analysis. In \autoref{sec:model}, we present our stellar modeling and parameter fitting framework. We summarize the results of our spectral analysis in \autoref{sec:results}, including constraints on spot temperatures, filling factors, and temporal variability. In \autoref{sec:discussion}, we contextualize our results using \Kepler{} and \TESS{} photometric data and chromospheric activity indicators for \hp{}, and compare them with previous studies of this system and other active K dwarfs. The implications of the stellar contamination are discussed in detail in \autoref{sec:ImplicationTransmSpectroscopy}. Finally, we summarize our main conclusions in \autoref{sec:conclusions}.

\section{Observations and Data Reduction}
\label{sec:data}

\begin{deluxetable*}{lccccccccc}[!ht]
\tablecaption{\label{tab:proposalID}\textit{Hubble} Observations of HAT-P-11}
\tablehead{
\colhead{Date} &
\colhead{Instrument} & 
\colhead{Disperser} &
\colhead{Proposal ID} &
\colhead{PI} & 
\colhead{Bandpass [\AA]} & 
\colhead{Exposures} & 
\colhead{Orbits} &
\colhead{Exp. Time [s]} & 
\colhead{Reference}
}
\startdata
2012-10-18 & WFC3 & G141 & 12449 &  D. Deming &  11450--16575 & 113 & 4 &  44.357 & F14  \\ 
2016-09-14 & WFC3 & G102 & 14793 & J. Bean &  8100--11350 & 99 & 4 & 81.089 & M18\\
2016-10-13 & WFC3 & G102 & 14793 & J. Bean &  8100--11350 & 99 & 4 & 81.089 & M18\\
2016-11-07 & WFC3 & G102 & 14793 & J. Bean &  8100--11350 & 99 & 4 & 81.089 & M18\\
2016-11-26 & WFC3 & G102 & 14793 & J. Bean &  8100--11350 & 99 & 4 & 81.089 & M18\\
2016-12-26 & WFC3 & G102 & 14793 & J. Bean &  8100--11350 & 99 & 4 & 81.089 & M18\\
\hline
2017-02-22 & STIS & G430L & 14767 & D. Sing & 3224--5678 & 82& 5 & 140.0 & Y19\\
2017-04-12 & STIS & G750L & 14767 &  D. Sing & 5305--10153 & 81 & 5 & 140.0 & Y19 \\
2017-05-26 & STIS & G430L & 14767 & D. Sing & 3224--5678 & 81 & 5 & 140.0 & Y19 \\
\enddata
\tablecomments{F14 -- \citet{fraine2014}; M18 -- \citet{mansfield2018}; Y19 -- \citet{chachan2019}.}
\end{deluxetable*}

Discovered via a ground-based transit survey \citep{bakos2010}, \hp{} has since been observed extensively with both space- and ground-based facilities, including \HST{}, \Spitzer, and \JWST{} (see \autoref{fig:TESSKeplerPhotometry}). 
In this work, we focus on the rich archive of \HST{} spectroscopy.
Here, we outline our methodology for extracting stellar spectra from these observations and summarize the additional datasets used in our analysis.

\subsection{\textit{HST} Spectroscopy}
\hp{} has been observed with \HST{} over multiple epochs using different instruments and grisms. These observations span 0.3--1.7\,$\micron$ using \stis{} and \wfcthree, each with two dispersers, as listed in \autoref{tab:proposalID}. A key advantage of \HST{} is its well-calibrated response function, which enables precise and accurate absolutely calibrated flux measurements (e.g., \citealt{wakeford2019}; \citealt{Garcia2022}; \citetalias{Narrett2024}). For each disperser, we conservatively restrict the wavelength range to exclude the edges of the transmissivity curve (see \autoref{tab:proposalID}). In the following, we describe the data reduction procedures employed for each instrument. 

\subsubsection{\wfcthree{} G102 and G141 Observations}
\HST{} observed \hp{} twice with \wfcthree{}/G141. One visit (2012-12-15) is excluded from both previous analyses \citep{chachan2019} and our analysis due to pronounced systematics. Additionally, five \wfcthree/G102 visits occurred over the course of three months in 2016, all of which are included in our analysis. Together, these data span a wavelength range of 0.8--1.6\,$\micron$ and have a spectral resolving power of $R{\sim}100{-}200$.

We extracted spectra for both G102 and G141 grisms using the \texttt{PACMAN} pipeline \citep{PACMAN}. Our procedure closely follows the methodology of \citetalias{Narrett2024}. The reduction yields a spectroscopic time series in terms of photon counts following background subtraction, dark correction, and cosmic ray corrections. To convert these into absolutely calibrated fluxes, we use the \HST{} sensitivity curves\footnote{\href{https://www.stsci.edu/hst/instrumentation/wfc3/documentation/grism-resources/wfc3-g102-calibrations}{G102 Sensitivity Curve} \& \href{https://www.stsci.edu/hst/instrumentation/wfc3/documentation/grism-resources/wfc3-g141-calibrations}{G141 Sensitivity Curve}} and account for the stellar radius and distance as well as the instrumental resolution and exposure time. 

Spectral extraction with \texttt{PACMAN} employs both box summation and optimal extraction techniques.  In typical fashion, \texttt{PACMAN} calibrates wavelength by the source position in the direct image. When extracting the spectra, telescope pointing drift can introduce minor errors in the wavelength calibration solution. 
In \wfcthree{}, each pixel corresponds to $\sim$25\,\AA, so small shifts of a few pixels may influence the wavelength calibration, though they typically do not exceed $\sim$100\,\AA.
To mitigate this, we applied a post-processing step to correct the wavelength calibration by comparing the observed photon counts with the expected flux count from a \phoenix{} NewEra model \citep{hauschildt2025}. This wavelength adjustment is performed at the count level, prior to the absolute flux conversion, which involves division by the wavelength-dependent sensitivity curve.

\wfcthree{} time-series data are also affected by charge trapping, which produces ramp-like systematics in the light curves \citep{zhou2017}. We model these effects using a parametric double-exponential model \citep{deWit2018}, which we then use to normalize the spectroscopic time series. The stellar spectra are extracted by computing the median of the out-of-transit exposures after masking the in-transit data. Uncertainties are estimated from the scatter in the spectroscopic time series and are generally larger than the formal Poisson noise. When combining multi-epoch observations for each instrument, the spectra are supersampled and interpolated onto a common wavelength grid, with uncertainties re-estimated from the scatter. 

\begin{deluxetable}{lcccc}
\tablecaption{\label{table:SystemParameters} System Properties of HAT-P-11}
\tablehead{
\colhead{Property} &
\colhead{Units} &
\colhead{Value} &
\colhead{Reference}
}
\startdata
TIC Identifier& - & 28230919 & S19\\
T$_{\rm eff}$ & K &  4778$^{+118}_{-107}$ & M17 \\
Fe/H & dex & 0.31$\pm$0.05 & B10\\
M  & M$_{\odot}$ & 0.770$^{+0.100}_{-0.076}$ & B10 \\
R  & R$_{\odot}$ & 0.760$^{+0.045}_{-0.051}$ & B10  \\
$\rho$ & gm/cm$^3$  & 2.474$^{+0.641}_{-0.590}$ & B10 \\
$\log$ g  & gm/cm$^3$ & 4.563$^{+0.093}_{-0.081}$ & S19 \\
Parallax & arcsec & 7.044$\pm$0.010 & G23 \\
Distance & pc & 141.965 $\pm$ 0.201 & G23 \\
Rotation Period & days & 29.2 & S11 \\
Spin-Orbital [$\Psi$] & deg & 106$^{+15}_{-11}$ & S11  \\
\hline
\multicolumn{3}{c}{Photometric Values}\\
\hline
B & & 10.556$\pm$0.141 & S19 \\
V & & 9.46$\pm$0.03 & S19 \\
J (2 MASS) & &7.608$\pm$0.029 & S06\\
H (2 MASS) & &7.131$\pm$0.021 & S06\\
K (2 MASS)& &7.009$\pm$0.020 & S06\\
\hline
\enddata
\tablecomments{ S19 -- \citet{stassun2019} $\mid$ M17 -- \citet{morris2017} $\mid$ B10 --   \citet{bakos2010} $\mid$ G23 -- \citet{gaiaDR3} $\mid$ S11 --  \citet{ojeda2011} $\mid$ S06 -- \citet{skrutskie2006} }
\end{deluxetable}

\subsubsection{\stis{} G430L and G750L Observations}
\stis{} observations of \hp{} were obtained using the G430L and G750L gratings. Two visits were gathered with G430L and one with G750L. Together, these data span a wavelength range of 0.30--1.03\,$\micron$ and have a spectral resolving power of $R{\sim}500{-}1000$. 
We processed all \stis{} observations using the \texttt{CALSTIS} pipeline, which produces 1D flux-calibrated spectra for each orbit.
The pipeline applies standard flat-fielding, wavelength calibration, and cosmic ray rejection steps. The G750L data show pronounced fringing at wavelengths ${>}7000$\,\AA. We mitigated this using contemporaneous flat field exposures and the \texttt{DEFRINGE} functionality within \texttt{CALSTIS} \citep{goudfrooij1998_Second}. 

Uncertainties are typically on the order of 1\%, estimated from visit-to-visit scatter whenever possible. Additional systematics may arise from several sources, including the instrument’s sensitivity drift, detector effects, and persistence. 
To evaluate if such effects impact our dataset, we compare overlapping wavelength regions among G430L, G750L, and G102 observations (see \autoref{fig:combinedSpectra}). Once the fluxes are appropriately scaled (which does not affect our astrophysical inferences, as shown in \citetalias{Narrett2024}),  and the resolution is matched, the continua among G430L, G750L, and G102 agree within $\sim$1\%, demonstrating that our typical uncertainties of 2\% adequately capture the typical noise sources.

\subsection{Supporting Data Sets}

\subsubsection{Photometry}
In addition to the \HST{} spectra, we incorporate calibrated broadband SED measurements in \textit{B} and \textit{V} bands \citep{stassun2019} as well as 2MASS \textit{J}, \textit{K} \& \textit{H} bands \citep{skrutskie2006}, together with the \Gaia{} DR3 parallax (\autoref{table:SystemParameters}). These measurements anchor the absolute stellar flux and provide additional constraints on global stellar parameters. Since these broadband data are well calibrated, we include them alongside the \HST{} spectra in our analysis without inflating their uncertainties or introducing vertical offsets.

Similarly, photometric timeseries probe stellar activity by showing the changes in the emergent surface flux \citep{shapiro2017}. \hp{} benefits particularly from extensive and precise long-term space-based photometry from \Kepler{} \citep{Borucki2010} and \TESS{} \citep{Ricker2015}. \Kepler{} observed \hp{} in short cadence mode for roughly 4 years from 2009 to 2013, while \TESS{} observed it across  11 sectors, which includes fast cadence mode of 20 seconds for the last 9 sectors (see \autoref{fig:TESSKeplerPhotometry}).

\begin{figure}[t!]
\centering
\includegraphics[width=\columnwidth]{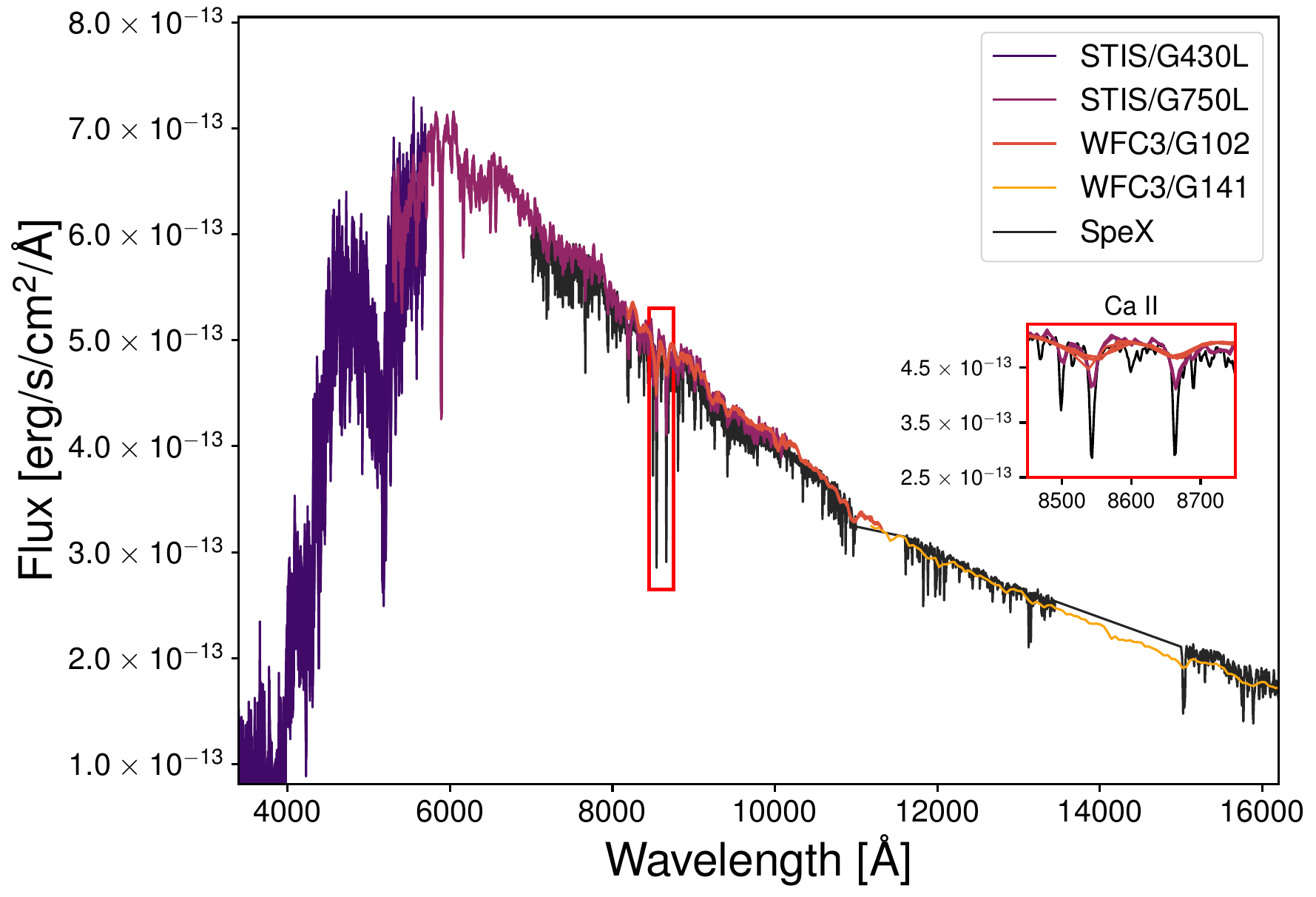}
\caption{Optical and near-infrared spectra of HAT-P-11. 
The \HST{} spectra of \hp{} were obtained with \stis{} (G430L, two visits; G750L, one visit) and \wfcthree{} (G102, five visits; G141, one visit), spanning 0.3--1.7\,$\micron$.
We show these data alongside an IRTF/\spex{} spectrum ($R{\sim}2000$ of \hp{}, which better resolves the \caii{} infrared triplet.
The \caii{} region is highlighted in the main panel and shown in the inset to illustrate the effect of spectral resolution in G750L, G102, and \spex{} data. 
All spectra are vertically scaled to match the \stis/G750L continuum level. 
After accounting for differences in resolution and wavelength sampling, the spectra from the various \HST{} modes are consistent at the 1--2\% level. \label{fig:combinedSpectra}}
\end{figure}

\subsubsection{IRTF/SpeX} 
We observed \hp{} on 21\,Jul\,2025 with the \spex{} spectrograph \citep{rayner2003} on the 3.2-m NASA Infrared Telescope Facility (IRTF).
Spectra were gathered using the short-wavelength cross-dispersed (SXD) mode and the $0\farcs3{\times}15''$ slit aligned to the parallactic angle, yielding 0.70--2.42\,$\micron$ coverage with a spectral resolving power of $R{\sim}2000$. Conditions were clear with seeing of $0\farcs7$. We collected six 60-s exposures in an ABBAB nodding pattern at an airmass of 1.2, followed by the standard SXD calibration sequence and observations of an A0\,V standard at a similar airmass. The data were reduced using Spextool V4.1 \citep{Cushing2004} with standard settings \citep[e.g.,][]{Delrez2022, Barkaoui2023, Ghachoui2023}. 
The final reduced spectrum has a median S/N of 400 per resolution element.

\section{Stellar Spectral Modeling}
\label{sec:model}

In this section, we describe our framework for modeling the disk-integrated stellar spectrum of \hp{}. 
We outline our parameterization of photospheric heterogeneity along with the adopted stellar atmosphere models, and our prior choices.
We then detail our fitting procedure, including a description of the likelihood function and inference methodology.

\subsection{Stellar Heterogeneity Model}
\label{sec:conceptStellarSuface}
The visible heterogeneous stellar surface can be represented as a linear combination of distinct spectral surface components, such as the quiet photosphere, cool spots, and hot faculae. In this work, we model the disk-integrated stellar flux as the sum of spectra from $N$ components with different effective temperatures:
\begin{equation}
F = \sum_{i=1}^{N} f_i F_i\,,
\end{equation}
where $F_i$ is the model spectrum of the $i^\mathrm{th}$ component and $f_i$ is its fractional coverage of the projected stellar disk (i.e., filling factor), subject to the normalization condition $\sum_{i}^{N} f_i = 1$.
All components are assumed to share $\log g$ and metallicity, differing only in effective temperature.

In this framework, discrepancies between observed spectra and a single-component model are interpreted as unresolved surface heterogeneity. 
We explore models with up to four surface components, each characterized by an independent temperature and covering fraction. 
We note that even when a single-component model is statistically preferred, multi-component fits remain informative.
In particular, two-component models provide an upper bound on the total spottedness compatible with the data, quantifying how much heterogeneous surface coverage can be tolerated before the model becomes inconsistent with observations.

Following \citetalias{Narrett2024}, we impose an ordering condition on the filling factors to prevent label degeneracies,
\begin{equation}
    f_{1}>f_{2}>\cdots>f_{N}\,, 
\end{equation}
and adopt a Dirichlet prior to enforce the normalization constraint. 
While this prior does not directly constrain the component temperatures, we typically expect the dominant component ($F_1$, parameterized by $T_1$ and $f_1$) to correspond to the quiescent photosphere, with additional components representing cooler spots and/or hotter faculae at smaller filling factors. Models with up to four components are physically motivated by the expected presence of quiescent photosphere, umbra, penumbra, and faculae. For a more detailed discussion of this framework, please see Section~3 of \citetalias{Narrett2024}.

\subsection{Stellar Atmosphere Models}
For our spectral fitting, we adopt the \phoenix{} NewEra stellar atmosphere models \citep{hauschildt2025}.
This grid spans a wide range of effective temperatures (sampled at ${\sim}100$\,K intervals) and metallicities up to $+0.5$\,dex, with $\log g$ sampled in 0.5\,dex steps. Model access and interpolation are handled through the \texttt{speclib} toolkit\footnote{\href{https://github.com/brackham/speclib}{https://github.com/brackham/speclib}} \citep{SPECLIB, Rackham2024}, which enables bilinear interpolation in effective temperature, $\log g$, and metallicity to generate spectra at arbitrary parameter values.

We adopt the \phoenix{} NewEra model grid uniformly across all datasets to ensure consistency. While specialized starspot atmosphere models would be desirable \citep[e.g.,][]{smitha2025}, \phoenix{} provides the broad wavelength and parameter coverage required for our analysis.
Our framework is readily extensible to future spot spectral model grids as they become available.

\subsection{Parameterization and Prior Choices}
\label{sec:priors}

The literature values we adopt for stellar parameters are listed in \autoref{table:SystemParameters}. We impose Gaussian priors on the stellar distance and radius and a truncated normal distribution on metallicity (\texttt{Fe/H}) centered on the value reported by \citet{bakos2010}, with the truncation set by the limits of the stellar model grid. Because surface gravity is only weakly constrained by our low-resolution spectra, we adopt a fixed value of $\log g = 4.563$ from \citet{stassun2019}.

The stellar distance and radius enter the model as multiplicative flux scalings. For \HST{} spectra, we additionally allow for an independent instrument-specific scaling factor to account for small systematic offsets (e.g., forward vs.\ reverse spatial scans; \citetalias{Narrett2024}). 
Absolutely calibrated broad photometry is not assigned such a scaling term, thus providing an anchor for the \HST{} fluxes.

To ensure consistency across instruments, model spectra are sampled on a uniform 1\,{\AA} wavelength grid, sufficient to resolve the instrumental line-spread function. We empirically determine and fix the effective convolution kernel for each instrument using a broad sample of stellar spectra from the HST Stellar Treasure Trove program. Given that the instrumental resolution is stable across visits, fixing these kernels reduces computational overhead.

\subsubsection{Single-Mode Fitting}
\hp{} exhibits variability on both rotational (29.2\,d) and multi-year magnetic-cycle timescales, as traced by the $\mathcal{R}_{HK}^{'}$ index \citep{morris2017CaHK}. Different observing epochs therefore likely sample distinct activity states. 

We began by fitting each epoch independently, starting with the five G102 visits obtained over three months. 
These spectra include prominent \caii{} infrared triplet                lines, which serve as diagnostics of the star’s activity. Given the stellar rotation period, these visits probe different rotational phases and thus potentially different spot configurations and covering fractions. 

We then performed a fit where we combined all available epochs for a given disperser to increase the signal-to-noise ratio. 
While stacking increases precision, intrinsically time-variable regions (e.g., the \caii{} triplet; \autoref{fig:combinedSpectra}) show increased scatter and are therefore down-weighted in the fit. This fit enables us to constrain persistent levels of the spottedness across epochs.

\subsubsection{Multi-Mode Fitting}
Fits to single instrumental modes are susceptible to degeneracies, as models of differing complexity can reproduce limited spectral regions equally well. However, when extrapolated to broader wavelength coverage---particularly into the optical---the predictions from single- and multi-component models diverge substantially. This motivates simultaneous fitting across all available instrument modes. 

A complication is that the multi-instrument dataset spans multiple epochs during which \hp{} likely exhibited varying activity levels. Consequently, the combined fit should be interpreted as providing an effective, time-averaged description of the stellar surface rather than a snapshot of a single activity state.  

In the joint retrieval, we allow for independent flux-scaling and error-inflation terms for each instrument (\autoref{eqn:ErrorInflation}).
When comparing the model to the broadband photometry fluxes, we adopt the mean of the epoch-specific filling factors as an effective, time-averaged description of the stellar surface. We also tested fits that allowed filling factors to vary between instruments while fitting for a consistent spot temperature.

\subsection{Likelihood Function and Inference Method}
We perform parameter estimation using dynamic nested sampling as implemented in \textsc{dynesty} \citep{speagle2020}. The sampler is initialized with 1000 live points, using the \texttt{rslice} method with bounding method set to \texttt{multis} to efficiently explore the posterior distribution. 
The maximum number of iterations was set to 500,000 to ensure convergence and adequate exploration of degenerate regions of parameter space.

To account for unmodeled systematics or underestimated uncertainties, we introduce an error-inflation (``jitter") term in the likelihood function. The log-likelihood is given by:
\begin{equation}
   \ln~\mathcal{L} = -\frac{1}{2}\sum_{\lambda}\left(\frac{(F_{\rm Obs, \lambda}-F_{\rm Model, \lambda})^2}{\sigma_\lambda^2}  + \rm{ln}(2 \pi \sigma^2_\lambda) \right)
\end{equation}
where the total model variance is modeled as:
\begin{equation}
    \sigma_\lambda^2 = \sigma_{\rm Photon,\lambda}^2 + \sigma_{\rm Jitter,\lambda}^2. 
\end{equation}

Following previous studies (\citetalias{Narrett2024}; \citealt{Rackham2024}), we parameterize this additional source of error as:
\begin{equation}
   \sigma_{\rm Jitter, \lambda} = f_{\rm var} F_{\rm Model, \lambda}
   \label{eqn:ErrorInflation}
\end{equation}
in which $f_{\rm var}$ is a free multiplicative scaling factor that inflates the model-dependent uncertainties, effectively introducing a heteroscedastic noise model in which the variance scales with the model-flux. This formulation provides a parsimonious way to absorb additional uncertainties.

\begin{figure*}[ht!]
\centering
\includegraphics[width=0.99\textwidth ]{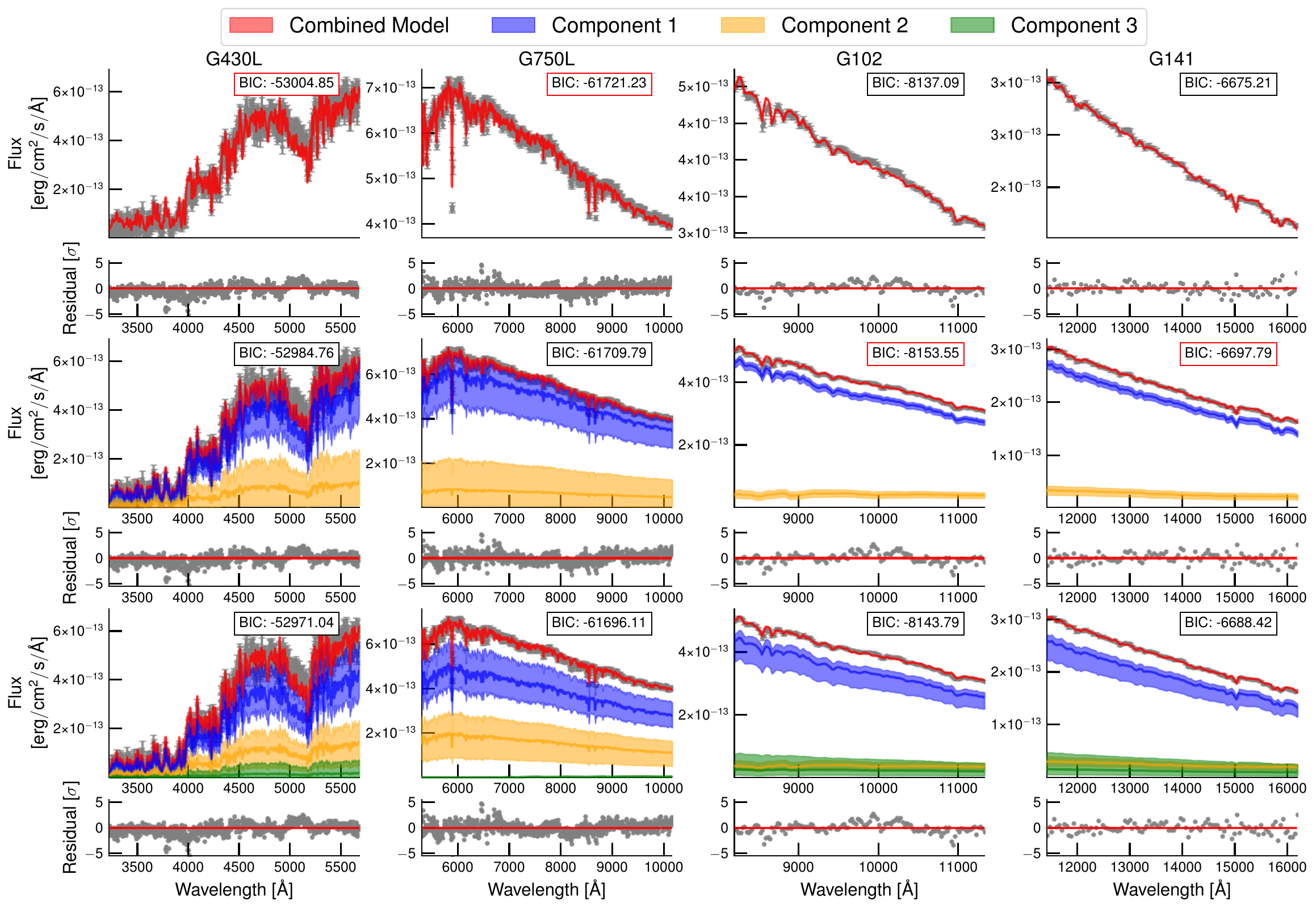}
\vspace{-5pt}
\caption{
Stellar spectral modeling of HAT-P-11 from blue to red wavelengths (\stis/G430L, \stis/G750L, \wfcthree/G102, and \wfcthree/G141).
Fits are performed on the averaged spectra for each mode, with uncertainties estimated from the data scatter.
In each pair of flux and residual panels, the top panel shows the observed spectrum (gray) and the best-fit combined model (red), with residuals displayed below. The first pair of rows shows the single-component model, the second pair the two-component model, and the third pair the three-component model. Contributions from the individual components are indicated by colored curves (blue, orange, and green), and their sum is shown in red. The Bayesian Information Criterion (BIC) for each fit is shown in the upper panels; the preferred model in each wavelength range is highlighted with a red box. \label{fig:multiComponentFit}}
\end{figure*}

\section{Results}
\label{sec:results}

Here, we present the major results of our analysis.  We first discuss the evidence for photospheric heterogeneity from the out-of-transit spectra of \hp. We then quantify the inferred spot temperatures and covering fractions, examine their temporal evolution, and finally assess the impact of multi-instrument-mode fitting and data-driven tests of spectral variability.

\begin{figure*}[ht!]
\centering
\includegraphics[width=\textwidth ]{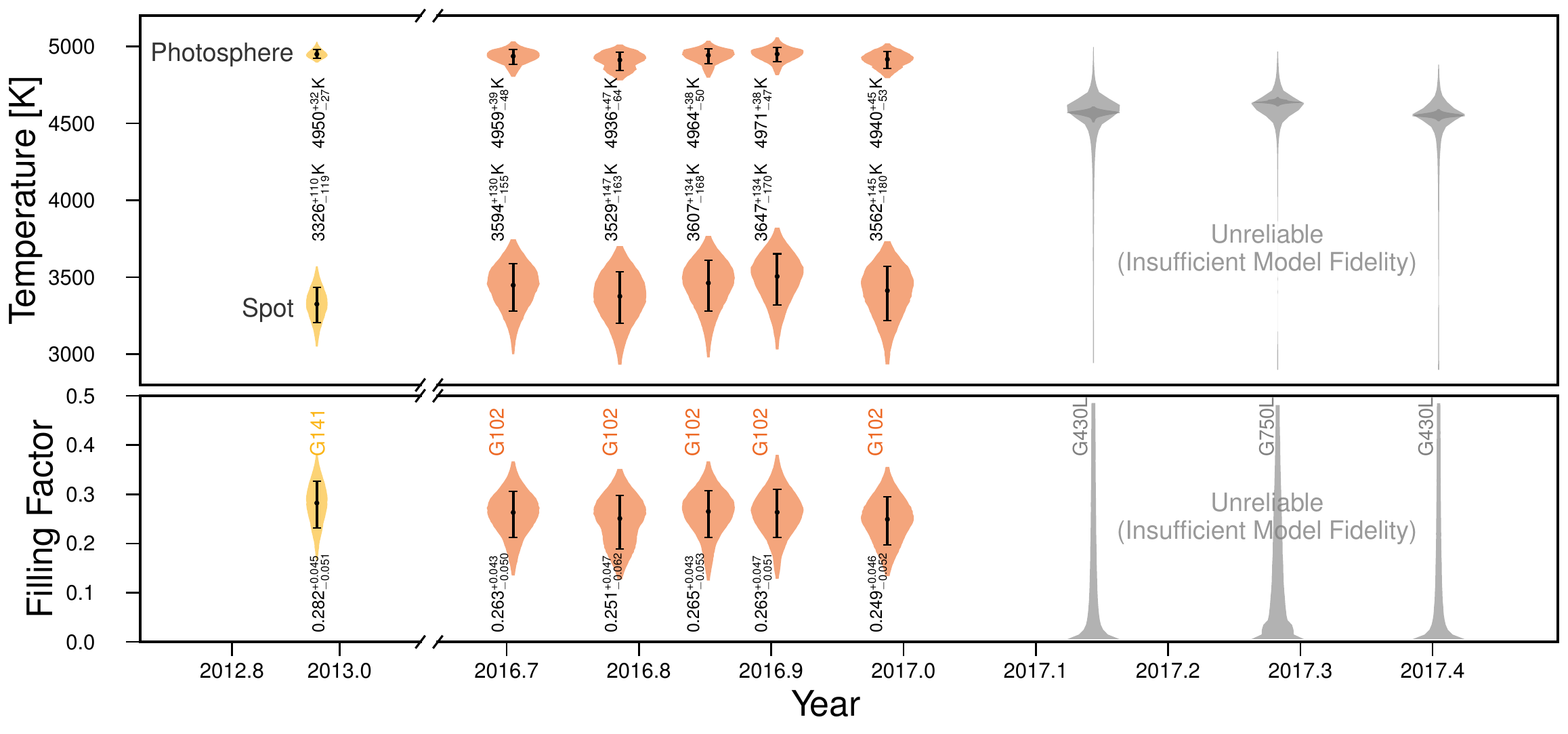}
\caption{Time series of stellar temperature components and spot filling factor from two-component spectral fits.
\textbf{Top:} Retrieved values of the photospheric temperature ($T_1$) and spot temperature ($T_2$) for each \HST{} epoch. 
For the \stis{} datasets (gray), the inferred photosphere and spot temperatures strongly overlap, indicating that the two-component model does not meaningfully separate distinct temperature components;
these results are not considered reliable due to significant residuals (see \autoref{sec:modelFidelity}) and are shown for completeness only.
\textbf{Bottom:} Corresponding inferences of the spot filling factor ($f_2$) as a function of epoch.  \label{fig:spotEvolution}}
\end{figure*}

\subsection{Evidence for Surface Heterogeneity}
Results from the single-instrument \HST{} spectra retrievals for up to three-component models are summarized in \autoref{table:NoPriors}. 
Based on Anderson--Darling (AD) statistics \citep{anderson1954}, we obtain statistically adequate fits for \wfcthree{} datasets (G102 and G141), but not for the \stis{} datasets (G430L and G750L). For G430L and G750L, AD values exceed 2, indicating departures from Gaussian residuals, whereas \wfcthree{} datasets yield AD values near or below 1, consistent with Gaussian noise at the 5\% level. This mismatch is also reflected in the large error-inflation factor required for G430L and G750L, which is more than two orders of magnitude higher than that of the \wfcthree{} datasets. The G430L retrieval additionally drives the metallicity to the upper prior bound (+0.50), in ${<}3\sigma$ tension with published estimates. 
We discuss these discrepancies further in \autoref{sec:modelFidelity}.

Model comparison using the Bayesian Information Criterion (BIC) show that the optical datasets (G430L and G750L) favor a single-component model, with the $\Delta\mathrm{BIC}$ value indicating little support for additional complexity. By contrast, both \wfcthree{} datasets (G102 and G141) strongly prefer a two-component model over a single-component model ($\Delta \mathrm{BIC}{\sim}15$ for G102 and $\Delta \mathrm{BIC}{\sim}22$ for G141). $\Delta$BIC values greater than 10 constitute strong evidence for the more complex model. Three-component models are not supported by the BIC for any dataset. Fits for each instrument mode, modeled up to three components, are shown in \autoref{fig:multiComponentFit}.

\subsection{Spot Temperatures and Covering Fractions}
For the combined G102 dataset, we infer a photospheric temperature of $T_1 = 4966^{+33}_{-41}$\,K, a spot temperature of  $T_2 = 3617^{+125}_{-149}$\,K, and a spot filling factor of $f_2 = 0.264^{+0.038}_{-0.043}$. 
For G141, we obtain $T_1 = 4950^{+32}_{-27}$\,K, $T_2 = 3326^{+110}_{-119}$\,K, and $f_2 = 0.282^{+0.045}_{-0.051}$. 
The retrieved photospheric temperatures are consistent between the two bandpasses and exceed literature effective temperature estimates, as would be expected if the literature values were averaging over a highly spotted photosphere. 
The retrieved spot temperatures differ modestly ($1.6\sigma$) between G102 and G141, and both datasets indicate substantial spot coverage at the ${\sim}25{-}30\%$ level. 

Although individual components in multi-component model are degenerate, the composite spectra are tightly constrained (\autoref{fig:multiComponentFit}), and the consistency between independent \wfcthree{} modes strengthens the case for significant photospheric heterogeneity.

\subsection{Temporal Evolution of Spot Properties}
\label{sec:temporalEvolution}
Restricting ourselves to the \wfcthree{} results, we next assess epoch-specific fits to assess temporal variability.
The G141 visit in late 2012 favors a two-component model with a photospheric temperature of $T_1 = 4950^{+32}_{-27}$\,K and a  cooler spot component of $T_2 = 3300^{+110}_{-119}$\,K, with a filling factor of $f_2 = 28.2^{+4.5}_{-5.1}\%$. 
Four years later, the five G102 observations in late 2016 similarly favor two-component solutions, with photospheric temperatures clustered near $\sim$4950\,K and spot temperatures between $\sim$3529--3647\,K (\autoref{fig:spotEvolution}). 
Across the roughly three-month sequence, the inferred filling factors remain remarkably stable ($\sim$25--27\%), suggesting persistent and substantial spot coverage rather than transient spot emergence.
The primary variability appears to be modest fluctuations in covering fraction (1--2\%) rather than large changes in spot temperature, validating our approach for the multi-epoch fits.

\subsection{Multi-Mode Fits}
\label{sec:multiInstrumentFit}
We again exclude the \stis{} datasets from the joint fits because the available photospheric models do not adequately reproduce those spectral regions and do not yield a reliable spot constraint (see \autoref{sec:modelFidelity}). 
We therefore perform joint fits using G102 and G141 only, allowing for independent filling factors but a common spot temperature.

The combined analysis strengthens the preference for a two-component model and does not support the addition of a third component.
A joint fit to G102+G141, assuming a common spot temperature but independent filling factors, yields tighter constraints:
$T_1\,{=}\,4979^{+20}_{-24}$\,K
and
$T_2\,{=}\,3412^{+77}_{-73}$\,K.
The inferred filling factors are
$f_{2,\mathrm{G102}} = 0.259^{+0.020}_{-0.023}$
and
$f_{2,\mathrm{G141}} = 0.331^{+0.041}_{-0.046}$.
As expected, the combined analysis sharpens the spot temperature constraint, with the joint value lying between those obtained from the individual fits.
The filling factors are consistent with the individual fits as well, reinforcing the inference of a persistent large areal coverage of spots contributing to the integrated disk spectra.

\subsection{Data-Driven Spectral Comparison}
\label{sec:dataDriven}

\begin{figure}[t!]
\centering
\includegraphics[width=\columnwidth]{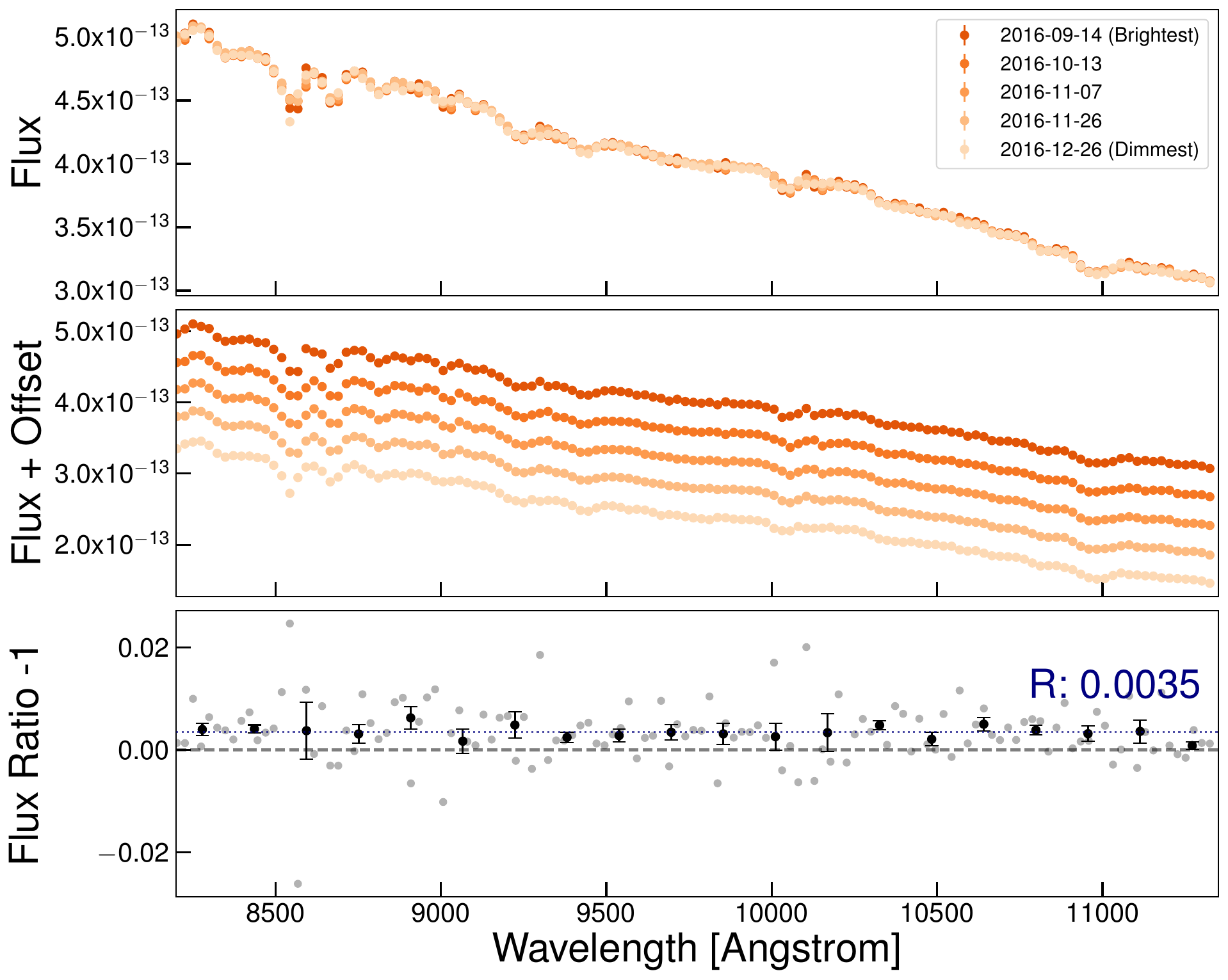}
\caption{
Comparison of five \wfcthree{}/G102 spectra obtained at different epochs.
\textbf{Top:} Extracted stellar flux for each epoch shown without scaling, demonstrating consistency in flux levels and spectral shape. 
\textbf{Middle:} The same spectra, vertically offset for clarity. 
\textbf{Bottom:} The ratio of the brightest epoch to the dimmest. 
The brightest spectrum is elevated by 0.35\% relative to the faintest, with no wavelength-dependent structure.
This uniform offset indicates instrumental stability and suggests that the stellar inhomogeneities remain relatively stable across five visits.
\label{fig:dataDriven}}
\end{figure}

We additionally tested for intrinsic spectral variability of \hp{} using the five \textit{HST}/\wfcthree{} G102 visits obtained over three months. To maximize sensitivity to potential changes, we compared spectra from the epochs with the highest and lowest median out-of-transit fluxes. For each visit, we computed median out-of-transit fluxes (\autoref{fig:dataDriven}) and identified the brightest and faintest epochs. We then constructed two median spectra by combining the five highest-flux exposures for the brightest epoch and the five lowest-flux exposures for the faintest epoch. This approach suppresses noise and reduces the impact of transient artifacts such as cosmic rays.

The resulting median spectra are nearly identical in spectral shape across the full G102 bandpass (\autoref{fig:dataDriven}). Although \hp{} exhibits $\sim$1\% flux modulation across these visits, no obvious wavelength-dependent structure is apparent at the precision of these data. The variability therefore appears largely achromatic, consistent with changes in surface coverage of features rather than large temperature-driven spectral differences. This interpretation aligns with our epoch-by-epoch retrievals (\autoref{sec:temporalEvolution}), which show stable photospheric and spot temperatures and attribute most of the variability to changes in filling factors.

The strongest absorption features in the \wfcthree{} spectra of \hp{} are the strong \caii{} infrared triplet absorption features at 8498\,\AA, 8542\,\AA, and 8662\,\AA, well-known diagnostics of stellar activity \citep{andretta2005}. These lines exhibited the largest apparent visit-to-visit differences (see \autoref{fig:combinedSpectra}). However, further analysis indicated that much of this variability arose from differences in spectral sampling between visits. The combined \caii{} equivalent widths show no significant mutual correlations or correlations with retrieved stellar surface properties within the measurement uncertainties. Similarly, the 2025 \spex{} spectrum which covers the \caii{} infrared triplet region (\autoref{fig:combinedSpectra}) at a resolution of $\sim$2000 does not show obvious signs of emission features, which sometimes are seen in active stars.

\section{Discussion}
\label{sec:discussion}
Our results reveal a picture of \hp{} as a highly spotted K dwarf whose near-infrared spectra require multiple components to model.
At the same time, the optical \stis{} spectra do not support multi-component models. Reconciling this discrepancy is essential both for interpreting the stellar surface properties and for assessing how stellar contamination affects the transmission spectrum of \hp{}\,b. 
We therefore first examine why \texttt{STIS} disfavors two-component models, then reassess the \wfcthree{} evidence for multi-component photospheres, place our results in the context of previous studies, and investigate other evidence of \hp{}'s stellar activity and its long-term evolution.

\subsection{Why Do \stis{} Results Not Favor Two-Component Models?}
\label{sec:modelFidelity}

The \stis{} spectra show no statistical preference for multi-component photospheric models.
This result should be interpreted in the broader context of wavelength-dependent model fidelity and the challenges associated with modeling contamination from cooler photospheric components \citep{Iyer2020, Rackham2024}. Because this work represents one of the first attempts to use wide-bandpass, absolutely calibrated spectra to probe stellar surface inhomogeneities, the inferred spot properties naturally depend on the underlying stellar atmosphere models employed.

\begin{figure}[t!]
\centering
\includegraphics[width=1\linewidth ]{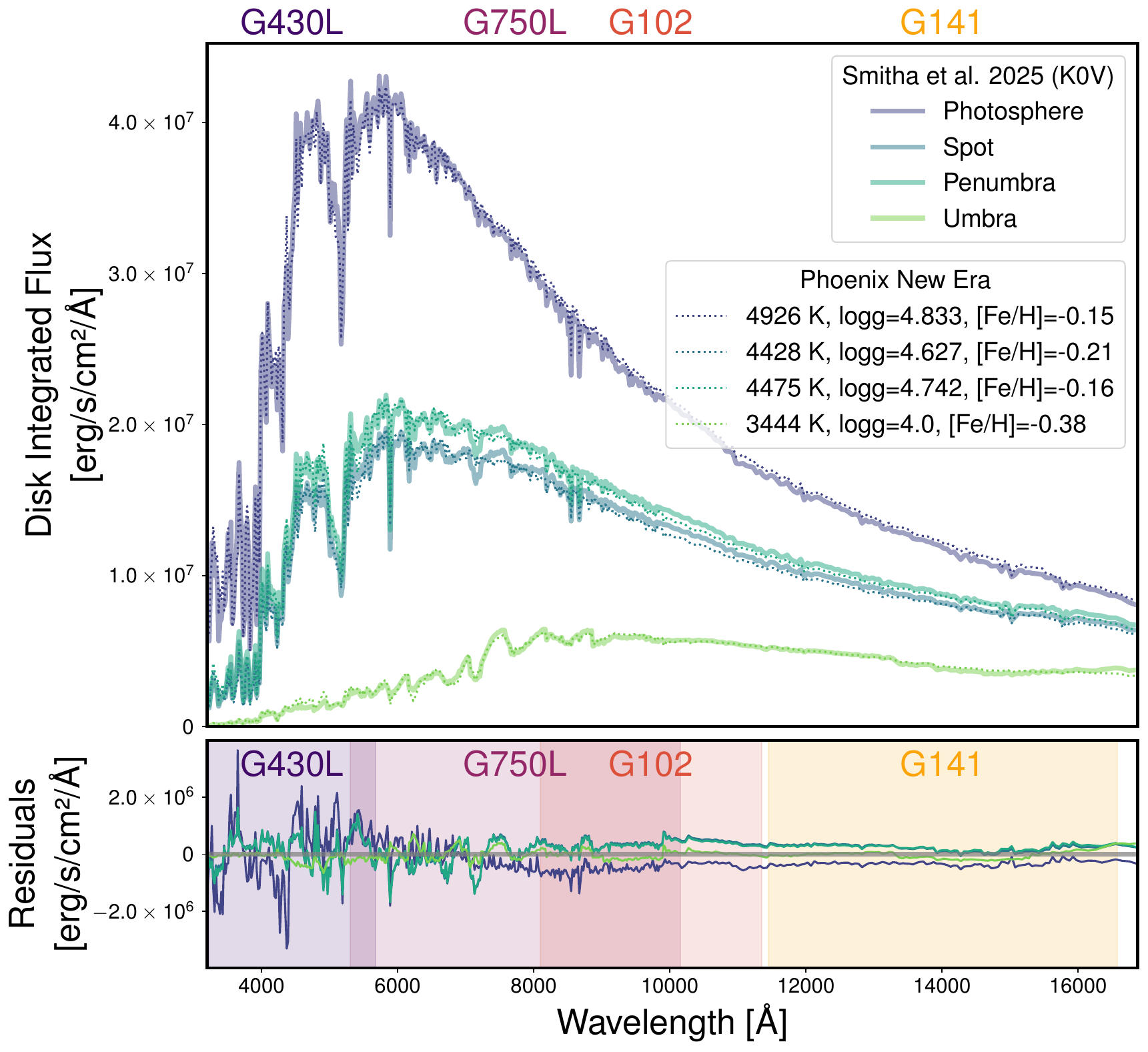}
\caption{Comparison between the first-principles-based models of \citet{smitha2025} and the best-fitting one-component stellar model from \texttt{PHOENIX} NewEra grid. 
\textbf{Top:} Disk-integrated flux spectra for the photosphere, spot, penumbra, and umbra components \citep[][solid lines]{smitha2025} alongside the corresponding PHOENIX models (dotted lines).
Agreement between models is strongest within the G102 and G141 bands, whereas larger discrepancies arise in the G430L and G750L regions. 
\textbf{Bottom:} Residuals between the two models.
The wavelength ranges of the \HST{} dispersers are shaded for reference.
\label{fig:spotModelComparison}
}
\end{figure}

To assess how well the \phoenix{} NewEra grid models reproduce realistic stellar atmospheres, we compared them against first-principles simulations of active-region spectral components for a K0 dwarf from \citet{smitha2025}, which is the closest available analog to \hp{}. In \autoref{fig:spotModelComparison}, the top panel shows the disk-integrated fluxes for the photosphere, spot, penumbra, and umbra components from the \citet{smitha2025} model (solid curves) and the corresponding \phoenix{} models (dotted curves). Overall, the models reproduce the large-scale continuum shape and the temperature ordering of the components reasonably well, confirming that the grid captures the broad spectral energy distribution (SED) of K dwarfs and their cooler surface components.

However, the residuals reveal wavelength-dependent discrepancies in the models, especially in the optical. The discrepancies are largest in the G430L region (roughly 3000--5700\,\AA), with secondary tension extending into the G750L region (5500--10000\,\AA). These tensions are not limited to reproducing spot spectra with \phoenix{} models. In fact, the differences between the photospheric models similarly show discrepancies in this wavelength regime exceeding the 2\% precision of the \stis{} data. These deviations arise because the optical regime is dominated by dense atomic and molecular line blanketing, which  suppresses the continuum, distorts line depths, and complicates the comparison of the pseudo-continuum between data and models. In addition, uncertain or incomplete line lists and NLTE effects disproportionately affect this regime. Such discrepancies between the model and the data have already been highlighted for M-dwarfs \citep[see their Figure 5]{iyer2023, iyer2025, wilson2025}, and we find that they likely extend to K dwarfs as well.

As a result, \stis{} data cannot robustly discriminate between one- and two-component models with the current stellar models. In contrast, model fidelity is substantially better in the \wfcthree{} near-infrared bands, where the evidence for multi-component photospheres is strongest.

\subsection{Robust Evidence for Multi-Component Photospheres from \wfcthree{}}
\begin{deluxetable*}{rcccccccc}[t]
\tablecaption{\label{table:TwoComponentTable} Retrieved Parameters from Two-component Fits to HAT-P-11 \HST{} Spectra}
\tablehead{
\colhead{Filter} &
\colhead{$T_{\rm 1}$ (K)} &
\colhead{$T_{\rm 2}$ (K)} & 
\colhead{$f_{\rm 2}$} &
\colhead{$R_{\rm Star}$ (R$_{\odot}$)} &
\colhead{$a_{\rm scale}$} &
\colhead{$\ln f_{\rm var}$} 
}
\startdata
G430 [Not Preferred] & 4625$^{+17}_{-17}$ & 4621$^{+73}_{-116}$ &  0.170$^{+0.218}_{-0.148}$ &  0.793$^{+0.006}_{-0.006}$ &  0.962$^{+0.017}_{-0.017}$ & -2.724$^{+0.043}_{-0.042}$\\
G750 [Not Preferred] & 4685$^{+12}_{-8}$ & 4662$^{+63}_{-87}$  &  0.121$^{+0.201}_{-0.101}$ &  0.779$^{+0.005}_{-0.005}$ & 1.055$^{+0.014}_{-0.015}$ & -4.300$^{+0.023}_{-0.024}$ \\
\textbf{G102} [\textbf{Preferred}] & 4966$^{+33}_{-41}$ & 3617$^{+125}_{-149}$ &  0.264$^{+0.038}_{-0.043}$ &  0.782$^{+0.007}_{-0.007}$ & 0.991$^{+0.013}_{-0.013}$ & -4.710$^{+0.069}_{-0.069}$\\
\textbf{G141} [\textbf{Preferred}] & 4950$^{+32}_{-27}$ & 3326$^{+110}_{-119}$ &  0.282$^{+0.045}_{-0.051}$ &  0.802$^{+0.015}_{-0.016}$ & 0.989$^{+0.014}_{-0.013}$ & -4.722$^{+0.073}_{-0.070}$ \\
\hline
\textbf{G102 + G141} [\textbf{Preferred}] & 4979$^{+20}_{-24}$ & 3412$^{+77}_{-73}$ &  0.259$^{+0.020}_{-0.023}$$^{\dagger}$ &  0.800$^{+0.011}_{-0.010}$ & 0.954$^{+0.024}_{-0.023}$$^{\dagger}$ & -4.709$^{+0.068}_{-0.074}$$^{\dagger}$ \\
 & -  & -   &  0.331$^{+0.041}_{-0.046}$$^{\dagger}$ & - &  1.005$^{+0.019}_{-0.018}$$^{\dagger}$ &-4.478$^{+0.074}_{-0.065}$$^{\dagger}$ \\
\enddata
\tablecomments{$^{\dagger}$For the combined G102+G141 fit, the first row lists the G102 parameters and the second row the G141 ones.}
\end{deluxetable*}
The \wfcthree{}/G102 and G141 retrievals consistently and significantly favor two-component models for \hp{} reported in \autoref{table:TwoComponentTable}. This preference is not limited to a single epoch or instrumental configuration, but instead persists over a multi-year baseline, indicating that the result is not an artifact of instrument systematics or local time variability. Moreover, the overall agreement between model and data across the full \wfcthree{} wavelength range (\autoref{fig:spotModelComparison}) supports the robustness of these fits.

For both \wfcthree{} gratings, the preferred two-component models yield a second-component temperature ($T_2$) that is approximately 1500\,K lower than that of the first component ($T_1$), consistent with umbral-like temperatures (cf.\ the K0 dwarf simulations of \citealt{smitha2025}). The temperature of the first component is slightly higher than the stellar effective temperature, as expected if the effective temperature were averaging over unresolved spot contributions. 

Comparing the best-fitting one- and two-component models, the improvement of the two-component models is not driven by isolated spectral features but by subtle, broadband differences across the near-infrared bandpass. Although these differences are comparable in scale to the observational uncertainties, they are coherent across wavelength.
Extending the analysis to optical wavelengths could, in principle, help discriminate between the one- and two-component models; however, as discussed in \autoref{sec:modelFidelity}, current stellar spectra models are less reliable in that regime, limiting the robustness of such a comparison.

\subsection{Comparison with Previous Studies}

Several past studies have examined the surface inhomogeneities of \hp{} using precise \Kepler{} photometry \citet{ojeda2011, beky2014b, morris2017}. In particular, the comprehensive analysis by \citet{morris2017} identified the active latitudes near ${\sim}16^\circ$ and inferred that 0.5--10\% of the transit chord covered by spots from transit to transit, with a median estimate of $3^{+6}_{-1}\%$ \citep{morris2017}. Although the inferred spot covering fractions from spot crossings exhibit a long tail toward higher values, these chord-based covering fractions are still substantially smaller than our disk-integrated estimates of 25.9$^{+2.0}_{-2.3}$\% for G102 and 33.1$^{+4.0}_{-4.6}$\% for G141 (see \autoref{table:combinedRetrievalG102G141}). 

However, constraints on stellar surface inhomogeneities derived from spot-crossing events are subject to their own unique biases and degeneracies. First, spot crossings probe only the transit chords, which for \hp{} spans a broad range of latitudes but, of course, does not sample the full stellar surface in a given transit. Moreover, despite the exquisite precision of \textit{Kepler} short-cadence data (58.85 s), the photometry often lacks the resolution needed to capture substructure within starspots. As a result, modeling spots as single circular features on the stellar surface is subject to strong degeneracies between their size and position \citep{morris2017}. Second, if \hp{} hosts a large population of small spots (similar to the ``polka-dotted'' configuration proposed for TRAPPIST-1 \citep{Rathcke2025}), their signals may fall below the photometric detection threshold. Consequently, spot-covering fractions ($f_2$) inferred from spot-crossing events should be interpreted as lower limits, as they primarily trace large, high-contrast structures \citep[e.g.,][]{murray2025}.

Using a different approach, \citet{morris2019} analyzed TiO molecular bands in high-resolution optical spectra and inferred a filling factor of 15$\pm$ 6\%. Their method decomposed the spectra into contributions from photospheric and spot components but fixed the spot temperature (assuming a TiO-forming temperature near 3300\,K) and relied on a narrow wavelength range (7052--7062\,\AA), much smaller than the \wfcthree{} bandpasses used here. In addition, their retrieval imposed a \Gaia{} color prior that disfavors large filling factors. For stars with $\sim$20--30\% spot coverage and several percent temporal variability, the nominal \Gaia{} photometric uncertainty of 0.006 may underestimate the level of spot-induced modulation. Moreover, as discussed in \autoref{sec:modelFidelity}, stellar spectral models in the optical remain imperfect, warranting caution when interpreting constraints derived in that regime. We introduce a relatively weak constraint in our fit by incorporating $B$- and $V$-band fluxes in our fit, which have larger uncertainties than \Gaia{} measurements. While stronger priors could similarly suppress large filling factors in our retrievals, doing so would risk introducing biases.

Estimates of the spot temperature contrast also vary in the literature. Simultaneous \textit{Kepler}--\textit{Spitzer} spot-crossing analyses by \citet{fraine2014} yielded a spot--photosphere temperature contrast of $900\pm300$ K\footnote{Error estimated from their Figure~1.}. By contrast, multi-band optical photometry by \citet{schutte2023} suggested a substantially smaller contrast of $\sim$250\,K. The latter study relies on ground-based optical photometry, where the precision and systematics are often comparable to the amplitudes of spot-crossing signals themselves, potentially limiting the robustness of the inferred contrasts. Additionally, the apparent temperature contrast depends on which portion of the spot is occulted during transit, with umbral crossings producing larger apparent contrasts than penumbral crossings.

\subsection{Evidence for Long-Term Variability in Stellar Activity}
\label{sec:variablity}

\begin{figure*}[ht!]
\centering
\includegraphics[width=\textwidth ]{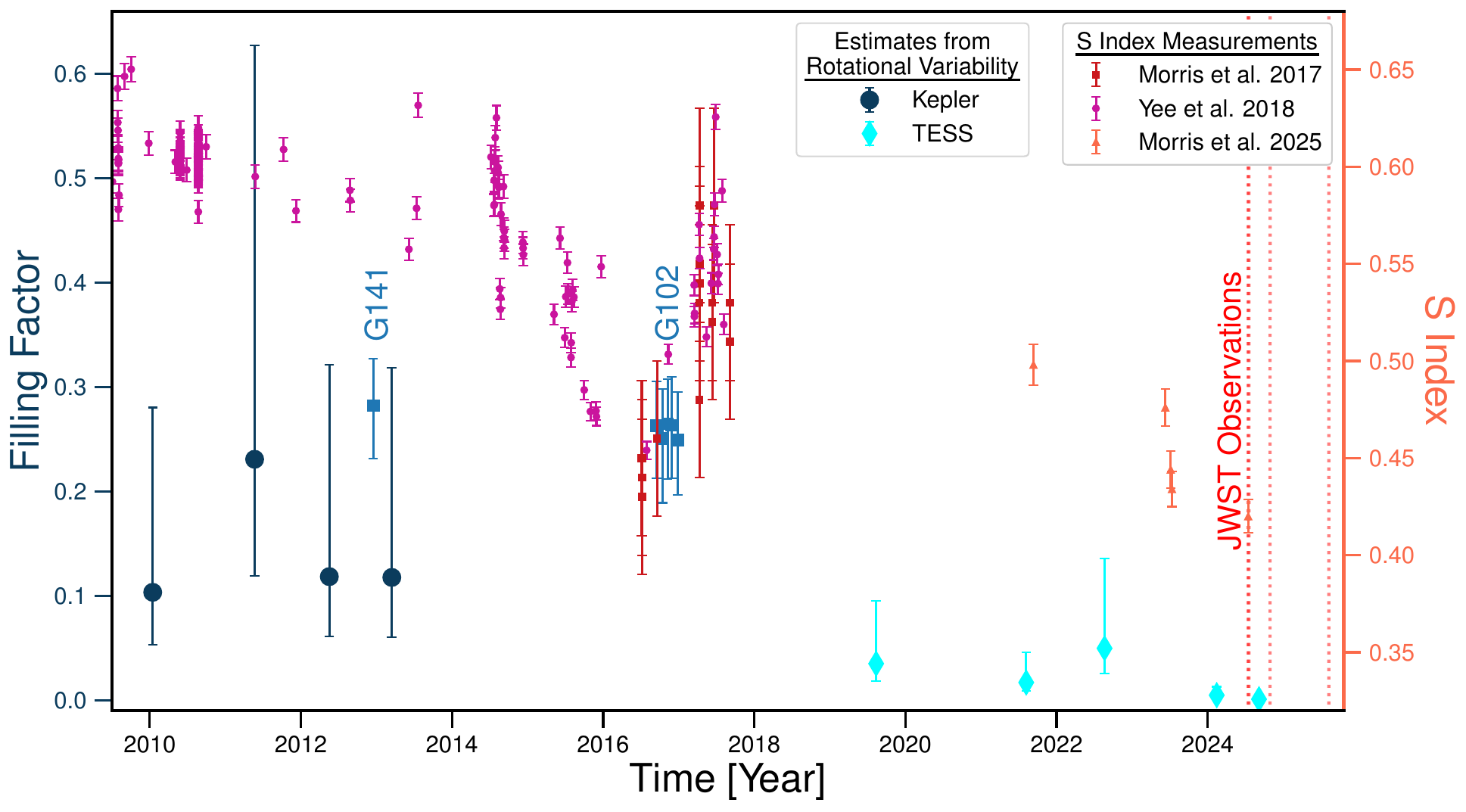}
\caption{ Spot filling factors of \hp{} as a function of time.
Filling factors derived from rotational variability in \Kepler{}  and \TESS{} light curves (using Equation 2 of \citealt{rackham2019}) are compared with values inferred from multi-component spectral fits to \HST{} G102 and G141 data. 
Independent chromospheric activity measurements (S-index values from the literature) are shown on the right axis.
The \Kepler{} era exhibits relatively high filling factors with substantial variability, whereas the \TESS{} epochs indicate a transition to markedly lower spot coverage. 
Red vertical dotted lines mark the epochs of \JWST{} observations. \label{fig:fillingFactorTimeSeries}}
\end{figure*}

Our spectral retrievals indicate different spot filling factors between the 2012 G141 and 2016 G102 observations (\autoref{fig:spotEvolution}), suggesting a possible secular decline in stellar activity over this period. Motivated by this trend, we examine independent activity diagnostics to assess whether they support a similar long-term evolution.

\subsubsection{Long-Term Evolution of the \caii{} H\&K index}

The strongest independent evidence for a magnetic activity cycle in \hp{} comes from long-term monitoring of the \caii{} H\&K S-index using Keck/HIRES and APO/ARCES spectroscopy \citep{morris2017CaHK}. These measurements show variability in both rotational and multiyear timescales. The data are consistent with a long-duration ($\geq$10 years) activity cycle in which \hp{} remained near an activity maximum during the Kepler era, declined toward a minimum around 2016, and began rising again by 2017.  This temporal evolution broadly mirrors our spectroscopic results: higher spot filling factors during the G141 epoch (near the \Kepler{} era) and lower values during the G102 observations in 2016 (see \autoref{fig:fillingFactorTimeSeries}). 

Given that \hp{} is a main-sequence K dwarf broadly similar to the Sun but substantially more active, we can also use empirical relations between chromospheric activity and spot coverage derived from solar observations to place our results in context. 
Applying the solar-calibrated relation of \citet[][Eq.~7]{sowmya2023} to the observed $S$-index measurements implies a spot coverage of $\sim$30--40\% during the activity maximum, in close agreement with our results. However, extrapolating such relations from the Sun to a cooler, more active K dwarf is non-trivial. Moreover, \caii{} H\&K emission predominantly traces chromospheric plage rather than photospheric spots \citep[e.g.,][]{Cretignier2024}; while plages and spots are correlated, the spot-to-faculae ratio is known to vary with activity level \citep{shapiro2014}, potentially biasing spot coverage estimates inferred from \caii{} H\&K.  

Recent $S$-index monitoring of \hp{} extends the \caii{} H\&K baseline and shows that the star approached one of its lowest measured activity levels around 2024 \citep{morris2025}. This low-activity phase coincides with the \JWST{} observations, suggesting those spectra were acquired near a relative minimum in stellar activity.

\subsubsection{Evidence from Photometric Time Series}
The four-year \Kepler{} light curve exhibits rotational modulation amplitudes ranging from ${\sim}1.44\%$ to ${\sim}2.15\%$, indicating temporal evolution of the stellar surface. \TESS{} extends the temporal baseline with eleven sectors of observation, albeit in a redder bandpass, with rotational amplitude ranging from $\sim$0.15\% to $\sim$0.97\%. 

Using the spot temperatures retrieved from our multi-component spectral fits, we estimate the wavelength-dependent spot contrast in the \Kepler{} and \TESS{} bandpasses. We obtain $\mathcal{C}_{\rm Kepler}$ = 0.933$\pm$0.011 and $\mathcal{C}_{\rm TESS}$ = 0.869$\pm$0.016, corresponding to a contrast ratio of 1.0735$\pm$0.0073 between the bands. For a K4 dwarf, this implies that only modest differences in rotational modulation amplitude are expected between the bluer \Kepler{} and redder \TESS{} bandpasses ($\sim$10\% changes).
The observed decline in modulation amplitude from \Kepler{} to \TESS{} (\autoref{fig:TESSKeplerPhotometry}), therefore, cannot be explained by bandpass effects alone.

To translate photometric amplitudes into spot covering fractions, we adopt the scaling relation of \citet[][Eq.~2]{rackham2019}, in which the rotational modulation amplitude scales approximately as the square root of the spot filling factor.  
We measure \Kepler{} rotational modulation amplitudes at the difference between the 5$^\mathrm{th}$ and 95$^\mathrm{th}$ percentile flux values within continuous observing windows separated by no more than 10\, days.
Using the contrast coefficient appropriate for K4 dwarfs ($C = 0.048 \pm 0.020$, defined as $F_\mathrm{spot}/F_\mathrm{phot}$ in that work), the peak-to-peak \Kepler{} amplitude of ${\sim}2\%$ implies a characteristic spot filling factor of order ${\sim}20\%$, albeit with a broad allowed range, given the uncertainty in $C$ driven by the intrinsic degeneracies of spot distributions. 
Nonetheless, this estimate is broadly consistent with our \wfcthree{} inferences.

Applying the same procedure to \TESS{} data reveals amplitudes of ${\sim}1\%$ in early sectors, decreasing to ${\sim}0.2\%$ in later sectors. 
Corresponding inferred spot filling factors are of the order of a few percent.
This photometric evolution is generally consistent with the behavior of the \caii{} H\&K index and supports the interpretation that the star entered a lower-activity state by 2024, before the \JWST{} observations.

\subsection{Forward Modeling of Rotational Modulation}

To further interpret the relationship between rotational modulation amplitude and spot coverage, we performed a suite of forward simulations using \texttt{spotter} \citep{Garcia2025}. We generated ensembles of synthetic stellar surfaces with randomly located spots and computed the resulting light curves. In this framework, each spot is modeled as a circular feature with a smooth, sigmoid-like edge rather than a hard boundary. We considered two characteristic spot sizes: small spots (angular radii $\leq$0.6$^\circ$) and medium spots ($0.6$--$3^\circ$), with sharpness drawn from a uniform distribution of $\mathcal{U}(10,50)$. Our simulated spot contrasts are representative of those expected in \Kepler{} bandpass ($\sim$0.93) based on our retrieved temperature. The resulting dispersion of the rotational amplitude is driven by the variability in spot configurations, which are an order of magnitude larger than the bandpass-dependent effects between \Kepler{} and \TESS. Thus, the same amplitude--filling factor relation remains applicable to \TESS{} data to a good approximation.

\begin{figure}[t]
\centering
\includegraphics[width=\columnwidth]{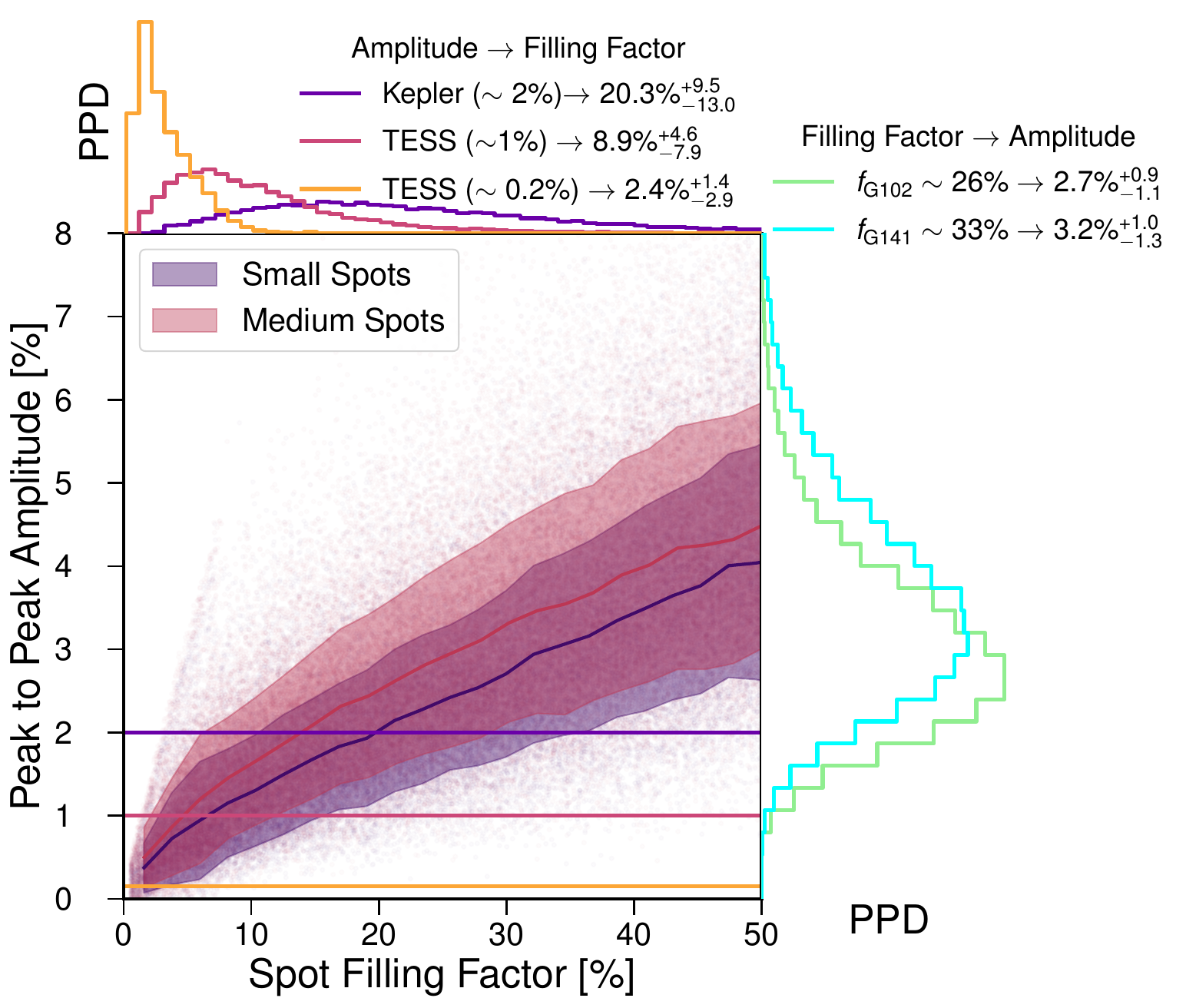}
\caption{ 
Peak-to-peak rotational modulation amplitude as a function of spot filling factor from ensembles of simulated stellar surface realizations. 
Colored curves and shaded regions show the median relation and dispersion for small-spot (purple) and medium-spot (red) populations using \texttt{spotter}. Horizontal lines mark rotational amplitudes from \Kepler{} ($\sim$2\%) and \TESS{} ($\sim$1\% and $\sim$0.2\%), and the distribution of the corresponding filling factor to produce such amplitudes is shown on the top. Similarly, the retrieved filling factors from \HST{}/\wfcthree{} ($f_{\rm G102}$$\sim$26\% and $f_{\rm G141}$$\sim$33\%) imply expected rotational amplitudes of $\sim$2--4\% in the \Kepler{} band. \label{fig:peak2PeakFillingFactor}}
\end{figure}

The simulated rotational amplitude shows a roughly square-root scaling with respect to filling factor for both small spots (angular radii $\leq$0.6$^\circ$) and medium spots ($0.6$--$3^\circ$), as shown in \autoref{fig:peak2PeakFillingFactor}. The observed \Kepler{} ($\sim$2\%) and early \TESS{} ($\sim$1\%) rotational amplitudes are thus consistent with the spot covering fractions inferred from the \HST/\wfcthree{} retrievals ($f_{\rm G102}$ $\sim$0.26, $f_{\rm G141}$ $\sim$0.33), which predict rotational modulations at the few-percent level during active phases. For rotational amplitudes $\gtrsim$1\%, the filling factor is only weakly constrained due to strong degeneracies with spot size and latitudinal distribution. In contrast, the very low amplitudes observed in later \TESS{} sectors tightly rule out large filling factors, indicating a markedly less spotted photosphere. 
   
\begin{figure*}[ht!]
\centering
\includegraphics[width=\textwidth]{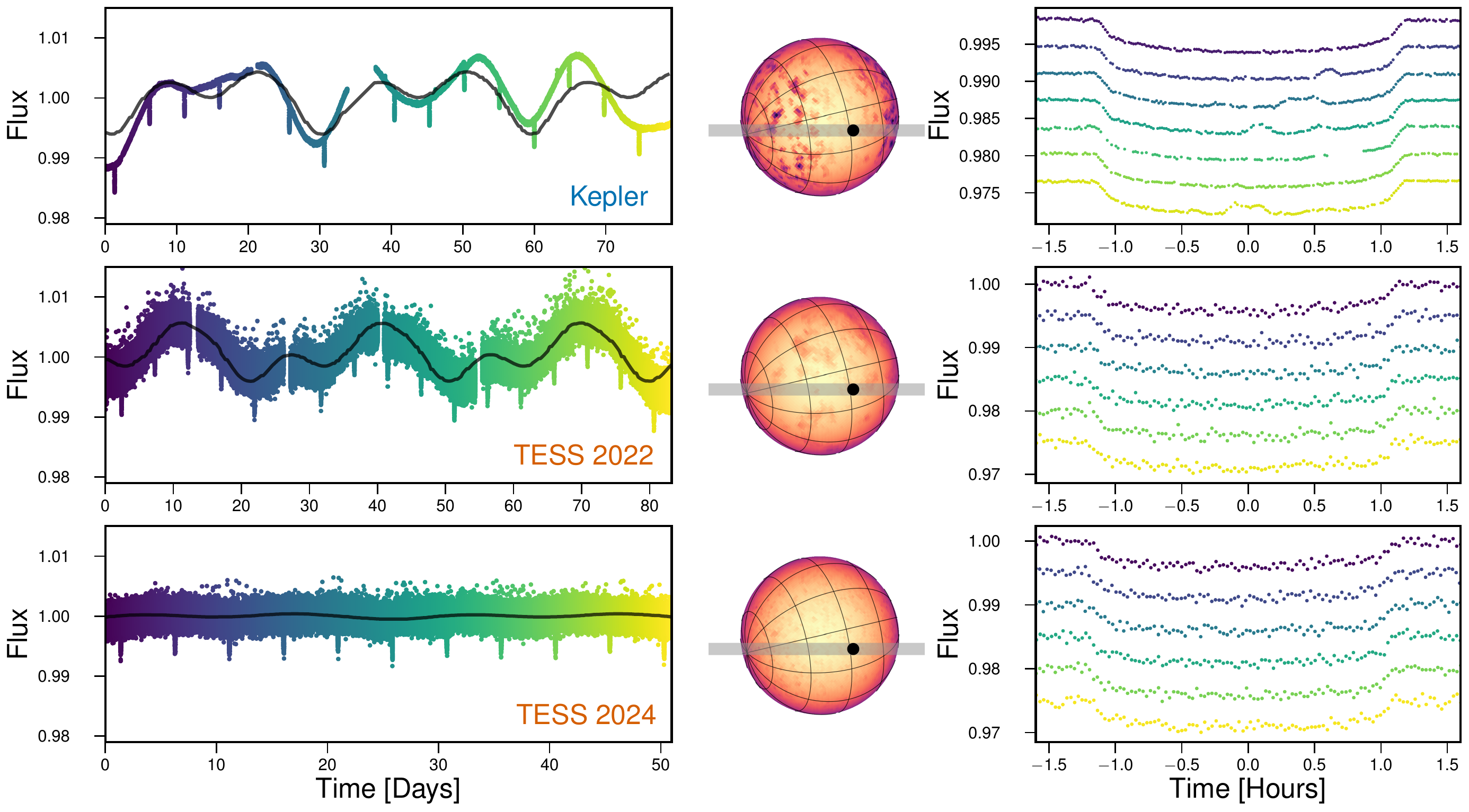}
\caption{ Evolution of stellar activity for HAT-P-11 across observing epochs. 
\textbf{Left:} Photometry during the \Kepler{} active phase and two \TESS{} epochs.
The \Kepler{} data show strong rotational modulation, while the 2022 and 2024 \TESS{} data show reduced variability and minimal modulation, respectively.
Black curves represent example forward-model realizations using \texttt{spotter} \citep{Garcia2025} that roughly model the spot configuration without accounting for temporal evolution. 
\textbf{Middle:} Representative stellar surface maps corresponding to the \texttt{spotter} light curves, illustrating decreasing spot coverage across epochs. 
\textbf{Right:} Vertically offset transit light curves from the same epochs (every other transit shown). Spot-crossing anomalies are prominent in the \Kepler{} data but not observed in the \TESS{} data, owing to reduced contrast and higher noise levels. The \TESS{} light curves are binned for clarity.\label{fig:stellarActivityEvolution}}
\end{figure*}

We also explored a light-curve inversion framework tailored to \hp{}. The system's well-constrained stellar obliquity, orbital geometry, and active latitudes ($16 \pm 5^\circ$; \citealt{morris2017}) allow us to generate maximum-likelihood surface realizations using a surface Gaussian process and Matérn-3/2 kernel. For each surface map, we computed the corresponding light curve and optimized the fit to the data using Nelder--Mead minimization. Owing to strong degeneracies inherent to the light curve inversion problem, the inferred maps in the middle panel for \autoref{fig:stellarActivityEvolution} are illustrative rather than unique solutions.
They demonstrate plausible spot distributions consistent with the light curves.

At the same time, visual inspection of the corresponding transit light curves (right column of \autoref{fig:stellarActivityEvolution}) reveals prominent spot-crossing anomalies in the \Kepler{} data but none in \TESS. Although \TESS{} photometry is less precise, spot crossings comparable in amplitude to those observed in \Kepler{} data should still be detectable. Their absence therefore supports a reduced level of spot-induced structure on \hp{} during the \TESS{} epoch relative to the \Kepler{} era.

\subsection{Comparison to Other K dwarfs}
\label{sec:comparison}
To place the inferred spot properties of \hp{} in context, we compare them with those of other active K-dwarf exoplanet hosts. A useful reference is HD\,189733 (K2V compared to K4V for \hp{}), which exhibits exceptionally high spot coverage from 38$\pm$4\% to 47$\pm$3\%, similarly inferred from \HST/\wfcthree{}/G141 spectra \citepalias{Narrett2024}. More broadly, mid-to-late K dwarfs in the \Kepler{} sample exhibit large photometric modulation amplitudes ${\sim}0.3{-}1.7\%$ \citep{McQuillan2014, rackham2019}, generally consistent with spot filling factors of tens of percent \citep{rackham2019}. In this context, the 20--30\% filling factors inferred for \hp{} fall within the upper range expected for active K dwarfs, but are not wholly unprecedented. Similarly, empirical relations predict $T_{\rm spot} {\sim} 3588$\,K for HAT-P-11 \citep{berdyugina2005}, with updated fits giving $3680 \pm 420$ K ($\Delta T = 1100 \pm 420$\,K; \citealt{herbst2021}), consistent with our inferred $T_2$.       

Turning to more precise and reliable \JWST{} observations, recent studies further indicate that substantial spot coverage is common among K-dwarf hosts. For example, the K2V hosts HAT-P-18 and WASP-52---both spectrally comparable to the K4V \hp{}---show prominent spot crossings in their transit observations \citep{tondreau2024, tondreau2025}. In the case of WASP-52, two distinct two-spot crossing events with different inferred temperatures were detected during a single \JWST{} transit, corresponding to a filling factor of 30$\pm$10\%.  Similarly, the slightly late-type K7V star GJ\,9827 shows evidence for spot coverage at the  $\sim$10\% level \citep{piaulet2025}. 

Taken together, these results indicate that \hp{} may not be exceptional among active K dwarfs in having a large spot filling factor. Rather, substantial photospheric heterogeneity appears to be a normal property of magnetically active K-dwarf planet hosts.

\subsection{Impact on the Stellar Radius}

\begin{figure}[t]
    \centering
    \includegraphics[width=\columnwidth]{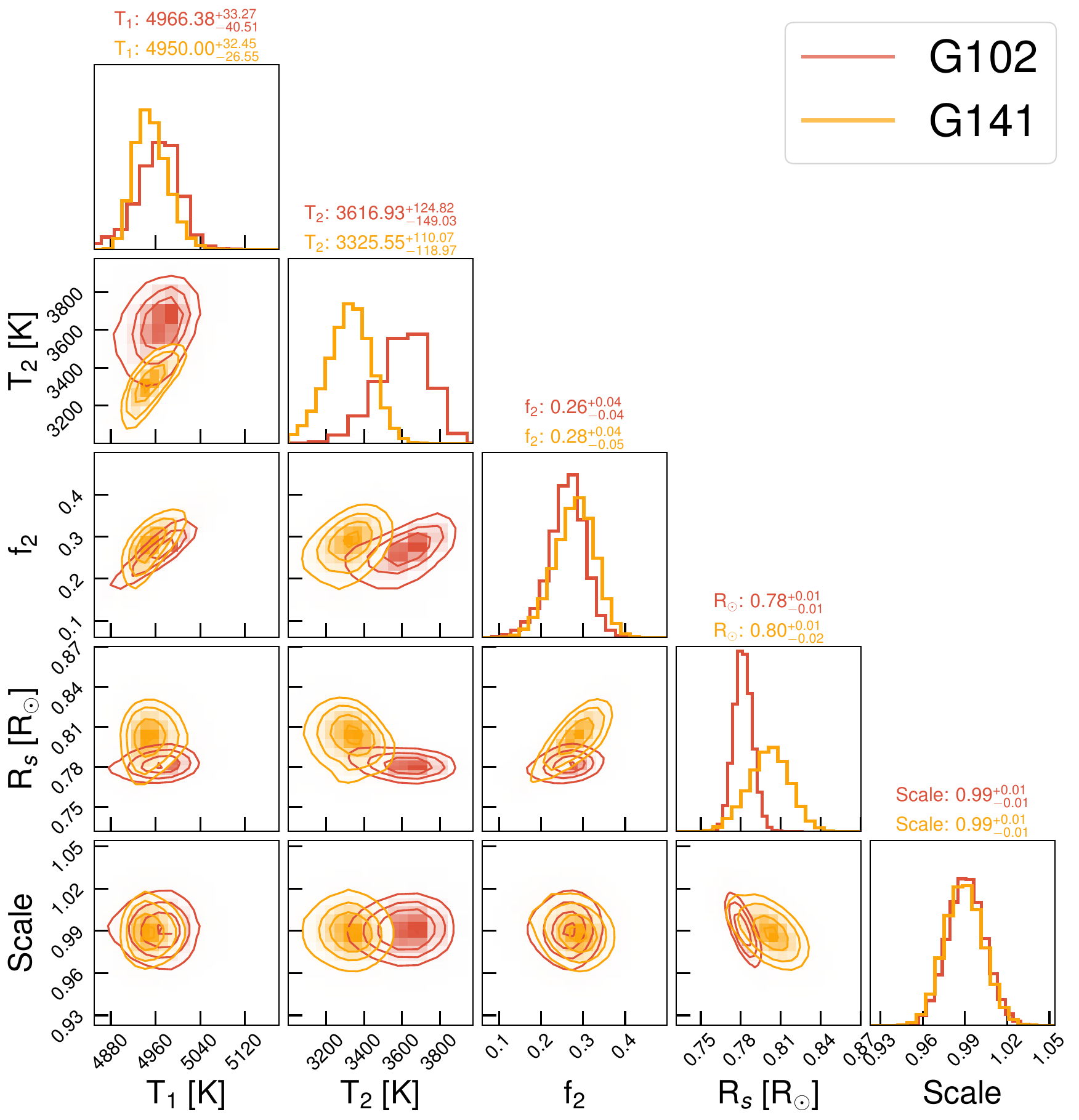}
    \caption{
    Corner plot illustrating parameter degeneracies in the two-component spectral fits to the \wfcthree{}/G102 (red) and G141 (orange) data.
    Shown are posterior distributions for the photospheric temperature ($T_1$), spot temperature ($T_2$), spot filling factor $f_2$, stellar radius $(R_\odot$), and the flux scale factor.  \label{fig:cornerPlotScalingFactor}}
\end{figure}

As described in \autoref{sec:priors}, we fit for the stellar radius and distance, adopting Gaussian priors from \citet{bakos2010} and \citet{gaiaDR3}, respectively. Broadband photometric fluxes constrain the absolute flux normalization of the \HST{} spectra, which depends jointly on the stellar radius, distance, and an overall scaling factor. Because the \Gaia{} distance is precisely known compared to stellar radius, the remaining freedom is largely captured by a degeneracy between the stellar radius and the flux scaling factor such that smaller radii can offset larger scale factors (and vice versa; see \autoref{fig:cornerPlotScalingFactor}).

In the literature, stellar radii are commonly estimated by fitting single-component models to broadband SED photometry, typically achieving uncertainties at the $\sim$5-10\% level \citep{stassun2019}. However, interferometric measurements of K- and M-dwarfs have often yielded radii systematically larger than those inferred from SED fits and theoretical predictions, with discrepancies on the order of a few percent and reaching $\sim$2$\sigma$ in some cases \citep[e.g.,][]{boyajian2012}.

While various explanations have been suggested for this, one plausible contributor to this discrepancy is neglecting surface heterogeneity, particularly for active stars. Multi-component photospheres redistribute flux across wavelength in a way that could systematically bias the radius inferred under the assumption of a single-component model. For \hp, our inferred stellar radius is consistent with the prior adopted from the literature \citep{bakos2010}. The resulting uncertainty is modestly reduced, reflecting the additional constraints provided by our data within our prior-informed framework.
Conceptually, our results suggest that larger spot filling factors typically tend to increase the inferred stellar radius (provided the photospheric temperature is well constrained), as cooler components would have to be compensated by a larger emitting area for the observed flux. To robustly establish the impact of multi-component models will require a broader population-level analysis and independent validation (e.g., via interferometric measurements), which is beyond the scope of this work, although our results provide a strong motivation for their consideration in inferring stellar radii.

\section{Implications for Transmission Spectroscopy}
\label{sec:ImplicationTransmSpectroscopy}

\begin{figure*}[t]
    \centering
    \includegraphics[width=\textwidth]{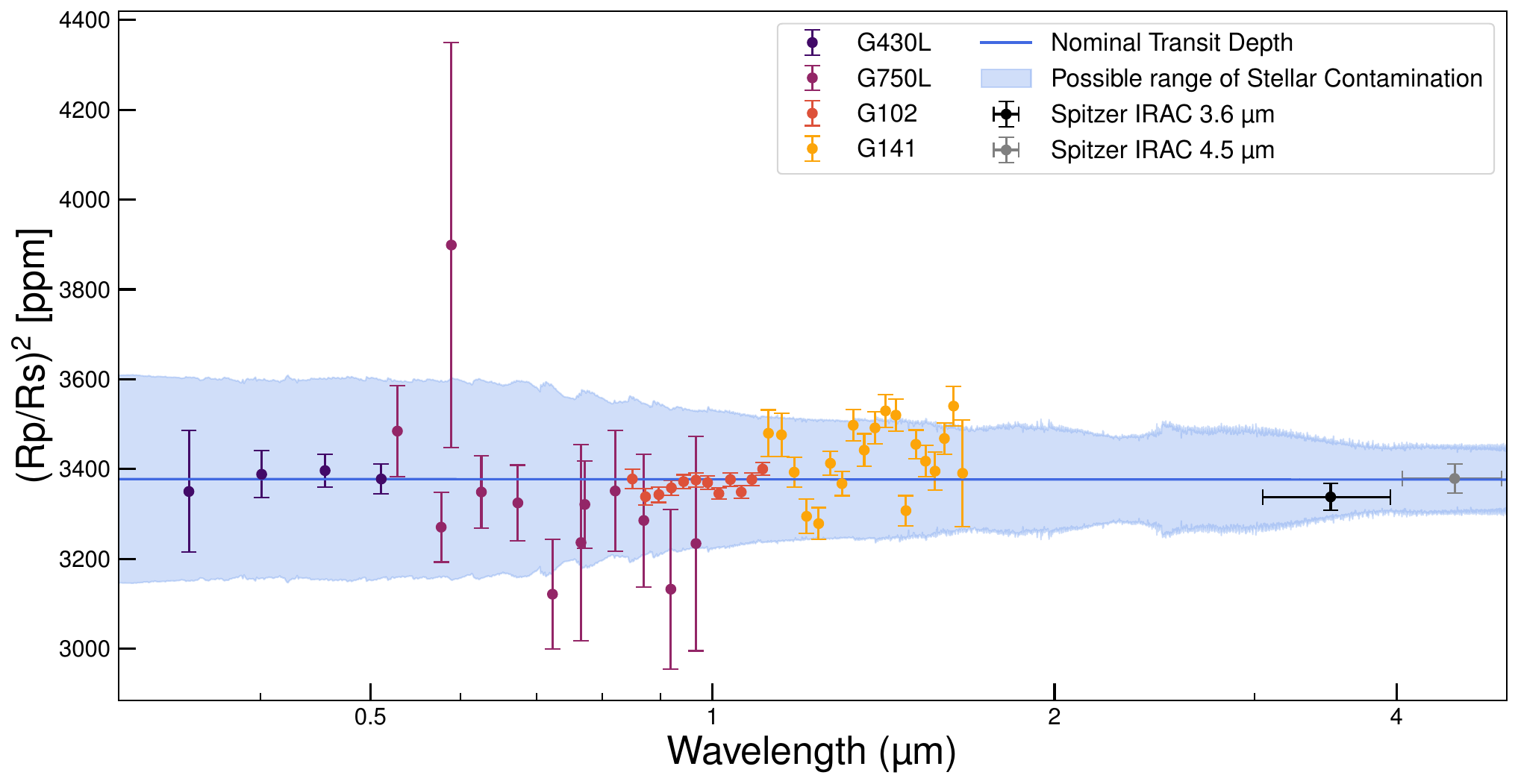}
    \caption{ \label{fig:epsilonTransmissionSpectra}
    Range of possible stellar contamination signals due to HAT-P-11's photospheric heterogeneity.
    The multi-instrument 0.3--5.0\,$\micron$ transmission spectrum of \hp{} \citep{chachan2019} is shown alongside the 68\% confidence interval for the expected contribution from stellar contamination (light blue region), assuming a nominal planetary transit depth (blue line).
    The amplitude of possible stellar contamination signals is comparable to the observed spectral features, including the $\sim$100\,ppm offset between the \HST{} and \Spitzer{} observations. 
    This implies that atmospheric retrievals that neglect stellar contamination are susceptible to significant biases.}
\end{figure*}

Observations of \hp{} have long shown a discrepancy between the transit depths measured with \HST{} and \Spitzer{}. In particular, \citet{fraine2014} reported a prominent $1.4\,\micron$ \ce{H2O} feature and argued that the spot--photosphere contrast would render starspots too warm to produce significant water absorption.
However, when combining broader wavelength coverage with \Spitzer{} data, \citet{chachan2019} found that reconciling the shallower Spitzer depths requires a $\sim$3\% brightening of the stellar disk under the assumption of $T_{\rm spot}=4500$~K, which they considered implausible given prior estimates of the stellar inhomogeneities of \hp. 
In contrast, \citet{cubillos2022} allowed inter-instrument offsets to vary within the retrieval framework and found that such offsets substantially impact the retrieved parameters for \hp{}.

Given our inference of substantial spot coverage, we model the effect of stellar contamination in this system.
Following \citet{rackham2018}, we express the observed transit depth as
\begin{align}
    D_{\lambda, \mathrm{obs}} &= \epsilon_\lambda D_{\rm True} 
\end{align}
where $D_{\rm True}$ is the intrinsic planetary transit depth and $\epsilon_\lambda$ is a wavelength-dependent contamination factor determined by the contrast between the transit-chord and disk-integrated stellar spectra.
Allowing spots to be present in the transit chord but at a different level than on the stellar disk, $\epsilon_\lambda$ is given by
\begin{align}
    \epsilon_\lambda &=
    \frac{
        (1-f_{\rm spot})S_{\lambda, \rm phot} + f_{\rm spot} S_{\lambda, \rm spot}
    }{
        (1-F_{\rm spot})S_{\lambda, \rm phot} + F_{\rm spot} S_{\lambda, \rm spot}
    } \\
     &= \dfrac{1 - f_{\rm spot} \left(1 - \dfrac{S_{\lambda, \rm spot}}{S_{\lambda, \rm phot}}\right)}{1 - F_{\rm spot} \left(1 - \dfrac{S_{\lambda, \rm spot}}{S_{\lambda, \rm phot}}\right)} \,,
\end{align}
where $f_{\mathrm{spot}}$ and $F_{\mathrm{spot}}$ are the transit-chord and disk-integrated spot filling factors, respectively, and $S_{\lambda,\mathrm{spot}}$ and $S_{\lambda,\mathrm{phot}}$ are the spectra of the spot and photosphere, respectively \citep{zhang2018}. 

Using spot temperatures and filling factors derived from our spectral retrievals and assuming the transit-chord and full-disk filling factors are independent realizations of the spot filling factor constraint, we compute the expected contamination signal across 0.3--5.0\,$\micron$ (\autoref{fig:epsilonTransmissionSpectra}).
The resulting contamination amplitudes are comparable to the observed spectral features, including the ${\sim}100$\,ppm offset between \HST{} and \Spitzer{}.
This implies that stellar heterogeneity alone could account for features previously attributed to planetary atmospheric structure.
While the detection of water vapor in the \wfcthree{}/G141 data remains robust---being driven primarily by a single high-quality epoch---the joint interpretation of multi-epoch, multi-instrument datasets without contemporaneous stellar constraints is vulnerable to bias.
When stellar contamination can be comparable in amplitude to planetary spectral features, retrieval outcomes depend sensitively on how stellar surface structure is treated.

\subsection{Re-evaluation of \HST{}+\Spitzer{} Results}
Before \JWST{}, transmission spectroscopy commonly relied on combining observations from multiple instruments or telescopes to extend spectral coverage \citep[e.g.,][]{chachan2019}.
This strategy was driven by instrumental bandpass limitations rather than scientific preference. 
However, stitching together data obtained at different epochs implicitly assumes a stable stellar photosphere, unless explicitly modeled otherwise.

In the analysis of \citet{chachan2019}, stellar contamination corrections were applied using strong priors on spot filling factor and fixed spot temperatures (4100--4500\,K). The adopted filling factor of 4.4\% during the G102 epoch was motivated by spot-crossing constraints from \citep{morris2017},  which, according to our analysis, likely underestimate the total disk-integrated heterogeneity. By contrast, a later independent analysis \citep{cubillos2022} treated inter-instrument offsets as free parameters, effectively allowing for changes in stellar brightness due to heterogeneity.  These differing treatments of unmodeled stellar contamination lead to inconsistent atmospheric inferences. 

Our results show that contamination signals of order 100\,ppm arise naturally from plausible spot configurations (\autoref{fig:epsilonTransmissionSpectra}). In this regime, unmodeled stellar heterogeneity can directly bias retrieved molecular abundances and atmospheric structure. The challenge is not the detection of spectral features themselves, but the reliable attribution of those features to planetary rather than stellar origins.

\subsection{Lessons for Future Observations}

The \JWST{} era has transformed transmission spectroscopy: molecular bands that once required stitching together data from multiple facilities to study can now be detected at exquisite precision within a single visit. Programs such as ongoing \JWST{} GO programs 4105 (PI: Lili Alderson), 2950 (PI: Ren Waters), and 5924 (PI: David Sing) illustrate this growing capability. However, this sharp increase in precision has also exposed significant limitations in earlier assumptions regarding stellar contamination. 
Our results demonstrate that stellar heterogeneity can imprint spurious spectral structure even beyond the 1--2~$\mu$m regime, challenging the common assumption that near- and mid-infrared transmission spectra are largely immune to stellar effects.

It is perhaps inevitable that the most compelling atmospheric signals are also the most challenging to interpret. As the field pushes toward detecting subtle features, such as weak molecular bands or hemispheric asymmetries \citep{espinoza2024}, the influence of stellar surface structure becomes increasingly non-negligible \citep{kostogryz2024, Kostogryz2025}. While the broad spectroscopic coverage of \JWST{} enables retrievals anchored by multiple molecular bands and continuum regions, our results show that stellar inhomogeneities can fundamentally limit the robustness of such inferences. Retrievals can recover atmospheric parameters only when stellar contamination is explicitly modeled; neglecting stellar heterogeneity leads to biased molecular abundances \citep{Iyer2020}. High-resolution cross-correlation techniques are somewhat more resilient to stellar contamination, but even these methods can be more challenging once realistic spot contrasts and spatial distributions are included \citep{Bourrier2024}.

Building on our ongoing efforts with \HST{} to characterize stellar surfaces for a set of high-priority exoplanet hosts, we will extend this work using out-of-transit \JWST{} observations as part of GO-AR-5370 (PI: Rackham)\footnote{\url{https://jwst-eots.github.io/}}. Jointly fitting for out-of-transit stellar spectra alongside transmission spectra will enable epoch-specific constraints on stellar heterogeneity, providing a pathway toward mitigating stellar contamination and achieving more robust atmospheric retrievals. Complementary photometric monitoring or activity diagnostics (such as \caii{} H\&K) could further identify optimal windows for transmission spectra even for active hosts.

\section{Conclusions}
\label{sec:conclusions}

We have presented a comprehensive analysis of the disk-integrated photospheric heterogeneity of \hp{} using multi-epoch, absolutely calibrated out-of-transit spectra from \HST{}/\stis{} and \HST{}/\wfcthree{}, spanning 0.3--1.7\,$\micron$.
By combining these data with broadband photometry, long-term \Kepler{} and \TESS{} monitoring, chromospheric activity indices, and forward modeling of stellar variability, we quantify the time-dependent photospheric heterogeneity of \hp{} and its impact on transmission spectroscopy.

Our main conclusions are as follows.

\begin{enumerate}
    \item The near-infrared spectra point to a heterogeneous photosphere.
    For all \HST{}/\wfcthree{} epochs, two-component photospheric models are strongly preferred over single-component fits, with $\Delta\mathrm{BIC}\sim10$ indicating robust evidence for a cooler surface component. The preferred solutions consist of a dominant component of $T_1\,\approx\,4980$\,K and a cooler component of $T_2\,\approx\,3400$\,K,
    with filling factors of $f_2\,\approx\,0.26{-}0.33$. The consistency across independent \wfcthree{} modes and epochs indicates that substantial photospheric heterogeneity is required to model the disk-integrated near-infrared spectrum.

    \item Inferences from optical spectra are limited by model fidelity challenges. The \stis{} G430L/G750L spectra do not statistically favor multi-component models. However, comparisons between available atmospheric models reveal ${>}2\%$ discrepancies in the optical, indicating that model uncertainties dominate at these wavelengths. The lack of a multi-component preference in the optical therefore likely reflects model fidelity limitations rather than evidence for a homogeneous photosphere, highlighting the need for targeted model improvements in the optical regime.

    \item HAT-P-11 was highly spotted during the \HST{} epochs. The inferred filling factors of 26--33\% exceed estimates derived from transit chord analyses, implying that a significant fraction of the spotted area lies outside the transit chord or consists of smaller-scale structures not resolved in spot-crossing events. These values, however, are more consistent with previous disk-integrated TiO-based analyses. The inferred temperature contrast ($\Delta T \approx 1500$\,K) is consistent with cool, umbra-like regions on active K dwarfs.
        
    \item Independent diagnostics confirm substantial temporal evolution in the spottedness of \hp{}. The \HST{} observations (2012--2016) occurred during a relatively active phase, coinciding with \Kepler{} rotational modulation amplitudes of ${\sim}2\%$ and elevated Ca\,\textsc{ii} H\&K emission. In contrast, later \TESS{} observations show reduced variability of ${\sim}1\%$ and as low as ${\sim}0.2\%$ in the final sector, accompanied by a decline in the S-index.
    By 2024, the implied spot coverage had decreased to only a few percent.
    The \JWST{} observations coincided with this lower-activity phase, and are thus expected to be less impacted by stellar contamination, though non-negligible biases remain possible with spot coverage at the $\sim$2--10\% level.

    \item The amplitude of plausible stellar contamination signals is comparable to planetary spectral features.
    Using the retrieved spot temperatures and filling factors, we compute contamination signals of order ${\sim}100$\,ppm across 0.3--5\,$\micron$.
    This is comparable to the amplitude of key transmission spectral features and to the previously reported \HST{}--\Spitzer{} offset.
    Atmospheric retrievals that neglect stellar heterogeneity are therefore susceptible to significant biases, particularly when combining multi-epoch or multi-instrument datasets.
\end{enumerate}

\hp{} provides a valuable testbed for modeling stellar surface heterogeneity using precise, space-based out-of-transit spectroscopy. Constraining the stellar surface properties during transit and propagating these constraints into transmission spectroscopy analyses enables more robust interpretations of planetary atmospheres.
The methodology developed here, combining absolute spectrophotometry, activity diagnostics, and forward contamination modeling, provides a framework that can be applied to other transiting planet hosts.

We are extending this approach through the HST Stellar Treasure Trove and the Eyes on the Stars Legacy Archival Programs, which are analyzing a broader sample of planet-hosting stars observed with both \HST{} and \JWST{} to enable population-level insights on stellar surface inhomogeneities.  Complementary long-term, high-resolution spectroscopic monitoring (e.g., the OWLS program; \citealt{morris2025}) and rotational modulation measurements from \TESS{} can further contextualize the state of the stellar photosphere at the time of transit observations. The recently launched \textit{Pandora SmallSat Mission} \citep{pandora}, designed specifically to constrain the properties of both active K- and M-dwarf host stars and their transiting exoplanets, will further provide critical context for interpreting transmission spectra \citep{rackham2026, Rotman2026}. 

While stellar contamination remains one of the central challenges for transmission spectroscopy, stellar surface characterization via space-based spectroscopy now offers a clear path towards robust atmospheric inferences in the \JWST{} era. The results presented here show further that for some active stars, there may be times when lower activity states provide better windows for transmission spectroscopy.

\vspace{5mm}
\facilities{ \textit{HST}, \textit{Kepler}, \textit{TESS}, }

\software{\texttt{speclib} \citep{SPECLIB},  \texttt{PACMAN} \citep{PACMAN}, spotter \citep{Garcia2025} } 

\section*{Acknowledgments}
The authors thank Prof.\ Sara Seager, Dr.\ Ana Glidden, Dr.\ Rachael C. Amaro, and Dr.\ Veronika Witzke for helpful discussions and constructive input on this work. 
Based on observations with the NASA/ESA Hubble Space Telescope obtained from the Mikulski Archive for Space Telescopes (MAST) at the Space Telescope Science Institute, which is operated by the Association of Universities for Research in Astronomy, Incorporated, under NASA contract NAS5-26555. Support for program number HST-AR-17551 was provided through a grant from the STScI under NASA contract NAS5-26555.
This paper includes data collected by the \Kepler{} and \TESS{} missions, obtained from the Mikulski Archive for Space Telescopes (MAST). Funding for the \Kepler{} and \TESS{} missions is provided by the NASA Science Mission Directorate. 
This material is based upon work supported by the National Aeronautics and Space Administration under Agreement No.\ 80NSSC21K0593 for the program ``Alien Earths''.
The results reported herein benefited from collaborations and/or information exchange within NASA’s Nexus for Exoplanet System Science (NExSS) research coordination network sponsored by NASA’s Science Mission Directorate.
This material is based upon work supported by the European Research Council (ERC) Synergy Grant under the European Union’s Horizon 2020 research and innovation program (grant No.\ 101118581---project REVEAL).

\bibliography{Bibliography}{}

@ARTICLE{PACMAN,
       author = {{Zieba}, Sebastian and {Kreidberg}, Laura},
        title = "{PACMAN: A pipeline to reduce and analyze Hubble Wide Field Camera 3 IR Grism data}",
      journal = {The Journal of Open Source Software},
         year = 2022,
        month = dec,
       volume = {7},
       number = {80},
          eid = {4838},
        pages = {4838},
          doi = {10.21105/joss.04838},
archivePrefix = {arXiv},
       eprint = {2212.11421},
 primaryClass = {astro-ph.IM},
       adsurl = {https://ui.adsabs.harvard.edu/abs/2022JOSS....7.4838Z}
}

@ARTICLE{Narrett2024,
       author = {{Narrett}, Isaac S. and {Rackham}, Benjamin V. and {de Wit}, Julien},
        title = "{Axisymmetric High Spot Coverage on Exoplanet Host HD 189733 A}",
      journal = {\aj},
         year = 2024,
        month = mar,
       volume = {167},
       number = {3},
          eid = {107},
        pages = {107},
          doi = {10.3847/1538-3881/ad1f6c},
       adsurl = {https://ui.adsabs.harvard.edu/abs/2024AJ....167..107N}
}

@ARTICLE{iyer2023,
       author = {{Iyer}, Aishwarya R. and {Line}, Michael R. and {Muirhead}, Philip S. and {Fortney}, Jonathan J. and {Gharib-Nezhad}, Ehsan},
        title = "{The SPHINX M-dwarf Spectral Grid. I. Benchmarking New Model Atmospheres to Derive Fundamental M-dwarf Properties}",
      journal = {\apj},
         year = 2023,
        month = feb,
       volume = {944},
       number = {1},
          eid = {41},
        pages = {41},
          doi = {10.3847/1538-4357/acabc2},
archivePrefix = {arXiv},
       eprint = {2206.12010},
 primaryClass = {astro-ph.SR},
       adsurl = {https://ui.adsabs.harvard.edu/abs/2023ApJ...944...41I}
}

@ARTICLE{zhang2018,
       author = {{Zhang}, Zhanbo and {Zhou}, Yifan and {Rackham}, Benjamin V. and {Apai}, D{\'a}niel},
        title = "{The Near-infrared Transmission Spectra of TRAPPIST-1 Planets b, c, d, e, f, and g and Stellar Contamination in Multi-epoch Transit Spectra}",
      journal = {\aj},
         year = 2018,
        month = oct,
       volume = {156},
       number = {4},
          eid = {178},
        pages = {178},
          doi = {10.3847/1538-3881/aade4f},
archivePrefix = {arXiv},
       eprint = {1802.02086},
 primaryClass = {astro-ph.EP},
       adsurl = {https://ui.adsabs.harvard.edu/abs/2018AJ....156..178Z}
}

@ARTICLE{rackham2019,
       author = {{Rackham}, Benjamin V. and {Apai}, D{\'a}niel and {Giampapa}, Mark S.},
        title = "{The Transit Light Source Effect. II. The Impact of Stellar Heterogeneity on Transmission Spectra of Planets Orbiting Broadly Sun-like Stars}",
      journal = {\aj},
         year = 2019,
        month = mar,
       volume = {157},
       number = {3},
          eid = {96},
        pages = {96},
          doi = {10.3847/1538-3881/aaf892},
archivePrefix = {arXiv},
       eprint = {1812.06184},
 primaryClass = {astro-ph.EP},
       adsurl = {https://ui.adsabs.harvard.edu/abs/2019AJ....157...96R}
}

@ARTICLE{deWit2018,
       author = {{de Wit}, Julien and {Wakeford}, Hannah R. and {Lewis}, Nikole K. and {Delrez}, Laetitia and {Gillon}, Micha{\"e}l and {Selsis}, Frank and {Leconte}, J{\'e}r{\'e}my and {Demory}, Brice-Olivier and {Bolmont}, Emeline and {Bourrier}, Vincent and {Burgasser}, Adam J. and {Grimm}, Simon and {Jehin}, Emmanu{\"e}l and {Lederer}, Susan M. and {Owen}, James E. and {Stamenkovi{\'c}}, Vlada and {Triaud}, Amaury H.~M.~J.},
        title = "{Atmospheric reconnaissance of the habitable-zone Earth-sized planets orbiting TRAPPIST-1}",
      journal = {Nature Astronomy},
         year = 2018,
        month = mar,
       volume = {2},
        pages = {214-219},
          doi = {10.1038/s41550-017-0374-z},
archivePrefix = {arXiv},
       eprint = {1802.02250},
 primaryClass = {astro-ph.EP},
       adsurl = {https://ui.adsabs.harvard.edu/abs/2018NatAs...2..214D}
}

@ARTICLE{morris2017,
       author = {{Morris}, Brett M. and {Hebb}, Leslie and {Davenport}, James R.~A. and {Rohn}, Graeme and {Hawley}, Suzanne L.},
        title = "{The Starspots of HAT-P-11: Evidence for a Solar-like Dynamo}",
      journal = {\apj},
         year = 2017,
        month = sep,
       volume = {846},
       number = {2},
          eid = {99},
        pages = {99},
          doi = {10.3847/1538-4357/aa8555},
archivePrefix = {arXiv},
       eprint = {1708.02583},
 primaryClass = {astro-ph.SR},
       adsurl = {https://ui.adsabs.harvard.edu/abs/2017ApJ...846...99M}
}

@misc{SPECLIB,
       author = {{Rackham}, Benjamin V.},
        title = "{speclib}",
         year = 2023,
        month = apr,
          eid = {10.5281/zenodo.7868050},
          doi = {10.5281/zenodo.7868050},
      version = {0.0-beta.0},
    publisher = {Zenodo},
       adsurl = {https://ui.adsabs.harvard.edu/abs/2023zndo...7868050R}
}

@ARTICLE{stassun2019,
       author = {{Stassun}, Keivan G. and {Oelkers}, Ryan J. and {Paegert}, Martin and {Torres}, Guillermo and {Pepper}, Joshua and {De Lee}, Nathan and {Collins}, Kevin and {Latham}, David W. and {Muirhead}, Philip S. and {Chittidi}, Jay and {Rojas-Ayala}, B{\'a}rbara and {Fleming}, Scott W. and {Rose}, Mark E. and {Tenenbaum}, Peter and {Ting}, Eric B. and {Kane}, Stephen R. and {Barclay}, Thomas and {Bean}, Jacob L. and {Brassuer}, C.~E. and {Charbonneau}, David and {Ge}, Jian and {Lissauer}, Jack J. and {Mann}, Andrew W. and {McLean}, Brian and {Mullally}, Susan and {Narita}, Norio and {Plavchan}, Peter and {Ricker}, George R. and {Sasselov}, Dimitar and {Seager}, S. and {Sharma}, Sanjib and {Shiao}, Bernie and {Sozzetti}, Alessandro and {Stello}, Dennis and {Vanderspek}, Roland and {Wallace}, Geoff and {Winn}, Joshua N.},
        title = "{The Revised TESS Input Catalog and Candidate Target List}",
      journal = {\aj},
         year = 2019,
        month = oct,
       volume = {158},
       number = {4},
          eid = {138},
        pages = {138},
          doi = {10.3847/1538-3881/ab3467},
archivePrefix = {arXiv},
       eprint = {1905.10694},
 primaryClass = {astro-ph.SR},
       adsurl = {https://ui.adsabs.harvard.edu/abs/2019AJ....158..138S}
}

@ARTICLE{rackham2017,
       author = {{Rackham}, Benjamin and {Espinoza}, N{\'e}stor and {Apai}, D{\'a}niel and {L{\'o}pez-Morales}, Mercedes and {Jord{\'a}n}, Andr{\'e}s and {Osip}, David J. and {Lewis}, Nikole K. and {Rodler}, Florian and {Fraine}, Jonathan D. and {Morley}, Caroline V. and {Fortney}, Jonathan J.},
        title = "{ACCESS I: An Optical Transmission Spectrum of GJ 1214b Reveals a Heterogeneous Stellar Photosphere}",
      journal = {\apj},
         year = 2017,
        month = jan,
       volume = {834},
       number = {2},
          eid = {151},
        pages = {151},
          doi = {10.3847/1538-4357/aa4f6c},
archivePrefix = {arXiv},
       eprint = {1612.00228},
 primaryClass = {astro-ph.EP},
       adsurl = {https://ui.adsabs.harvard.edu/abs/2017ApJ...834..151R}
}

@ARTICLE{rackham2018,
       author = {{Rackham}, Benjamin V. and {Apai}, D{\'a}niel and {Giampapa}, Mark S.},
        title = "{The Transit Light Source Effect: False Spectral Features and Incorrect Densities for M-dwarf Transiting Planets}",
      journal = {\apj},
         year = 2018,
        month = feb,
       volume = {853},
       number = {2},
          eid = {122},
        pages = {122},
          doi = {10.3847/1538-4357/aaa08c},
archivePrefix = {arXiv},
       eprint = {1711.05691},
 primaryClass = {astro-ph.EP},
       adsurl = {https://ui.adsabs.harvard.edu/abs/2018ApJ...853..122R}
}

@ARTICLE{chachan2019,
       author = {{Chachan}, Yayaati and {Knutson}, Heather A. and {Gao}, Peter and {Kataria}, Tiffany and {Wong}, Ian and {Henry}, Gregory W. and {Benneke}, Bjorn and {Zhang}, Michael and {Barstow}, Joanna and {Bean}, Jacob L. and {Mikal-Evans}, Thomas and {Lewis}, Nikole K. and {Mansfield}, Megan and {L{\'o}pez-Morales}, Mercedes and {Nikolov}, Nikolay and {Sing}, David K. and {Wakeford}, Hannah},
        title = "{A Hubble PanCET Study of HAT-P-11b: A Cloudy Neptune with a Low Atmospheric Metallicity}",
      journal = {\aj},
         year = 2019,
        month = dec,
       volume = {158},
       number = {6},
          eid = {244},
        pages = {244},
          doi = {10.3847/1538-3881/ab4e9a},
archivePrefix = {arXiv},
       eprint = {1910.07523},
 primaryClass = {astro-ph.EP},
       adsurl = {https://ui.adsabs.harvard.edu/abs/2019AJ....158..244C}
}

@ARTICLE{mansfield2018,
       author = {{Mansfield}, Megan and {Bean}, Jacob L. and {Oklop{\v{c}}i{\'c}}, Antonija and {Kreidberg}, Laura and {D{\'e}sert}, Jean-Michel and {Kempton}, Eliza M. -R. and {Line}, Michael R. and {Fortney}, Jonathan J. and {Henry}, Gregory W. and {Mallonn}, Matthias and {Stevenson}, Kevin B. and {Dragomir}, Diana and {Allart}, Romain and {Bourrier}, Vincent},
        title = "{Detection of Helium in the Atmosphere of the Exo-Neptune HAT-P-11b}",
      journal = {\apjl},
         year = 2018,
        month = dec,
       volume = {868},
       number = {2},
          eid = {L34},
        pages = {L34},
          doi = {10.3847/2041-8213/aaf166},
archivePrefix = {arXiv},
       eprint = {1812.02214},
 primaryClass = {astro-ph.EP},
       adsurl = {https://ui.adsabs.harvard.edu/abs/2018ApJ...868L..34M}
}

@ARTICLE{fraine2014,
       author = {{Fraine}, Jonathan and {Deming}, Drake and {Benneke}, Bjorn and {Knutson}, Heather and {Jord{\'a}n}, Andr{\'e}s and {Espinoza}, N{\'e}stor and {Madhusudhan}, Nikku and {Wilkins}, Ashlee and {Todorov}, Kamen},
        title = "{Water vapour absorption in the clear atmosphere of a Neptune-sized exoplanet}",
      journal = {\nat},
         year = 2014,
        month = sep,
       volume = {513},
       number = {7519},
        pages = {526-529},
          doi = {10.1038/nature13785},
archivePrefix = {arXiv},
       eprint = {1409.8349},
 primaryClass = {astro-ph.EP},
       adsurl = {https://ui.adsabs.harvard.edu/abs/2014Natur.513..526F}
}

@ARTICLE{cubillos2022,
       author = {{Cubillos}, Patricio E. and {Harrington}, Joseph and {Blecic}, Jasmina and {Himes}, Michael D. and {Rojo}, Patricio M. and {Loredo}, Thomas J. and {Lust}, Nate B. and {Challener}, Ryan C. and {Foster}, Austin J. and {Stemm}, Madison M. and {Foster}, Andrew S.~D. and {Blumenthal}, Sarah D.},
        title = "{An Open-source Bayesian Atmospheric Radiative Transfer (BART) Code. II. The TRANSIT Radiative Transfer Module and Retrieval of HAT-P-11b}",
      journal = {\psj},
         year = 2022,
        month = apr,
       volume = {3},
       number = {4},
          eid = {81},
        pages = {81},
          doi = {10.3847/PSJ/ac348b},
archivePrefix = {arXiv},
       eprint = {2104.12524},
 primaryClass = {astro-ph.EP},
       adsurl = {https://ui.adsabs.harvard.edu/abs/2022PSJ.....3...81C}
}

@ARTICLE{basilicata2024,
       author = {{Basilicata}, M. and {Giacobbe}, P. and {Bonomo}, A.~S. and {Scandariato}, G. and {Brogi}, M. and {Singh}, V. and {Di Paola}, A. and {Mancini}, L. and {Sozzetti}, A. and {Lanza}, A.~F. and {Cubillos}, P.~E. and {Damasso}, M. and {Desidera}, S. and {Biazzo}, K. and {Bignamini}, A. and {Borsa}, F. and {Cabona}, L. and {Carleo}, I. and {Ghedina}, A. and {Guilluy}, G. and {Maggio}, A. and {Mainella}, G. and {Micela}, G. and {Molinari}, E. and {Molinaro}, M. and {Nardiello}, D. and {Pedani}, M. and {Pino}, L. and {Poretti}, E. and {Southworth}, J. and {Stangret}, M. and {Turrini}, D.},
        title = "{The GAPS Programme at TNG. LV. Multiple molecular species in the atmosphere of HAT-P-11 b and review of the HAT-P-11 planetary system}",
      journal = {\aap},
         year = 2024,
        month = jun,
       volume = {686},
          eid = {A127},
        pages = {A127},
          doi = {10.1051/0004-6361/202347659},
archivePrefix = {arXiv},
       eprint = {2403.01527},
 primaryClass = {astro-ph.EP},
       adsurl = {https://ui.adsabs.harvard.edu/abs/2024A&A...686A.127B}
}

@ARTICLE{gaiaDR3,
       author = {{Gaia Collaboration} and {Vallenari}, A. and {Brown}, A.~G.~A. and {Prusti}, T. and {de Bruijne}, J.~H.~J. and {Arenou}, F. and {Babusiaux}, C. and {Biermann}, M. and {Creevey}, O.~L. and {Ducourant}, C. and {Evans}, D.~W. and {Eyer}, L. and {Guerra}, R. and {Hutton}, A. and {Jordi}, C. and {Klioner}, S.~A. and {Lammers}, U.~L. and {Lindegren}, L. and {Luri}, X. and {Mignard}, F. and {Panem}, C. and {Pourbaix}, D. and {Randich}, S. and {Sartoretti}, P. and {Soubiran}, C. and {Tanga}, P. and {Walton}, N.~A. and {Bailer-Jones}, C.~A.~L. and {Bastian}, U. and {Drimmel}, R. and {Jansen}, F. and {Katz}, D. and {Lattanzi}, M.~G. and {van Leeuwen}, F. and {Bakker}, J. and {Cacciari}, C. and {Casta{\~n}eda}, J. and {De Angeli}, F. and {Fabricius}, C. and {Fouesneau}, M. and {Fr{\'e}mat}, Y. and {Galluccio}, L. and {Guerrier}, A. and {Heiter}, U. and {Masana}, E. and {Messineo}, R. and {Mowlavi}, N. and {Nicolas}, C. and {Nienartowicz}, K. and {Pailler}, F. and {Panuzzo}, P. and {Riclet}, F. and {Roux}, W. and {Seabroke}, G.~M. and {Sordo}, R. and {Th{\'e}venin}, F. and {Gracia-Abril}, G. and {Portell}, J. and {Teyssier}, D. and {Altmann}, M. and {Andrae}, R. and {Audard}, M. and {Bellas-Velidis}, I. and {Benson}, K. and {Berthier}, J. and {Blomme}, R. and {Burgess}, P.~W. and {Busonero}, D. and {Busso}, G. and {C{\'a}novas}, H. and {Carry}, B. and {Cellino}, A. and {Cheek}, N. and {Clementini}, G. and {Damerdji}, Y. and {Davidson}, M. and {de Teodoro}, P. and {Nu{\~n}ez Campos}, M. and {Delchambre}, L. and {Dell'Oro}, A. and {Esquej}, P. and {Fern{\'a}ndez-Hern{\'a}ndez}, J. and {Fraile}, E. and {Garabato}, D. and {Garc{\'\i}a-Lario}, P. and {Gosset}, E. and {Haigron}, R. and {Halbwachs}, J. -L. and {Hambly}, N.~C. and {Harrison}, D.~L. and {Hern{\'a}ndez}, J. and {Hestroffer}, D. and {Hodgkin}, S.~T. and {Holl}, B. and {Jan{\ss}en}, K. and {Jevardat de Fombelle}, G. and {Jordan}, S. and {Krone-Martins}, A. and {Lanzafame}, A.~C. and {L{\"o}ffler}, W. and {Marchal}, O. and {Marrese}, P.~M. and {Moitinho}, A. and {Muinonen}, K. and {Osborne}, P. and {Pancino}, E. and {Pauwels}, T. and {Recio-Blanco}, A. and {Reyl{\'e}}, C. and {Riello}, M. and {Rimoldini}, L. and {Roegiers}, T. and {Rybizki}, J. and {Sarro}, L.~M. and {Siopis}, C. and {Smith}, M. and {Sozzetti}, A. and {Utrilla}, E. and {van Leeuwen}, M. and {Abbas}, U. and {{\'A}brah{\'a}m}, P. and {Abreu Aramburu}, A. and {Aerts}, C. and {Aguado}, J.~J. and {Ajaj}, M. and {Aldea-Montero}, F. and {Altavilla}, G. and {{\'A}lvarez}, M.~A. and {Alves}, J. and {Anders}, F. and {Anderson}, R.~I. and {Anglada Varela}, E. and {Antoja}, T. and {Baines}, D. and {Baker}, S.~G. and {Balaguer-N{\'u}{\~n}ez}, L. and {Balbinot}, E. and {Balog}, Z. and {Barache}, C. and {Barbato}, D. and {Barros}, M. and {Barstow}, M.~A. and {Bartolom{\'e}}, S. and {Bassilana}, J. -L. and {Bauchet}, N. and {Becciani}, U. and {Bellazzini}, M. and {Berihuete}, A. and {Bernet}, M. and {Bertone}, S. and {Bianchi}, L. and {Binnenfeld}, A. and {Blanco-Cuaresma}, S. and {Blazere}, A. and {Boch}, T. and {Bombrun}, A. and {Bossini}, D. and {Bouquillon}, S. and {Bragaglia}, A. and {Bramante}, L. and {Breedt}, E. and {Bressan}, A. and {Brouillet}, N. and {Brugaletta}, E. and {Bucciarelli}, B. and {Burlacu}, A. and {Butkevich}, A.~G. and {Buzzi}, R. and {Caffau}, E. and {Cancelliere}, R. and {Cantat-Gaudin}, T. and {Carballo}, R. and {Carlucci}, T. and {Carnerero}, M.~I. and {Carrasco}, J.~M. and {Casamiquela}, L. and {Castellani}, M. and {Castro-Ginard}, A. and {Chaoul}, L. and {Charlot}, P. and {Chemin}, L. and {Chiaramida}, V. and {Chiavassa}, A. and {Chornay}, N. and {Comoretto}, G. and {Contursi}, G. and {Cooper}, W.~J. and {Cornez}, T. and {Cowell}, S. and {Crifo}, F. and {Cropper}, M. and {Crosta}, M. and {Crowley}, C. and {Dafonte}, C. and {Dapergolas}, A. and {David}, M. and {David}, P. and {de Laverny}, P. and {De Luise}, F. and {De March}, R. and {De Ridder}, J. and {de Souza}, R. and {de Torres}, A. and {del Peloso}, E.~F. and {del Pozo}, E. and {Delbo}, M. and {Delgado}, A. and {Delisle}, J. -B. and {Demouchy}, C. and {Dharmawardena}, T.~E. and {Di Matteo}, P. and {Diakite}, S. and {Diener}, C. and {Distefano}, E. and {Dolding}, C. and {Edvardsson}, B. and {Enke}, H. and {Fabre}, C. and {Fabrizio}, M. and {Faigler}, S. and {Fedorets}, G. and {Fernique}, P. and {Fienga}, A. and {Figueras}, F. and {Fournier}, Y. and {Fouron}, C. and {Fragkoudi}, F. and {Gai}, M. and {Garcia-Gutierrez}, A. and {Garcia-Reinaldos}, M. and {Garc{\'\i}a-Torres}, M. and {Garofalo}, A. and {Gavel}, A. and {Gavras}, P. and {Gerlach}, E. and {Geyer}, R. and {Giacobbe}, P. and {Gilmore}, G. and {Girona}, S. and {Giuffrida}, G. and {Gomel}, R. and {Gomez}, A. and {Gonz{\'a}lez-N{\'u}{\~n}ez}, J. and {Gonz{\'a}lez-Santamar{\'\i}a}, I. and {Gonz{\'a}lez-Vidal}, J.~J. and {Granvik}, M. and {Guillout}, P. and {Guiraud}, J. and {Guti{\'e}rrez-S{\'a}nchez}, R. and {Guy}, L.~P. and {Hatzidimitriou}, D. and {Hauser}, M. and {Haywood}, M. and {Helmer}, A. and {Helmi}, A. and {Sarmiento}, M.~H. and {Hidalgo}, S.~L. and {Hilger}, T. and {H{\l}adczuk}, N. and {Hobbs}, D. and {Holland}, G. and {Huckle}, H.~E. and {Jardine}, K. and {Jasniewicz}, G. and {Jean-Antoine Piccolo}, A. and {Jim{\'e}nez-Arranz}, {\'O}. and {Jorissen}, A. and {Juaristi Campillo}, J. and {Julbe}, F. and {Karbevska}, L. and {Kervella}, P. and {Khanna}, S. and {Kontizas}, M. and {Kordopatis}, G. and {Korn}, A.~J. and {K{\'o}sp{\'a}l}, {\'A}. and {Kostrzewa-Rutkowska}, Z. and {Kruszy{\'n}ska}, K. and {Kun}, M. and {Laizeau}, P. and {Lambert}, S. and {Lanza}, A.~F. and {Lasne}, Y. and {Le Campion}, J. -F. and {Lebreton}, Y. and {Lebzelter}, T. and {Leccia}, S. and {Leclerc}, N. and {Lecoeur-Taibi}, I. and {Liao}, S. and {Licata}, E.~L. and {Lindstr{\o}m}, H.~E.~P. and {Lister}, T.~A. and {Livanou}, E. and {Lobel}, A. and {Lorca}, A. and {Loup}, C. and {Madrero Pardo}, P. and {Magdaleno Romeo}, A. and {Managau}, S. and {Mann}, R.~G. and {Manteiga}, M. and {Marchant}, J.~M. and {Marconi}, M. and {Marcos}, J. and {Marcos Santos}, M.~M.~S. and {Mar{\'\i}n Pina}, D. and {Marinoni}, S. and {Marocco}, F. and {Marshall}, D.~J. and {Martin Polo}, L. and {Mart{\'\i}n-Fleitas}, J.~M. and {Marton}, G. and {Mary}, N. and {Masip}, A. and {Massari}, D. and {Mastrobuono-Battisti}, A. and {Mazeh}, T. and {McMillan}, P.~J. and {Messina}, S. and {Michalik}, D. and {Millar}, N.~R. and {Mints}, A. and {Molina}, D. and {Molinaro}, R. and {Moln{\'a}r}, L. and {Monari}, G. and {Mongui{\'o}}, M. and {Montegriffo}, P. and {Montero}, A. and {Mor}, R. and {Mora}, A. and {Morbidelli}, R. and {Morel}, T. and {Morris}, D. and {Muraveva}, T. and {Murphy}, C.~P. and {Musella}, I. and {Nagy}, Z. and {Noval}, L. and {Oca{\~n}a}, F. and {Ogden}, A. and {Ordenovic}, C. and {Osinde}, J.~O. and {Pagani}, C. and {Pagano}, I. and {Palaversa}, L. and {Palicio}, P.~A. and {Pallas-Quintela}, L. and {Panahi}, A. and {Payne-Wardenaar}, S. and {Pe{\~n}alosa Esteller}, X. and {Penttil{\"a}}, A. and {Pichon}, B. and {Piersimoni}, A.~M. and {Pineau}, F. -X. and {Plachy}, E. and {Plum}, G. and {Poggio}, E. and {Pr{\v{s}}a}, A. and {Pulone}, L. and {Racero}, E. and {Ragaini}, S. and {Rainer}, M. and {Raiteri}, C.~M. and {Rambaux}, N. and {Ramos}, P. and {Ramos-Lerate}, M. and {Re Fiorentin}, P. and {Regibo}, S. and {Richards}, P.~J. and {Rios Diaz}, C. and {Ripepi}, V. and {Riva}, A. and {Rix}, H. -W. and {Rixon}, G. and {Robichon}, N. and {Robin}, A.~C. and {Robin}, C. and {Roelens}, M. and {Rogues}, H.~R.~O. and {Rohrbasser}, L. and {Romero-G{\'o}mez}, M. and {Rowell}, N. and {Royer}, F. and {Ruz Mieres}, D. and {Rybicki}, K.~A. and {Sadowski}, G. and {S{\'a}ez N{\'u}{\~n}ez}, A. and {Sagrist{\`a} Sell{\'e}s}, A. and {Sahlmann}, J. and {Salguero}, E. and {Samaras}, N. and {Sanchez Gimenez}, V. and {Sanna}, N. and {Santove{\~n}a}, R. and {Sarasso}, M. and {Schultheis}, M. and {Sciacca}, E. and {Segol}, M. and {Segovia}, J.~C. and {S{\'e}gransan}, D. and {Semeux}, D. and {Shahaf}, S. and {Siddiqui}, H.~I. and {Siebert}, A. and {Siltala}, L. and {Silvelo}, A. and {Slezak}, E. and {Slezak}, I. and {Smart}, R.~L. and {Snaith}, O.~N. and {Solano}, E. and {Solitro}, F. and {Souami}, D. and {Souchay}, J. and {Spagna}, A. and {Spina}, L. and {Spoto}, F. and {Steele}, I.~A. and {Steidelm{\"u}ller}, H. and {Stephenson}, C.~A. and {S{\"u}veges}, M. and {Surdej}, J. and {Szabados}, L. and {Szegedi-Elek}, E. and {Taris}, F. and {Taylor}, M.~B. and {Teixeira}, R. and {Tolomei}, L. and {Tonello}, N. and {Torra}, F. and {Torra}, J. and {Torralba Elipe}, G. and {Trabucchi}, M. and {Tsounis}, A.~T. and {Turon}, C. and {Ulla}, A. and {Unger}, N. and {Vaillant}, M.~V. and {van Dillen}, E. and {van Reeven}, W. and {Vanel}, O. and {Vecchiato}, A. and {Viala}, Y. and {Vicente}, D. and {Voutsinas}, S. and {Weiler}, M. and {Wevers}, T. and {Wyrzykowski}, {\L}. and {Yoldas}, A. and {Yvard}, P. and {Zhao}, H. and {Zorec}, J. and {Zucker}, S. and {Zwitter}, T.},
        title = "{Gaia Data Release 3. Summary of the content and survey properties}",
      journal = {\aap},
         year = 2023,
        month = jun,
       volume = {674},
          eid = {A1},
        pages = {A1},
          doi = {10.1051/0004-6361/202243940},
archivePrefix = {arXiv},
       eprint = {2208.00211},
 primaryClass = {astro-ph.GA},
       adsurl = {https://ui.adsabs.harvard.edu/abs/2023A&A...674A...1G}
}

@ARTICLE{shapiro2014,
       author = {{Shapiro}, A.~I. and {Solanki}, S.~K. and {Krivova}, N.~A. and {Schmutz}, W.~K. and {Ball}, W.~T. and {Knaack}, R. and {Rozanov}, E.~V. and {Unruh}, Y.~C.},
        title = "{Variability of Sun-like stars: reproducing observed photometric trends}",
      journal = {\aap},
         year = 2014,
        month = sep,
       volume = {569},
          eid = {A38},
        pages = {A38},
          doi = {10.1051/0004-6361/201323086},
archivePrefix = {arXiv},
       eprint = {1406.2383},
 primaryClass = {astro-ph.SR},
       adsurl = {https://ui.adsabs.harvard.edu/abs/2014A&A...569A..38S}
}

@ARTICLE{schutte2023,
       author = {{Schutte}, Maria C. and {Hebb}, Leslie and {Wisniewski}, John P. and {Ca{\~n}as}, Caleb I. and {Libby-Roberts}, Jessica E. and {Lin}, Andrea S.~J. and {Robertson}, Paul and {Stef{\'a}nsson}, Gumundur},
        title = "{Measuring the Temperature of Starspots from Multi-filter Photometry}",
      journal = {\aj},
         year = 2023,
        month = sep,
       volume = {166},
       number = {3},
          eid = {92},
        pages = {92},
          doi = {10.3847/1538-3881/ace59c},
archivePrefix = {arXiv},
       eprint = {2307.16015},
 primaryClass = {astro-ph.SR},
       adsurl = {https://ui.adsabs.harvard.edu/abs/2023AJ....166...92S}
}

@ARTICLE{ojeda2011,
       author = {{Sanchis-Ojeda}, Roberto and {Winn}, Joshua N.},
        title = "{Starspots, Spin-Orbit Misalignment, and Active Latitudes in the HAT-P-11 Exoplanetary System}",
      journal = {\apj},
         year = 2011,
        month = dec,
       volume = {743},
       number = {1},
          eid = {61},
        pages = {61},
          doi = {10.1088/0004-637X/743/1/61},
archivePrefix = {arXiv},
       eprint = {1107.2920},
 primaryClass = {astro-ph.EP},
       adsurl = {https://ui.adsabs.harvard.edu/abs/2011ApJ...743...61S}
}

@ARTICLE{berdyugina2005,
       author = {{Berdyugina}, Svetlana V.},
        title = "{Starspots: A Key to the Stellar Dynamo}",
      journal = {Living Reviews in Solar Physics},
         year = 2005,
        month = dec,
       volume = {2},
       number = {1},
          eid = {8},
        pages = {8},
          doi = {10.12942/lrsp-2005-8},
       adsurl = {https://ui.adsabs.harvard.edu/abs/2005LRSP....2....8B}
}

@ARTICLE{herbst2021,
       author = {{Herbst}, Konstantin and {Papaioannou}, Athanasios and {Airapetian}, Vladimir S. and {Atri}, Dimitra},
        title = "{From Starspots to Stellar Coronal Mass Ejections{\textemdash}Revisiting Empirical Stellar Relations}",
      journal = {\apj},
         year = 2021,
        month = feb,
       volume = {907},
       number = {2},
          eid = {89},
        pages = {89},
          doi = {10.3847/1538-4357/abcc04},
archivePrefix = {arXiv},
       eprint = {2011.03761},
 primaryClass = {astro-ph.SR},
       adsurl = {https://ui.adsabs.harvard.edu/abs/2021ApJ...907...89H}
}

@ARTICLE{smitha2025,
       author = {{Smitha}, H.~N. and {Shapiro}, Alexander I. and {Witzke}, Veronika and {Kostogryz}, Nadiia M. and {Unruh}, Yvonne C. and {Bhatia}, Tanayveer S. and {Cameron}, Robert and {Seager}, Sara and {Solanki}, Sami K.},
        title = "{First Calculations of Starspot Spectra Based on 3D Radiative Magnetohydrodynamics Simulations}",
      journal = {\apjl},
         year = 2025,
        month = jan,
       volume = {978},
       number = {1},
          eid = {L13},
        pages = {L13},
          doi = {10.3847/2041-8213/ad9aaa},
archivePrefix = {arXiv},
       eprint = {2411.14056},
 primaryClass = {astro-ph.SR},
       adsurl = {https://ui.adsabs.harvard.edu/abs/2025ApJ...978L..13S}
}

@ARTICLE{speagle2020,
       author = {{Speagle}, Joshua S.},
        title = "{DYNESTY: a dynamic nested sampling package for estimating Bayesian posteriors and evidences}",
      journal = {\mnras},
         year = 2020,
        month = apr,
       volume = {493},
       number = {3},
        pages = {3132-3158},
          doi = {10.1093/mnras/staa278},
archivePrefix = {arXiv},
       eprint = {1904.02180},
 primaryClass = {astro-ph.IM},
       adsurl = {https://ui.adsabs.harvard.edu/abs/2020MNRAS.493.3132S}
}

@ARTICLE{yee2018,
       author = {{Yee}, Samuel W. and {Petigura}, Erik A. and {Fulton}, Benjamin J. and {Knutson}, Heather A. and {Batygin}, Konstantin and {Bakos}, G{\'a}sp{\'a}r {\'A}. and {Hartman}, Joel D. and {Hirsch}, Lea A. and {Howard}, Andrew W. and {Isaacson}, Howard and {Kosiarek}, Molly R. and {Sinukoff}, Evan and {Weiss}, Lauren M.},
        title = "{HAT-P-11: Discovery of a Second Planet and a Clue to Understanding Exoplanet Obliquities}",
      journal = {\aj},
         year = 2018,
        month = jun,
       volume = {155},
       number = {6},
          eid = {255},
        pages = {255},
          doi = {10.3847/1538-3881/aabfec},
archivePrefix = {arXiv},
       eprint = {1805.09352},
 primaryClass = {astro-ph.EP},
       adsurl = {https://ui.adsabs.harvard.edu/abs/2018AJ....155..255Y}
}

@ARTICLE{morris2019,
       author = {{Morris}, Brett M. and {Curtis}, Jason L. and {Sakari}, Charli and {Hawley}, Suzanne L. and {Agol}, Eric},
        title = "{Stellar Properties of Active G and K Stars: Exploring the Connection between Starspots and Chromospheric Activity}",
      journal = {\aj},
         year = 2019,
        month = sep,
       volume = {158},
       number = {3},
          eid = {101},
        pages = {101},
          doi = {10.3847/1538-3881/ab2e04},
archivePrefix = {arXiv},
       eprint = {1907.00423},
 primaryClass = {astro-ph.SR},
       adsurl = {https://ui.adsabs.harvard.edu/abs/2019AJ....158..101M}
}

@ARTICLE{bakos2010,
       author = {{Bakos}, G. {\'A}. and {Torres}, G. and {P{\'a}l}, A. and {Hartman}, J. and {Kov{\'a}cs}, G{\'e}za and {Noyes}, R.~W. and {Latham}, D.~W. and {Sasselov}, D.~D. and {Sip{\H{o}}cz}, B. and {Esquerdo}, G.~A. and {Fischer}, D.~A. and {Johnson}, J.~A. and {Marcy}, G.~W. and {Butler}, R.~P. and {Isaacson}, H. and {Howard}, A. and {Vogt}, S. and {Kov{\'a}cs}, G{\'a}bor and {Fernandez}, J. and {Mo{\'o}r}, A. and {Stefanik}, R.~P. and {L{\'a}z{\'a}r}, J. and {Papp}, I. and {S{\'a}ri}, P.},
        title = "{HAT-P-11b: A Super-Neptune Planet Transiting a Bright K Star in the Kepler Field}",
      journal = {\apj},
         year = 2010,
        month = feb,
       volume = {710},
       number = {2},
        pages = {1724-1745},
          doi = {10.1088/0004-637X/710/2/1724},
archivePrefix = {arXiv},
       eprint = {0901.0282},
 primaryClass = {astro-ph.EP},
       adsurl = {https://ui.adsabs.harvard.edu/abs/2010ApJ...710.1724B}
}

@ARTICLE{Moran2023,
       author = {{Moran}, Sarah E. and {Stevenson}, Kevin B. and {Sing}, David K. and {MacDonald}, Ryan J. and {Kirk}, James and {Lustig-Yaeger}, Jacob and {Peacock}, Sarah and {Mayorga}, L.~C. and {Bennett}, Katherine A. and {L{\'o}pez-Morales}, Mercedes and {May}, E.~M. and {Rustamkulov}, Zafar and {Valenti}, Jeff A. and {Adams Redai}, J{\'e}a I. and {Alam}, Munazza K. and {Batalha}, Natasha E. and {Fu}, Guangwei and {Gonzalez-Quiles}, Junellie and {Highland}, Alicia N. and {Kruse}, Ethan and {Lothringer}, Joshua D. and {Ortiz Ceballos}, Kevin N. and {Sotzen}, Kristin S. and {Wakeford}, Hannah R.},
        title = "{High Tide or Riptide on the Cosmic Shoreline? A Water-rich Atmosphere or Stellar Contamination for the Warm Super-Earth GJ 486b from JWST Observations}",
      journal = {\apjl},
         year = 2023,
        month = may,
       volume = {948},
       number = {1},
          eid = {L11},
        pages = {L11},
          doi = {10.3847/2041-8213/accb9c},
archivePrefix = {arXiv},
       eprint = {2305.00868},
 primaryClass = {astro-ph.EP},
       adsurl = {https://ui.adsabs.harvard.edu/abs/2023ApJ...948L..11M}
}

@ARTICLE{Lim2023,
       author = {{Lim}, Olivia and {Benneke}, Bj{\"o}rn and {Doyon}, Ren{\'e} and {MacDonald}, Ryan J. and {Piaulet}, Caroline and {Artigau}, {\'E}tienne and {Coulombe}, Louis-Philippe and {Radica}, Michael and {L'Heureux}, Alexandrine and {Albert}, Lo{\"\i}c and {Rackham}, Benjamin V. and {de Wit}, Julien and {Salhi}, Salma and {Roy}, Pierre-Alexis and {Flagg}, Laura and {Fournier-Tondreau}, Marylou and {Taylor}, Jake and {Cook}, Neil J. and {Lafreni{\`e}re}, David and {Cowan}, Nicolas B. and {Kaltenegger}, Lisa and {Rowe}, Jason F. and {Espinoza}, N{\'e}stor and {Dang}, Lisa and {Darveau-Bernier}, Antoine},
        title = "{Atmospheric Reconnaissance of TRAPPIST-1 b with JWST/NIRISS: Evidence for Strong Stellar Contamination in the Transmission Spectra}",
      journal = {\apjl},
         year = 2023,
        month = sep,
       volume = {955},
       number = {1},
          eid = {L22},
        pages = {L22},
          doi = {10.3847/2041-8213/acf7c4},
archivePrefix = {arXiv},
       eprint = {2309.07047},
 primaryClass = {astro-ph.EP},
       adsurl = {https://ui.adsabs.harvard.edu/abs/2023ApJ...955L..22L}
}

@ARTICLE{Radica2025,
       author = {{Radica}, Michael and {Piaulet-Ghorayeb}, Caroline and {Taylor}, Jake and {Coulombe}, Louis-Philippe and {Benneke}, Bj{\"o}rn and {Albert}, Loic and {Artigau}, {\'E}tienne and {Cowan}, Nicolas B. and {Doyon}, Ren{\'e} and {Lafreni{\`e}re}, David and {L'Heureux}, Alexandrine and {Lim}, Olivia},
        title = "{Promise and Peril: Stellar Contamination and Strict Limits on the Atmosphere Composition of TRAPPIST-1 c from JWST NIRISS Transmission Spectra}",
      journal = {\apjl},
         year = 2025,
        month = jan,
       volume = {979},
       number = {1},
          eid = {L5},
        pages = {L5},
          doi = {10.3847/2041-8213/ada381},
archivePrefix = {arXiv},
       eprint = {2409.19333},
 primaryClass = {astro-ph.EP},
       adsurl = {https://ui.adsabs.harvard.edu/abs/2025ApJ...979L...5R}
}

@ARTICLE{Rathcke2025,
       author = {{Rathcke}, Alexander D. and {Buchhave}, Lars A. and {Wit}, Julien de and {Rackham}, Benjamin V. and {August}, Prune C. and {Diamond-Lowe}, Hannah and {Mendon{\c{C}}a}, Jo{\~a}o M. and {Bello-Arufe}, Aaron and {L{\'o}pez-Morales}, Mercedes and {Kitzmann}, Daniel and {Heng}, Kevin},
        title = "{Stellar Contamination Correction Using Back-to-back Transits of TRAPPIST-1 b and c}",
      journal = {\apjl},
         year = 2025,
        month = jan,
       volume = {979},
       number = {1},
          eid = {L19},
        pages = {L19},
          doi = {10.3847/2041-8213/ada5c7},
archivePrefix = {arXiv},
       eprint = {2412.16541},
 primaryClass = {astro-ph.EP},
       adsurl = {https://ui.adsabs.harvard.edu/abs/2025ApJ...979L..19R}
}

@ARTICLE{Rackham2023,
       author = {{Rackham}, Benjamin V. and {Espinoza}, N{\'e}stor and {Berdyugina}, Svetlana V. and {Korhonen}, Heidi and {MacDonald}, Ryan J. and {Montet}, Benjamin T. and {Morris}, Brett M. and {Oshagh}, Mahmoudreza and {Shapiro}, Alexander I. and {Unruh}, Yvonne C. and {Quintana}, Elisa V. and {Zellem}, Robert T. and {Apai}, D{\'a}niel and {Barclay}, Thomas and {Barstow}, Joanna K. and {Bruno}, Giovanni and {Carone}, Ludmila and {Casewell}, Sarah L. and {Cegla}, Heather M. and {Criscuoli}, Serena and {Fischer}, Catherine and {Fournier}, Damien and {Giampapa}, Mark S. and {Giles}, Helen and {Iyer}, Aishwarya and {Kopp}, Greg and {Kostogryz}, Nadiia M. and {Krivova}, Natalie and {Mallonn}, Matthias and {McGruder}, Chima and {Molaverdikhani}, Karan and {Newton}, Elisabeth R. and {Panja}, Mayukh and {Peacock}, Sarah and {Reardon}, Kevin and {Roettenbacher}, Rachael M. and {Scandariato}, Gaetano and {Solanki}, Sami and {Stassun}, Keivan G. and {Steiner}, Oskar and {Stevenson}, Kevin B. and {Tregloan-Reed}, Jeremy and {Valio}, Adriana and {Wedemeyer}, Sven and {Welbanks}, Luis and {Yu}, Jie and {Alam}, Munazza K. and {Davenport}, James R.~A. and {Deming}, Drake and {Dong}, Chuanfei and {Ducrot}, Elsa and {Fisher}, Chloe and {Gilbert}, Emily and {Kostov}, Veselin and {L{\'o}pez-Morales}, Mercedes and {Line}, Mike and {Mo{\v{c}}nik}, Teo and {Mullally}, Susan and {Paudel}, Rishi R. and {Ribas}, Ignasi and {Valenti}, Jeff A.},
        title = "{The effect of stellar contamination on low-resolution transmission spectroscopy: needs identified by NASA's Exoplanet Exploration Program Study Analysis Group 21}",
      journal = {RAS Techniques and Instruments},
         year = 2023,
        month = jan,
       volume = {2},
       number = {1},
        pages = {148-206},
          doi = {10.1093/rasti/rzad009},
archivePrefix = {arXiv},
       eprint = {2201.09905},
 primaryClass = {astro-ph.IM},
       adsurl = {https://ui.adsabs.harvard.edu/abs/2023RASTI...2..148R}
}

@ARTICLE{Rackham2024,
       author = {{Rackham}, Benjamin V. and {de Wit}, Julien},
        title = "{Toward Robust Corrections for Stellar Contamination in JWST Exoplanet Transmission Spectra}",
      journal = {\aj},
         year = 2024,
        month = aug,
       volume = {168},
       number = {2},
          eid = {82},
        pages = {82},
          doi = {10.3847/1538-3881/ad5833},
archivePrefix = {arXiv},
       eprint = {2303.15418},
 primaryClass = {astro-ph.EP},
       adsurl = {https://ui.adsabs.harvard.edu/abs/2024AJ....168...82R}
}

@ARTICLE{Pont2008,
       author = {{Pont}, F. and {Knutson}, H. and {Gilliland}, R.~L. and {Moutou}, C. and {Charbonneau}, D.},
        title = "{Detection of atmospheric haze on an extrasolar planet: the 0.55-1.05 {\ensuremath{\mu}}m transmission spectrum of HD 189733b with the HubbleSpaceTelescope}",
      journal = {\mnras},
         year = 2008,
        month = mar,
       volume = {385},
       number = {1},
        pages = {109-118},
          doi = {10.1111/j.1365-2966.2008.12852.x},
archivePrefix = {arXiv},
       eprint = {0712.1374},
 primaryClass = {astro-ph},
       adsurl = {https://ui.adsabs.harvard.edu/abs/2008MNRAS.385..109P}
}

@ARTICLE{Sing2011,
       author = {{Sing}, D.~K. and {Pont}, F. and {Aigrain}, S. and {Charbonneau}, D. and {D{\'e}sert}, J. -M. and {Gibson}, N. and {Gilliland}, R. and {Hayek}, W. and {Henry}, G. and {Knutson}, H. and {Lecavelier Des Etangs}, A. and {Mazeh}, T. and {Shporer}, A.},
        title = "{Hubble Space Telescope transmission spectroscopy of the exoplanet HD 189733b: high-altitude atmospheric haze in the optical and near-ultraviolet with STIS}",
      journal = {\mnras},
         year = 2011,
        month = sep,
       volume = {416},
       number = {2},
        pages = {1443-1455},
          doi = {10.1111/j.1365-2966.2011.19142.x},
archivePrefix = {arXiv},
       eprint = {1103.0026},
 primaryClass = {astro-ph.EP},
       adsurl = {https://ui.adsabs.harvard.edu/abs/2011MNRAS.416.1443S}
}

@ARTICLE{McCullough2014,
       author = {{McCullough}, P.~R. and {Crouzet}, N. and {Deming}, D. and {Madhusudhan}, N.},
        title = "{Water Vapor in the Spectrum of the Extrasolar Planet HD 189733b. I. The Transit}",
      journal = {\apj},
         year = 2014,
        month = aug,
       volume = {791},
       number = {1},
          eid = {55},
        pages = {55},
          doi = {10.1088/0004-637X/791/1/55},
archivePrefix = {arXiv},
       eprint = {1407.2462},
 primaryClass = {astro-ph.SR},
       adsurl = {https://ui.adsabs.harvard.edu/abs/2014ApJ...791...55M}
}

@ARTICLE{Espinoza2019,
       author = {{Espinoza}, N{\'e}stor and {Rackham}, Benjamin V. and {Jord{\'a}n}, Andr{\'e}s and {Apai}, D{\'a}niel and {L{\'o}pez-Morales}, Mercedes and {Osip}, David J. and {Grimm}, Simon L. and {Hoeijmakers}, Jens and {Wilson}, Paul A. and {Bixel}, Alex and {McGruder}, Chima and {Rodler}, Florian and {Weaver}, Ian and {Lewis}, Nikole K. and {Fortney}, Jonathan J. and {Fraine}, Jonathan},
        title = "{ACCESS: a featureless optical transmission spectrum for WASP-19b from Magellan/IMACS}",
      journal = {\mnras},
         year = 2019,
        month = jan,
       volume = {482},
       number = {2},
        pages = {2065-2087},
          doi = {10.1093/mnras/sty2691},
archivePrefix = {arXiv},
       eprint = {1807.10652},
 primaryClass = {astro-ph.EP},
       adsurl = {https://ui.adsabs.harvard.edu/abs/2019MNRAS.482.2065E}
}

@MISC{goudfrooij1998_Second,
       author = {{Goudfrooij}, Paul and {Christensen}, Jennifer A.},
        title = "{STIS Near-IR Fringing. III. A Tutorial on the Use of the IRAF Tasks}",
 howpublished = {STIS Instrument Science Report 98-29, 14 pages},
         year = 1998,
        month = nov,
        pages = {29},
       adsurl = {https://ui.adsabs.harvard.edu/abs/1998stis.rept...29G}
}

@ARTICLE{hauschildt2025,
       author = {{Hauschildt}, P.~H. and {Barman}, T. and {Baron}, E. and {Aufdenberg}, J.~P. and {Schweitzer}, A.},
        title = "{The NewEra model grid}",
      journal = {\aap},
         year = 2025,
        month = jun,
       volume = {698},
          eid = {A47},
        pages = {A47},
          doi = {10.1051/0004-6361/202554171},
archivePrefix = {arXiv},
       eprint = {2504.17597},
 primaryClass = {astro-ph.SR},
       adsurl = {https://ui.adsabs.harvard.edu/abs/2025A&A...698A..47H}
}

@ARTICLE{beky2014b,
       author = {{B{\'e}ky}, Bence and {Kipping}, David M. and {Holman}, Matthew J.},
        title = "{SPOTROD: a semi-analytic model for transits of spotted stars}",
      journal = {\mnras},
         year = 2014,
        month = aug,
       volume = {442},
       number = {4},
        pages = {3686-3699},
          doi = {10.1093/mnras/stu1061},
archivePrefix = {arXiv},
       eprint = {1407.4465},
 primaryClass = {astro-ph.EP},
       adsurl = {https://ui.adsabs.harvard.edu/abs/2014MNRAS.442.3686B}
}

@ARTICLE{andretta2005,
       author = {{Andretta}, V. and {Bus{\`a}}, I. and {Gomez}, M.~T. and {Terranegra}, L.},
        title = "{The Ca II Infrared Triplet as a stellar activity diagnostic . I. Non-LTE photospheric profiles and definition of the R$_{IRT}$ indicator}",
      journal = {\aap},
         year = 2005,
        month = feb,
       volume = {430},
        pages = {669-677},
          doi = {10.1051/0004-6361:20041745},
       adsurl = {https://ui.adsabs.harvard.edu/abs/2005A&A...430..669A}
}

@ARTICLE{rayner2003,
       author = {{Rayner}, J.~T. and {Toomey}, D.~W. and {Onaka}, P.~M. and {Denault}, A.~J. and {Stahlberger}, W.~E. and {Vacca}, W.~D. and {Cushing}, M.~C. and {Wang}, S.},
        title = "{SpeX: A Medium-Resolution 0.8-5.5 Micron Spectrograph and Imager for the NASA Infrared Telescope Facility}",
      journal = {\pasp},
         year = 2003,
        month = mar,
       volume = {115},
       number = {805},
        pages = {362-382},
          doi = {10.1086/367745},
       adsurl = {https://ui.adsabs.harvard.edu/abs/2003PASP..115..362R}
}

@ARTICLE{morris2017CaHK,
       author = {{Morris}, Brett M. and {Hawley}, Suzanne L. and {Hebb}, Leslie and {Sakari}, Charli and {Davenport}, James. R.~A. and {Isaacson}, Howard and {Howard}, Andrew W. and {Montet}, Benjamin T. and {Agol}, Eric},
        title = "{Chromospheric Activity of HAT-P-11: An Unusually Active Planet-hosting K Star}",
      journal = {\apj},
         year = 2017,
        month = oct,
       volume = {848},
       number = {1},
          eid = {58},
        pages = {58},
          doi = {10.3847/1538-4357/aa8cca},
archivePrefix = {arXiv},
       eprint = {1709.03913},
 primaryClass = {astro-ph.SR},
       adsurl = {https://ui.adsabs.harvard.edu/abs/2017ApJ...848...58M}
}

@ARTICLE{shapiro2017,
       author = {{Shapiro}, A.~I. and {Solanki}, S.~K. and {Krivova}, N.~A. and {Cameron}, R.~H. and {Yeo}, K.~L. and {Schmutz}, W.~K.},
        title = "{The nature of solar brightness variations}",
      journal = {Nature Astronomy},
         year = 2017,
        month = aug,
       volume = {1},
        pages = {612-616},
          doi = {10.1038/s41550-017-0217-y},
archivePrefix = {arXiv},
       eprint = {1711.04156},
 primaryClass = {astro-ph.SR},
       adsurl = {https://ui.adsabs.harvard.edu/abs/2017NatAs...1..612S}
}

@ARTICLE{wilson2025,
       author = {{Wilson}, David J. and {Froning}, Cynthia S. and {Duvvuri}, Girish M. and {Youngblood}, Allison and {France}, Kevin and {Brown}, Alexander and {Schneider}, P. Christian and {Berta-Thompson}, Zachory and {Buccino}, Andrea P. and {Linsky}, Jeffrey and {Loyd}, R.~O. Parke and {Miguel}, Yamila and {Newton}, Elisabeth and {Pineda}, J. Sebastian and {Redfield}, Seth and {Roberge}, Aki and {Rugheimer}, Sarah and {Vieytes}, Mariela C.},
        title = "{The Mega-MUSCLES Treasury Survey: X-Ray to Infrared Spectral Energy Distributions of a Representative Sample of M Dwarfs}",
      journal = {\apj},
         year = 2025,
        month = jan,
       volume = {978},
       number = {1},
          eid = {85},
        pages = {85},
          doi = {10.3847/1538-4357/ad9251},
archivePrefix = {arXiv},
       eprint = {2411.07394},
 primaryClass = {astro-ph.EP},
       adsurl = {https://ui.adsabs.harvard.edu/abs/2025ApJ...978...85W}
}

@ARTICLE{iyer2025,
       author = {{Iyer}, Aishwarya R. and {Line}, Michael R. and {Muirhead}, Philip S. and {Fortney}, Jonathan J. and {Faherty}, Jacqueline K.},
        title = "{The SPHINX M dwarf Spectral Grid. II. New Model Atmospheres and Spectra to Derive Fundamental Properties of mid-to-late type M-dwarfs}",
      journal = {arXiv e-prints},
         year = 2025,
        month = dec,
          eid = {arXiv:2512.02269},
        pages = {arXiv:2512.02269},
          doi = {10.48550/arXiv.2512.02269},
archivePrefix = {arXiv},
       eprint = {2512.02269},
 primaryClass = {astro-ph.SR},
       adsurl = {https://ui.adsabs.harvard.edu/abs/2025arXiv251202269I}
}

@ARTICLE{kostogryz2024,
       author = {{Kostogryz}, Nadiia M. and {Shapiro}, Alexander I. and {Witzke}, Veronika and {Cameron}, Robert H. and {Gizon}, Laurent and {Krivova}, Natalie A. and {Ludwig}, Hans-G. and {Maxted}, Pierre F.~L. and {Seager}, Sara and {Solanki}, Sami K. and {Valenti}, Jeff},
        title = "{Magnetic origin of the discrepancy between stellar limb-darkening models and observations}",
      journal = {Nature Astronomy},
         year = 2024,
        month = jul,
       volume = {8},
        pages = {929-937},
          doi = {10.1038/s41550-024-02252-5},
       adsurl = {https://ui.adsabs.harvard.edu/abs/2024NatAs...8..929K}
}

@ARTICLE{wakeford2019,
       author = {{Wakeford}, H.~R. and {Lewis}, N.~K. and {Fowler}, J. and {Bruno}, G. and {Wilson}, T.~J. and {Moran}, S.~E. and {Valenti}, J. and {Batalha}, N.~E. and {Filippazzo}, J. and {Bourrier}, V. and {H{\"o}rst}, S.~M. and {Lederer}, S.~M. and {de Wit}, J.},
        title = "{Disentangling the Planet from the Star in Late-Type M Dwarfs: A Case Study of TRAPPIST-1g}",
      journal = {\aj},
         year = 2019,
        month = jan,
       volume = {157},
       number = {1},
          eid = {11},
        pages = {11},
          doi = {10.3847/1538-3881/aaf04d},
archivePrefix = {arXiv},
       eprint = {1811.04877},
 primaryClass = {astro-ph.EP},
       adsurl = {https://ui.adsabs.harvard.edu/abs/2019AJ....157...11W}
}

@ARTICLE{murray2025,
       author = {{Murray}, C.~A. and {Berta-Thompson}, Z.},
        title = "{Quantifying the Impact of Starspot-Crossing Events on Retrieved Parameters from Transit Lightcurves}",
      journal = {arXiv e-prints},
         year = 2025,
        month = nov,
          eid = {arXiv:2511.03045},
        pages = {arXiv:2511.03045},
          doi = {10.48550/arXiv.2511.03045},
archivePrefix = {arXiv},
       eprint = {2511.03045},
 primaryClass = {astro-ph.EP},
       adsurl = {https://ui.adsabs.harvard.edu/abs/2025arXiv251103045M}
}

@ARTICLE{sowmya2023,
       author = {{Sowmya}, K. and {Shapiro}, A.~I. and {Rouppe van der Voort}, L.~H.~M. and {Krivova}, N.~A. and {Solanki}, S.~K.},
        title = "{Modeling Stellar Ca II H and K Emission Variations: Spot Contribution to the S-index}",
      journal = {\apjl},
         year = 2023,
        month = oct,
       volume = {956},
       number = {1},
          eid = {L10},
        pages = {L10},
          doi = {10.3847/2041-8213/acf92a},
archivePrefix = {arXiv},
       eprint = {2309.03690},
 primaryClass = {astro-ph.SR},
       adsurl = {https://ui.adsabs.harvard.edu/abs/2023ApJ...956L..10S}
}

@ARTICLE{boyajian2012,
       author = {{Boyajian}, Tabetha S. and {von Braun}, Kaspar and {van Belle}, Gerard and {McAlister}, Harold A. and {ten Brummelaar}, Theo A. and {Kane}, Stephen R. and {Muirhead}, Philip S. and {Jones}, Jeremy and {White}, Russel and {Schaefer}, Gail and {Ciardi}, David and {Henry}, Todd and {L{\'o}pez-Morales}, Mercedes and {Ridgway}, Stephen and {Gies}, Douglas and {Jao}, Wei-Chun and {Rojas-Ayala}, B{\'a}rbara and {Parks}, J. Robert and {Sturmann}, Laszlo and {Sturmann}, Judit and {Turner}, Nils H. and {Farrington}, Chris and {Goldfinger}, P.~J. and {Berger}, David H.},
        title = "{Stellar Diameters and Temperatures. II. Main-sequence K- and M-stars}",
      journal = {\apj},
         year = 2012,
        month = oct,
       volume = {757},
       number = {2},
          eid = {112},
        pages = {112},
          doi = {10.1088/0004-637X/757/2/112},
archivePrefix = {arXiv},
       eprint = {1208.2431},
 primaryClass = {astro-ph.SR},
       adsurl = {https://ui.adsabs.harvard.edu/abs/2012ApJ...757..112B}
}

@ARTICLE{morris2025,
       author = {{Morris}, Brett M. and {Hebb}, Leslie and {Hawley}, Suzanne L. and {Jones}, Kathryn and {Romney}, Jake},
        title = "{OWLS. I. The Olin Wilson Legacy Survey}",
      journal = {\apj},
         year = 2025,
        month = sep,
       volume = {990},
       number = {2},
          eid = {113},
        pages = {113},
          doi = {10.3847/1538-4357/adeca5},
archivePrefix = {arXiv},
       eprint = {2507.07330},
 primaryClass = {astro-ph.SR},
       adsurl = {https://ui.adsabs.harvard.edu/abs/2025ApJ...990..113M}
}

@ARTICLE{piaulet2025,
       author = {{Piaulet-Ghorayeb}, Caroline and {Benneke}, Bj{\"o}rn and {Radica}, Michael and {Raul}, Eshan and {Coulombe}, Louis-Philippe and {Ahrer}, Eva-Maria and {Kubyshkina}, Daria and {Howard}, Ward S. and {Krissansen-Totton}, Joshua and {MacDonald}, Ryan J. and {Roy}, Pierre-Alexis and {Louca}, Amy and {Christie}, Duncan and {Fournier-Tondreau}, Marylou and {Allart}, Romain and {Miguel}, Yamila and {Schlichting}, Hilke E. and {Welbanks}, Luis and {Cadieux}, Charles and {Dorn}, Caroline and {Evans-Soma}, Thomas M. and {Fortney}, Jonathan J. and {Pierrehumbert}, Raymond and {Lafreni{\`e}re}, David and {Acu{\~n}a}, Lorena and {Komacek}, Thaddeus and {Innes}, Hamish and {Beatty}, Thomas G. and {Cloutier}, Ryan and {Doyon}, Ren{\'e} and {Gagnebin}, Anna and {Gapp}, Cyril and {Knutson}, Heather A.},
        title = "{JWST/NIRISS Reveals the Water-rich ``Steam World'' Atmosphere of GJ 9827 d}",
      journal = {\apjl},
         year = 2024,
        month = oct,
       volume = {974},
       number = {1},
          eid = {L10},
        pages = {L10},
          doi = {10.3847/2041-8213/ad6f00},
archivePrefix = {arXiv},
       eprint = {2410.03527},
 primaryClass = {astro-ph.EP},
       adsurl = {https://ui.adsabs.harvard.edu/abs/2024ApJ...974L..10P}
}

@ARTICLE{tondreau2024,
       author = {{Fournier-Tondreau}, Marylou and {MacDonald}, Ryan J. and {Radica}, Michael and {Lafreni{\`e}re}, David and {Welbanks}, Luis and {Piaulet}, Caroline and {Coulombe}, Louis-Philippe and {Allart}, Romain and {Morel}, Kim and {Artigau}, {\'E}tienne and {Albert}, Lo{\"\i}c and {Lim}, Olivia and {Doyon}, Ren{\'e} and {Benneke}, Bj{\"o}rn and {Rowe}, Jason F. and {Darveau-Bernier}, Antoine and {Cowan}, Nicolas B. and {Lewis}, Nikole K. and {Cook}, Neil J. and {Flagg}, Laura and {Genest}, Fr{\'e}d{\'e}ric and {Pelletier}, Stefan and {Johnstone}, Doug and {Dang}, Lisa and {Kaltenegger}, Lisa and {Taylor}, Jake and {Turner}, Jake D.},
        title = "{Near-infrared transmission spectroscopy of HAT-P-18 b with NIRISS: Disentangling planetary and stellar features in the era of JWST}",
      journal = {\mnras},
         year = 2024,
        month = feb,
       volume = {528},
       number = {2},
        pages = {3354-3377},
          doi = {10.1093/mnras/stad3813},
archivePrefix = {arXiv},
       eprint = {2310.14950},
 primaryClass = {astro-ph.EP},
       adsurl = {https://ui.adsabs.harvard.edu/abs/2024MNRAS.528.3354F}
}

@ARTICLE{tondreau2025,
       author = {{Fournier-Tondreau}, Marylou and {Pan}, Yanbo and {Morel}, Kim and {Lafreni{\`e}re}, David and {MacDonald}, Ryan J. and {Coulombe}, Louis-Philippe and {Allart}, Romain and {Albert}, Lo{\"\i}c and {Radica}, Michael and {Piaulet-Ghorayeb}, Caroline and {Roy}, Pierre-Alexis and {Pelletier}, Stefan and {Dang}, Lisa and {Doyon}, Ren{\'e} and {Benneke}, Bj{\"o}rn and {Cowan}, Nicolas B. and {Darveau-Bernier}, Antoine and {Lim}, Olivia and {Artigau}, {\'E}tienne and {Johnstone}, Doug and {Kaltenegger}, Lisa and {Taylor}, Jake and {Flagg}, Laura},
        title = "{Transmission spectroscopy of WASP-52 b with JWST NIRISS: water and helium atmospheric absorption, alongside prominent star-spot crossings}",
      journal = {\mnras},
         year = 2025,
        month = may,
       volume = {539},
       number = {1},
        pages = {422-438},
          doi = {10.1093/mnras/staf489},
archivePrefix = {arXiv},
       eprint = {2412.17072},
 primaryClass = {astro-ph.EP},
       adsurl = {https://ui.adsabs.harvard.edu/abs/2025MNRAS.539..422F}
}

@ARTICLE{cretignier2024,
       author = {{Cretignier}, M. and {Pietrow}, A.~G.~M. and {Aigrain}, S.},
        title = "{Stellar surface information from the Ca II H\&K lines - I. Intensity profiles of the solar activity components}",
      journal = {\mnras},
         year = 2024,
        month = jan,
       volume = {527},
       number = {2},
        pages = {2940-2962},
          doi = {10.1093/mnras/stad3292},
archivePrefix = {arXiv},
       eprint = {2310.15926},
 primaryClass = {astro-ph.SR},
       adsurl = {https://ui.adsabs.harvard.edu/abs/2024MNRAS.527.2940C}
}

@INPROCEEDINGS{pandora,
       author = {{Quintana}, Elisa V. and {Dotson}, Jessie L. and {Col{\'o}n}, Knicole D. and {Barclay}, Thomas and {Supsinskas}, Pete and {Karburn}, Jordan and {Apai}, D{\'a}niel and {Hedges}, Christina and {Rackham}, Benjamin V. and {Rowe}, Jason F. and {Allen}, Natalie H. and {Bonney}, Paul and {Cano}, Samuel and {Christiansen}, Jessie L. and {Ciardi}, David and {Espinoza}, N{\'e}stor and {Foote}, Trevor O. and {Gilbert}, Emily A. and {Greene}, Thomas P. and {Hoffman}, Kelsey and {Hord}, Benjamin J. and {Iyer}, Aishwarya and {Kesseli}, Aurora and {Kostov}, Veselin B. and {Lewis}, Nikole K. and {Logsdon}, Sarah E. and {Mann}, Andrew W. and {Mansfield}, Megan and {Mason}, James and {Morris}, Brett M. and {Mosby}, Gregory and {Mullally}, Susan E. and {Newton}, Elisabeth R. and {Nguyen}, Fuda and {Schlieder}, Joshua E. and {Stevenson}, Kevin B. and {Wiser}, Lindsey S. and {Youngblood}, Allison and {Zellem}, Robert T.},
        title = "{The Pandora SmallSat: multiwavelength characterization of exoplanets and their host stars}",
    booktitle = {Space Telescopes and Instrumentation 2024: Optical, Infrared, and Millimeter Wave},
         year = 2024,
       editor = {{Coyle}, Laura E. and {Matsuura}, Shuji and {Perrin}, Marshall D.},
       series = {Society of Photo-Optical Instrumentation Engineers (SPIE) Conference Series},
       volume = {13092},
        month = aug,
          eid = {1309214},
        pages = {1309214},
          doi = {10.1117/12.3020633},
       adsurl = {https://ui.adsabs.harvard.edu/abs/2024SPIE13092E..14Q}
}

@ARTICLE{Garcia2022,
       author = {{Garcia}, L.~J. and {Moran}, S.~E. and {Rackham}, B.~V. and {Wakeford}, H.~R. and {Gillon}, M. and {de Wit}, J. and {Lewis}, N.~K.},
        title = "{HST/WFC3 transmission spectroscopy of the cold rocky planet TRAPPIST-1h}",
      journal = {\aap},
         year = 2022,
        month = sep,
       volume = {665},
          eid = {A19},
        pages = {A19},
          doi = {10.1051/0004-6361/202142603},
archivePrefix = {arXiv},
       eprint = {2203.13698},
 primaryClass = {astro-ph.EP},
       adsurl = {https://ui.adsabs.harvard.edu/abs/2022A&A...665A..19G}
}

@ARTICLE{Garcia2025,
       author = {{Garcia}, Lionel and {Rackham}, Benjamin and {Panwar}, Vatsal},
        title = "{spotter: Hardware-Accelerated Forward Models of Pixelated Stars}",
      journal = {The Journal of Open Source Software},
         year = 2025,
        month = dec,
       volume = {10},
       number = {116},
          eid = {8305},
        pages = {8305},
          doi = {10.21105/joss.08305},
       adsurl = {https://ui.adsabs.harvard.edu/abs/2025JOSS...10.8305G}
}

@ARTICLE{Pinhas2018,
       author = {{Pinhas}, Arazi and {Rackham}, Benjamin V. and {Madhusudhan}, Nikku and {Apai}, D{\'a}niel},
        title = "{Retrieval of planetary and stellar properties in transmission spectroscopy with AURA}",
      journal = {\mnras},
         year = 2018,
        month = nov,
       volume = {480},
       number = {4},
        pages = {5314-5331},
          doi = {10.1093/mnras/sty2209},
archivePrefix = {arXiv},
       eprint = {1808.10017},
 primaryClass = {astro-ph.EP},
       adsurl = {https://ui.adsabs.harvard.edu/abs/2018MNRAS.480.5314P}
}

@ARTICLE{Iyer2020,
       author = {{Iyer}, Aishwarya R. and {Line}, Michael R.},
        title = "{The Influence of Stellar Contamination on the Interpretation of Near-infrared Transmission Spectra of Sub-Neptune Worlds around M-dwarfs}",
      journal = {\apj},
         year = 2020,
        month = feb,
       volume = {889},
       number = {2},
          eid = {78},
        pages = {78},
          doi = {10.3847/1538-4357/ab612e},
archivePrefix = {arXiv},
       eprint = {1912.04389},
 primaryClass = {astro-ph.EP},
       adsurl = {https://ui.adsabs.harvard.edu/abs/2020ApJ...889...78I}
}

@ARTICLE{Cushing2004,
       author = {{Cushing}, Michael C. and {Vacca}, William D. and {Rayner}, John T.},
        title = "{Spextool: A Spectral Extraction Package for SpeX, a 0.8-5.5 Micron Cross-Dispersed Spectrograph}",
      journal = {\pasp},
         year = 2004,
        month = apr,
       volume = {116},
       number = {818},
        pages = {362-376},
          doi = {10.1086/382907},
       adsurl = {https://ui.adsabs.harvard.edu/abs/2004PASP..116..362C}
}

@ARTICLE{Barkaoui2023,
       author = {{Barkaoui}, K. and {Timmermans}, M. and {Soubkiou}, A. and {Rackham}, B.~V. and {Burgasser}, A.~J. and {Chouqar}, J. and {Pozuelos}, F.~J. and {Collins}, K.~A. and {Howell}, S.~B. and {Simcoe}, R. and {Melis}, C. and {Stassun}, K.~G. and {Tregloan-Reed}, J. and {Cointepas}, M. and {Gillon}, M. and {Bonfils}, X. and {Furlan}, E. and {Gnilka}, C.~L. and {Almenara}, J.~M. and {Alonso}, R. and {Benkhaldoun}, Z. and {Bonavita}, M. and {Bouchy}, F. and {Burdanov}, A. and {Chinchilla}, P. and {Davoudi}, F. and {Delrez}, L. and {Demangeon}, O. and {Dominik}, M. and {Demory}, B.-O. and {de Wit}, J. and {Dransfield}, G. and {Ducrot}, E. and {Fukui}, A. and {Hinse}, T.~C. and {Hooton}, M.~J. and {Jehin}, E. and {Jenkins}, J.~M. and {J{\o}rgensen}, U.~G. and {Latham}, D.~W. and {Garcia}, L. and {Carrazco-Gaxiola}, S. and {Ghachoui}, M. and {G{\'o}mez Maqueo Chew}, Y. and {G{\"u}nther}, M.~N. and {McCormac}, J. and {Murgas}, F. and {Murray}, C.~A. and {Narita}, N. and {Niraula}, P. and {Pedersen}, P.~P. and {Queloz}, D. and {Rebolo-L{\'o}pez}, R. and {Ricker}, G. and {Sabin}, L. and {Sajadian}, S. and {Schanche}, N. and {Schwarz}, R.~P. and {Seager}, S. and {Sebastian}, D. and {Sefako}, R. and {Sohy}, S. and {Southworth}, J. and {Srdoc}, G. and {Thompson}, S.~J. and {Triaud}, A.~H.~M.~J. and {Vanderspek}, R. and {Wells}, R.~D. and {Winn}, J.~N. and {Z{\'u}{\~n}iga-Fern{\'a}ndez}, S.},
        title = "{TOI-2084 b and TOI-4184 b: Two new sub-Neptunes around M dwarf stars}",
      journal = {\aap},
         year = 2023,
        month = sep,
       volume = {677},
          eid = {A38},
        pages = {A38},
          doi = {10.1051/0004-6361/202346838},
archivePrefix = {arXiv},
       eprint = {2306.15095},
 primaryClass = {astro-ph.EP},
       adsurl = {https://ui.adsabs.harvard.edu/abs/2023A&A...677A..38B}
}

@ARTICLE{Ghachoui2023,
       author = {{Ghachoui}, M. and {Soubkiou}, A. and {Wells}, R.~D. and {Rackham}, B.~V. and {Triaud}, A.~H.~M.~J. and {Sebastian}, D. and {Giacalone}, S. and {Stassun}, K.~G. and {Ciardi}, D.~R. and {Collins}, K.~A. and {Liu}, A. and {G{\'o}mez Maqueo Chew}, Y. and {Gillon}, M. and {Benkhaldoun}, Z. and {Delrez}, L. and {Eastman}, J.~D. and {Demangeon}, O. and {Barkaoui}, K. and {Burdanov}, A. and {Demory}, B.-O. and {de Wit}, J. and {Dransfield}, G. and {Ducrot}, E. and {Garcia}, L. and {G{\'o}mez-Mu{\~n}oz}, M.~A. and {Hooton}, M.~J. and {Jehin}, E. and {Murray}, C.~A. and {Pedersen}, P.~P. and {Pozuelos}, F.~J. and {Queloz}, D. and {Sabin}, L. and {Schanche}, N. and {Timmermans}, M. and {Gonzales}, E.~J. and {Dressing}, C.~D. and {Aganze}, C. and {Burgasser}, A.~J. and {Gerasimov}, R. and {Hsu}, C. and {Theissen}, C.~A. and {Charbonneau}, D. and {Jenkins}, J.~M. and {Latham}, D.~W. and {Ricker}, G. and {Seager}, S. and {Shporer}, A. and {Twicken}, J.~D. and {Vanderspek}, R. and {Winn}, J.~N. and {Collins}, K.~I. and {Fukui}, A. and {Gan}, T. and {Narita}, N. and {Schwarz}, R.~P.},
        title = "{TESS discovery of a super-Earth orbiting the M-dwarf star TOI-1680}",
      journal = {\aap},
         year = 2023,
        month = sep,
       volume = {677},
          eid = {A31},
        pages = {A31},
          doi = {10.1051/0004-6361/202347040},
archivePrefix = {arXiv},
       eprint = {2307.05368},
 primaryClass = {astro-ph.EP},
       adsurl = {https://ui.adsabs.harvard.edu/abs/2023A&A...677A..31G}
}

@ARTICLE{Delrez2022,
       author = {{Delrez}, L. and {Murray}, C.~A. and {Pozuelos}, F.~J. and {Narita}, N. and {Ducrot}, E. and {Timmermans}, M. and {Watanabe}, N. and {Burgasser}, A.~J. and {Hirano}, T. and {Rackham}, B.~V. and {Stassun}, K.~G. and {Van Grootel}, V. and {Aganze}, C. and {Cointepas}, M. and {Howell}, S. and {Kaltenegger}, L. and {Niraula}, P. and {Sebastian}, D. and {Almenara}, J.~M. and {Barkaoui}, K. and {Baycroft}, T.~A. and {Bonfils}, X. and {Bouchy}, F. and {Burdanov}, A. and {Caldwell}, D.~A. and {Charbonneau}, D. and {Ciardi}, D.~R. and {Collins}, K.~A. and {Daylan}, T. and {Demory}, B.-O. and {de Wit}, J. and {Dransfield}, G. and {Fajardo-Acosta}, S.~B. and {Fausnaugh}, M. and {Fukui}, A. and {Furlan}, E. and {Garcia}, L.~J. and {Gnilka}, C.~L. and {G{\'o}mez Maqueo Chew}, Y. and {G{\'o}mez-Mu{\~n}oz}, M.~A. and {G{\"u}nther}, M.~N. and {Harakawa}, H. and {Heng}, K. and {Hooton}, M.~J. and {Hori}, Y. and {Ikoma}, M. and {Jehin}, E. and {Jenkins}, J.~M. and {Kagetani}, T. and {Kawauchi}, K. and {Kimura}, T. and {Kodama}, T. and {Kotani}, T. and {Krishnamurthy}, V. and {Kudo}, T. and {Kunovac}, V. and {Kusakabe}, N. and {Latham}, D.~W. and {Littlefield}, C. and {McCormac}, J. and {Melis}, C. and {Mori}, M. and {Murgas}, F. and {Palle}, E. and {Pedersen}, P.~P. and {Queloz}, D. and {Ricker}, G. and {Sabin}, L. and {Schanche}, N. and {Schroffenegger}, U. and {Seager}, S. and {Shiao}, B. and {Sohy}, S. and {Standing}, M.~R. and {Tamura}, M. and {Theissen}, C.~A. and {Thompson}, S.~J. and {Triaud}, A.~H.~M.~J. and {Vanderspek}, R. and {Vievard}, S. and {Wells}, R.~D. and {Winn}, J.~N. and {Zou}, Y. and {Z{\'u}{\~n}iga-Fern{\'a}ndez}, S. and {Gillon}, M.},
        title = "{Two temperate super-Earths transiting a nearby late-type M dwarf}",
      journal = {\aap},
         year = 2022,
        month = nov,
       volume = {667},
          eid = {A59},
        pages = {A59},
          doi = {10.1051/0004-6361/202244041},
archivePrefix = {arXiv},
       eprint = {2209.02831},
 primaryClass = {astro-ph.EP},
       adsurl = {https://ui.adsabs.harvard.edu/abs/2022A&A...667A..59D}
}

@ARTICLE{McQuillan2014,
       author = {{McQuillan}, A. and {Mazeh}, T. and {Aigrain}, S.},
        title = "{Rotation Periods of 34,030 Kepler Main-sequence Stars: The Full Autocorrelation Sample}",
      journal = {\apjs},
         year = 2014,
        month = apr,
       volume = {211},
       number = {2},
          eid = {24},
        pages = {24},
          doi = {10.1088/0067-0049/211/2/24},
archivePrefix = {arXiv},
       eprint = {1402.5694},
 primaryClass = {astro-ph.SR},
       adsurl = {https://ui.adsabs.harvard.edu/abs/2014ApJS..211...24M}
}

@ARTICLE{skrutskie2006,
       author = {{Skrutskie}, M.~F. and {Cutri}, R.~M. and {Stiening}, R. and {Weinberg}, M.~D. and {Schneider}, S. and {Carpenter}, J.~M. and {Beichman}, C. and {Capps}, R. and {Chester}, T. and {Elias}, J. and {Huchra}, J. and {Liebert}, J. and {Lonsdale}, C. and {Monet}, D.~G. and {Price}, S. and {Seitzer}, P. and {Jarrett}, T. and {Kirkpatrick}, J.~D. and {Gizis}, J.~E. and {Howard}, E. and {Evans}, T. and {Fowler}, J. and {Fullmer}, L. and {Hurt}, R. and {Light}, R. and {Kopan}, E.~L. and {Marsh}, K.~A. and {McCallon}, H.~L. and {Tam}, R. and {Van Dyk}, S. and {Wheelock}, S.},
        title = "{The Two Micron All Sky Survey (2MASS)}",
      journal = {\aj},
         year = 2006,
        month = feb,
       volume = {131},
       number = {2},
        pages = {1163-1183},
          doi = {10.1086/498708},
       adsurl = {https://ui.adsabs.harvard.edu/abs/2006AJ....131.1163S}
}

@ARTICLE{rackham2026,
       author = {{Rackham}, Benjamin V. and {Iyer}, Aishwarya R. and {Apai}, D{\'a}niel and {McGill}, Peter and {Rotman}, Yoav and {Col{\'o}n}, Knicole D. and {Morris}, Brett M. and {Gilbert}, Emily A. and {Quintana}, Elisa V. and {Dotson}, Jessie L. and {Barclay}, Thomas and {Supsinskas}, Pete and {Karburn}, Jordan and {Hedges}, Christina and {Rowe}, Jason F. and {Ciardi}, David R. and {Christiansen}, Jessie L. and {Foote}, Trevor O. and {Greene}, Thomas P. and {Hoffman}, Kelsey and {Holcomb}, Rae and {Kesseli}, Aurora Y. and {Kostov}, Veselin B. and {Lewis}, Nikole K. and {Mason}, James P. and {Mosby}, Gregory and {Mullally}, Susan E. and {Schlieder}, Joshua E. and {Weiner Mansfield}, Megan and {Welbanks}, Luis and {Youngblood}, Allison},
        title = "{NASA's $\textit{Pandora SmallSat Mission}$: Simulating the Impact of Stellar Photospheric Heterogeneity and Its Correction}",
      journal = {arXiv e-prints},
         year = 2026,
        month = mar,
          eid = {arXiv:2603.04519},
        pages = {arXiv:2603.04519},
archivePrefix = {arXiv},
       eprint = {2603.04519},
 primaryClass = {astro-ph.EP},
       adsurl = {https://ui.adsabs.harvard.edu/abs/2026arXiv260304519R}
}

@ARTICLE{zhou2017,
       author = {{Zhou}, Yifan and {Apai}, D{\'a}niel and {Lew}, Ben W.~P. and {Schneider}, Glenn},
        title = "{A Physical Model-based Correction for Charge Traps in the Hubble Space Telescope{\textquoteright}s Wide Field Camera 3 Near-IR Detector and Its Applications to Transiting Exoplanets and Brown Dwarfs}",
      journal = {\aj},
         year = 2017,
        month = jun,
       volume = {153},
       number = {6},
          eid = {243},
        pages = {243},
          doi = {10.3847/1538-3881/aa6481},
archivePrefix = {arXiv},
       eprint = {1703.01301},
 primaryClass = {astro-ph.IM},
       adsurl = {https://ui.adsabs.harvard.edu/abs/2017AJ....153..243Z}
}

@article{anderson1954,
  author = {Anderson, T. W. and Darling, D. A.},
  title = {A Test of Goodness of Fit},
  journal = {Journal of the American Statistical Association},
  volume = {49},
  pages = {765--769},
  year = {1954},
  doi = {10.1080/01621459.1954.10501232}
}

@ARTICLE{espinoza2024,
       author = {{Espinoza}, N{\'e}stor and {Steinrueck}, Maria E. and {Kirk}, James and {MacDonald}, Ryan J. and {Savel}, Arjun B. and {Arnold}, Kenneth and {Kempton}, Eliza M.-R. and {Murphy}, Matthew M. and {Carone}, Ludmila and {Zamyatina}, Maria and {Lewis}, David A. and {Samra}, Dominic and {Kiefer}, Sven and {Rauscher}, Emily and {Christie}, Duncan and {Mayne}, Nathan and {Helling}, Christiane and {Rustamkulov}, Zafar and {Parmentier}, Vivien and {May}, Erin M. and {Carter}, Aarynn L. and {Zhang}, Xi and {L{\'o}pez-Morales}, Mercedes and {Allen}, Natalie and {Blecic}, Jasmina and {Decin}, Leen and {Mancini}, Luigi and {Molaverdikhani}, Karan and {Rackham}, Benjamin V. and {Palle}, Enric and {Tsai}, Shang-Min and {Ahrer}, Eva-Maria and {Bean}, Jacob L. and {Crossfield}, Ian J.~M. and {Haegele}, David and {H{\'e}brard}, Eric and {Kreidberg}, Laura and {Powell}, Diana and {Schneider}, Aaron D. and {Welbanks}, Luis and {Wheatley}, Peter and {Brahm}, Rafael and {Crouzet}, Nicolas},
        title = "{Inhomogeneous terminators on the exoplanet WASP-39 b}",
      journal = {\nat},
         year = 2024,
        month = aug,
       volume = {632},
       number = {8027},
        pages = {1017-1020},
          doi = {10.1038/s41586-024-07768-4},
archivePrefix = {arXiv},
       eprint = {2407.10294},
 primaryClass = {astro-ph.EP},
       adsurl = {https://ui.adsabs.harvard.edu/abs/2024Natur.632.1017E}
}

@ARTICLE{Rotman2026,
       author = {{Rotman}, Yoav and {McGill}, Peter and {Welbanks}, Luis and {Rackham}, Benjamin V. and {Iyer}, Aishwarya and {Apai}, Daniel and {Line}, Michael R. and {Quintana}, Elisa V. and {Dotson}, Jessie L. and {Colon}, Knicole D. and {Barclay}, Thomas and {Hedges}, Christina and {Rowe}, Jason F. and {Gilbert}, Emily A. and {Morris}, Brett M. and {Christiansen}, Jessie L. and {Foote}, Trevor O. and {Garcia Soto}, Aylin and {Greene}, Thomas P. and {Hoffman}, Kelsey and {Hord}, Benjamin J. and {Kesseli}, Aurora Y. and {Kostov}, Veselin B. and {Weiner Mansfield}, Megan and {Wiser}, Lindsey S.},
        title = "{NASA's Pandora SmallSat Mission: Simulated Modeling and Retrieval of Near-Infrared Exoplanet Transmission Spectra}",
      journal = {arXiv e-prints},
         year = 2026,
        month = mar,
          eid = {arXiv:2603.04488},
        pages = {arXiv:2603.04488},
          doi = {10.48550/arXiv.2603.04488},
archivePrefix = {arXiv},
       eprint = {2603.04488},
 primaryClass = {astro-ph.EP},
       adsurl = {https://ui.adsabs.harvard.edu/abs/2026arXiv260304488R}
}

@ARTICLE{Borucki2010,
       author = {{Borucki}, William J. and {Koch}, David and {Basri}, Gibor and {Batalha}, Natalie and {Brown}, Timothy and {Caldwell}, Douglas and {Caldwell}, John and {Christensen-Dalsgaard}, J{\o}rgen and {Cochran}, William D. and {DeVore}, Edna and {Dunham}, Edward W. and {Dupree}, Andrea K. and {Gautier}, Thomas N. and {Geary}, John C. and {Gilliland}, Ronald and {Gould}, Alan and {Howell}, Steve B. and {Jenkins}, Jon M. and {Kondo}, Yoji and {Latham}, David W. and {Marcy}, Geoffrey W. and {Meibom}, S{\o}ren and {Kjeldsen}, Hans and {Lissauer}, Jack J. and {Monet}, David G. and {Morrison}, David and {Sasselov}, Dimitar and {Tarter}, Jill and {Boss}, Alan and {Brownlee}, Don and {Owen}, Toby and {Buzasi}, Derek and {Charbonneau}, David and {Doyle}, Laurance and {Fortney}, Jonathan and {Ford}, Eric B. and {Holman}, Matthew J. and {Seager}, Sara and {Steffen}, Jason H. and {Welsh}, William F. and {Rowe}, Jason and {Anderson}, Howard and {Buchhave}, Lars and {Ciardi}, David and {Walkowicz}, Lucianne and {Sherry}, William and {Horch}, Elliott and {Isaacson}, Howard and {Everett}, Mark E. and {Fischer}, Debra and {Torres}, Guillermo and {Johnson}, John Asher and {Endl}, Michael and {MacQueen}, Phillip and {Bryson}, Stephen T. and {Dotson}, Jessie and {Haas}, Michael and {Kolodziejczak}, Jeffrey and {Van Cleve}, Jeffrey and {Chandrasekaran}, Hema and {Twicken}, Joseph D. and {Quintana}, Elisa V. and {Clarke}, Bruce D. and {Allen}, Christopher and {Li}, Jie and {Wu}, Haley and {Tenenbaum}, Peter and {Verner}, Ekaterina and {Bruhweiler}, Frederick and {Barnes}, Jason and {Prsa}, Andrej},
        title = "{Kepler Planet-Detection Mission: Introduction and First Results}",
      journal = {Science},
         year = 2010,
        month = feb,
       volume = {327},
       number = {5968},
        pages = {977},
          doi = {10.1126/science.1185402},
       adsurl = {https://ui.adsabs.harvard.edu/abs/2010Sci...327..977B}
}

@ARTICLE{Ricker2015,
       author = {{Ricker}, George R. and {Winn}, Joshua N. and {Vanderspek}, Roland and {Latham}, David W. and {Bakos}, G{\'a}sp{\'a}r {\'A}. and {Bean}, Jacob L. and {Berta-Thompson}, Zachory K. and {Brown}, Timothy M. and {Buchhave}, Lars and {Butler}, Nathaniel R. and {Butler}, R. Paul and {Chaplin}, William J. and {Charbonneau}, David and {Christensen-Dalsgaard}, J{\o}rgen and {Clampin}, Mark and {Deming}, Drake and {Doty}, John and {De Lee}, Nathan and {Dressing}, Courtney and {Dunham}, Edward W. and {Endl}, Michael and {Fressin}, Francois and {Ge}, Jian and {Henning}, Thomas and {Holman}, Matthew J. and {Howard}, Andrew W. and {Ida}, Shigeru and {Jenkins}, Jon M. and {Jernigan}, Garrett and {Johnson}, John Asher and {Kaltenegger}, Lisa and {Kawai}, Nobuyuki and {Kjeldsen}, Hans and {Laughlin}, Gregory and {Levine}, Alan M. and {Lin}, Douglas and {Lissauer}, Jack J. and {MacQueen}, Phillip and {Marcy}, Geoffrey and {McCullough}, Peter R. and {Morton}, Timothy D. and {Narita}, Norio and {Paegert}, Martin and {Palle}, Enric and {Pepe}, Francesco and {Pepper}, Joshua and {Quirrenbach}, Andreas and {Rinehart}, Stephen A. and {Sasselov}, Dimitar and {Sato}, Bun'ei and {Seager}, Sara and {Sozzetti}, Alessandro and {Stassun}, Keivan G. and {Sullivan}, Peter and {Szentgyorgyi}, Andrew and {Torres}, Guillermo and {Udry}, Stephane and {Villasenor}, Joel},
        title = "{Transiting Exoplanet Survey Satellite (TESS)}",
      journal = {Journal of Astronomical Telescopes, Instruments, and Systems},
         year = 2015,
        month = jan,
       volume = {1},
          eid = {014003},
        pages = {014003},
          doi = {10.1117/1.JATIS.1.1.014003},
       adsurl = {https://ui.adsabs.harvard.edu/abs/2015JATIS...1a4003R}
}

@ARTICLE{Kostogryz2025,
       author = {{Kostogryz}, N. and {Shapiro}, A.~I. and {Carone}, L. and {Gizon}, L. and {Helling}, Ch. and {Kiefer}, S. and {Mercier}, S. and {Seager}, S. and {Solanki}, S.~K. and {Unruh}, Y. and {de Wit}, J. and {Witzke}, V.},
        title = "{The Effect of Stellar Magnetic Activity on Measurements of Morning and Evening Asymmetry of Planetary Terminator}",
      journal = {\apjl},
         year = 2025,
        month = aug,
       volume = {989},
       number = {1},
          eid = {L6},
        pages = {L6},
          doi = {10.3847/2041-8213/adf0fb},
archivePrefix = {arXiv},
       eprint = {2507.16451},
 primaryClass = {astro-ph.SR},
       adsurl = {https://ui.adsabs.harvard.edu/abs/2025ApJ...989L...6K}
}

@ARTICLE{Bourrier2024,
       author = {{Bourrier}, V. and {Delisle}, J.-B. and {Lovis}, C. and {Cegla}, H.~M. and {Cretignier}, M. and {Allart}, R. and {Al Moulla}, K. and {Tavella}, S. and {Attia}, M. and {Mounzer}, D. and {Vaulato}, V. and {Steiner}, M. and {Vrignaud}, T. and {Mercier}, S. and {Dumusque}, X. and {Ehrenreich}, D. and {Seidel}, J.~V. and {Wyttenbach}, A. and {Dethier}, W. and {Pepe}, F.},
        title = "{The ANTARESS workflow: I. Optimal extraction of spatially resolved stellar spectra with high-resolution transit spectroscopy}",
      journal = {\aap},
         year = 2024,
        month = nov,
       volume = {691},
          eid = {A113},
        pages = {A113},
          doi = {10.1051/0004-6361/202449203},
archivePrefix = {arXiv},
       eprint = {2407.19012},
 primaryClass = {astro-ph.EP},
       adsurl = {https://ui.adsabs.harvard.edu/abs/2024A&A...691A.113B}
}
\bibliographystyle{aasjournal}

\appendix

This appendix provides summary tables of the stellar spectral modeling results described in the main text.
\autoref{table:combinedRetrievalG102G141} presents the results of the simultaneous retrieval of the \wfcthree{} G102 and G141 spectra. We report posterior medians and credible intervals for the stellar temperatures, filling factors, and nuisance parameters for one-, two-, and three-component models, along with model comparison metrics.
\autoref{table:NoPriors} summarizes the corresponding multi-component fits to the individual \stis{} (G430L, G750L) and \wfcthree{} (G102, G141) datasets.

\begin{deluxetable*}{rccccc}[h]
\tabletypesize{\footnotesize}  
\tablecaption{\label{table:combinedRetrievalG102G141} A retrieval for simultaneous fit of HAT-P-11 with G102 and G141 data.}
\tablehead{
\colhead{\textbf{Param.}} &
\colhead{\textbf{Units}} &
\colhead{\textbf{Prior}} & 
\colhead{\textbf{1 Comp}} & 
\colhead{\textbf{2 Comp}} &
\colhead{\textbf{3 Comp}} 
}
\startdata
\multicolumn{6}{c}{\textbf{With No Prior}} \\ 
$T_1$ & K & $\mathcal{U}$[2300-6500] & $4779_{-10}^{+10}$ &$\mathbf{4979_{-24}^{+20}}$ & $4981_{-31}^{+26}$ \\
$T_2$ & K & $\mathcal{U}$[2300-6500] &   - &$\mathbf{3413_{-73}^{+77}}$ & $3402_{-88}^{+89}$ \\
$T_3$ & K & $\mathcal{U}$[2300-6500] &  - & - & $4924_{-314}^{+167}$ \\
$f_{1, {\mathrm G102}}$ & & $\mathrm{Dirichlet}^{\dagger}$  &    - &$\mathbf{0.741_{-0.020}^{+0.023}}$ & $0.667_{-0.075}^{+0.052}$ \\
$f_{2, {\mathrm G102}}$ & & $\mathrm{Dirichlet}^{\dagger}$  &    - &$\mathbf{0.259_{-0.020}^{+0.023}}$ & $0.253_{-0.031}^{+0.026}$ \\
$f_{3\, {\mathrm G102}}$ & - & $\mathrm{Dirichlet}^{\dagger}$  &    - & - & $0.073_{-0.050}^{+0.070}$   \\
$f_{1, {\mathrm G141}}$ & & $\mathrm{Dirichlet}^{\dagger}$  &    - & $\mathbf{0.669_{-0.041}^{+0.046}}$ &  $0.569_{-0.101}^{+0.088}$ \\
$f_{2, {\mathrm G141}}$ & - & $\mathrm{Dirichlet}^{\dagger}$  & - & $\mathbf{0.331_{-0.046}^{+0.041}}$ & $0.322_{-0.050}^{+0.054}$ \\
$f_{3, {\mathrm G141}}$ & - & $\mathrm{Dirichlet}^{\dagger}$  &    - & - & $0.105_{-0.066}^{+0.095}$  \\
R$_*$ & R$_\odot$ & $\mathcal{N}$({0.760,0.048}) & 0.675$^{+0.004}_{-0.004}$   & $\mathbf{0.800^{+0.011}_{-0.010}}$ & 0.797$^{+0.011}_{-0.010}$  \\
Fe/H & - & $\mathcal{U}_{[-2,+0.5]}$(0.31, 0.05) & 0.358$^{+0.033}_{-0.033}$ & $\mathbf{0.344^{+0.030}_{-0.026}}$ & 0.331$^{+0.036}_{-0.035}$  \\
ln $f_{\rm var,\,G102}$ & - & $\mathcal{U}$[-10-0] & $-4.982_{-0.053}^{+0.055}$ &$\mathbf{-4.709^{+0.068}_{-0.074}}$ & $-4.722_{-0.070}^{+0.073}$  \\
ln $f_{\rm var,\, G141}$ & - & $\mathcal{U}$[-10-0] & $-4.098_{-0.080}^{+0.083}$ &$\mathbf{-4.478_{-0.065}^{+0.074}}$ &$-4.462_{-0.071}^{+0.074}$  \\
a$_{\rm G102}$ & - & $\mathcal{U}$[0.5-2] & 1.253$^{+0.008}_{-0.008}$  & $\mathbf{0.954^{+0.024}_{-0.023}}$  &  0.966$^{+0.024}_{-0.023}$ \\
a$_{\rm G141}$ & - & $\mathcal{U}$[0.5-2] & 1.224$^{+0.009}_{-0.009}$ & $\mathbf{1.005^{+0.019}_{-0.018}}$  & 1.013$^{+0.022}_{-0.021}$ \\
$\Delta$BIC & - & - & 336.0 & $\mathbf{\checkmark}$ & 16.0 \\
AD Stat. & - & - & 0.818 & 1.198 & 1.264 \\
\enddata
\tablecomments{$^\dagger$A Dirichlet prior was used for filling factor such  that $f_1>f_2>f_3$ and $\Sigma f_i = 1$}
\end{deluxetable*}
\begin{deluxetable*}{rccccc}[htp]
\tabletypesize{\footnotesize}  
\tablecaption{\label{table:NoPriors} Multi-components Fits to \hp{} spectra.}
\tablehead{
\colhead{\textbf{Param.}} &
\colhead{\textbf{Units}} &
\colhead{\textbf{Prior}} & 
\colhead{\textbf{1 Comp}} & 
\colhead{\textbf{2 Comp}} &
\colhead{\textbf{3 Comp}} 
}
\startdata
\multicolumn{6}{c}{\textbf{STIS G430L}} \\ 
$T_1$ & K & $\mathcal{U}$[2300-6500] & {$\mathbf{4626_{-9}^{+10}}$} &$4625_{-17}^{+17}$ &$4624_{-20}^{+21}$ \\
$T_2$ & K & $\mathcal{U}$[2300-6500] &    &$4621_{-116}^{+73}$ &$4627_{-65}^{+60}$ \\
$T_3$ & K & $\mathcal{U}$[2300-6500] &  - & - &$4583_{-462}^{+140}$ \\
$f_1$ & - & $\mathrm{Dirichlet}^{\dagger}$ &    - &$0.830_{-0.218}^{+0.148}$ &$0.708_{-0.174}^{+0.186}$ \\
$f_2$ & - & $\mathrm{Dirichlet}^{\dagger}$ &  - &$0.170_{-0.148}^{+0.218}$ &$0.232_{-0.149}^{+0.146}$ \\
$f_3$ & - & $\mathrm{Dirichlet}^{\dagger}$ &    - & - &$0.032_{-0.027}^{+0.088}$ \\
$R_*$ & R$_\odot$ & $\mathcal{N}$({0.760,0.048}) &  $\mathbf{0.793_{-0.006}^{+0.006}}$ &$0.793_{-0.006}^{+0.006}$ &$0.794_{-0.006}^{+0.006}$ \\
Fe/H & - & $\mathcal{N}_{[-2,+0.5]}$(0.31, 0.05) & $\mathbf{0.498_{-0.004}^{+0.002}}$ &$0.498_{-0.003}^{+0.002}$ &$0.498_{-0.003}^{+0.002}$ \\
ln $f_{\rm var}$ & - & $\mathcal{U}$[-10-0] & $\mathbf{-2.725_{-0.043}^{+0.043}}$ &$-2.724_{-0.042}^{+0.043}$ &$-2.724_{-0.042}^{+0.042}$ \\
$\Delta$BIC & - & - & $\mathbf{\checkmark}$ & 20.0 & 33.0 \\
AD Stat. & - & - & $\mathbf{2.641}$ &   2.662 & 2.646 \\
\hline
\multicolumn{6}{c}{\textbf{STIS G750L}} \\ 
$T_1$ & K & $\mathcal{U}$[2300-6500] & $\mathbf{4683_{-3}^{+3}}$ &$4685_{-8}^{+12}$ &$4701_{-12}^{+16}$ \\
$T_2$ & K & $\mathcal{U}$[2300-6500] &   - &$4662_{-87}^{+63}$ &$4698_{-27}^{+29}$ \\
$T_3$ & K & $\mathcal{U}$[2300-6500] &  - & - &$3044_{-33}^{+146}$ \\
$f_1$ & - & $\mathrm{Dirichlet}^{\dagger}$ &    - &$0.879_{-0.201}^{+0.101}$ &$0.684_{-0.141}^{+0.151}$ \\
$f_2$ & - & $\mathrm{Dirichlet}^{\dagger}$ &  - &$0.121_{-0.101}^{+0.201}$ &$0.271_{-0.147}^{+0.134}$ \\
$f_3$ & - & $\mathrm{Dirichlet}^{\dagger}$ &    - & - &$0.044_{-0.018}^{+0.020}$ \\
$R_*$ & R$_\odot$ & $\mathcal{N}$({0.760,0.048}) &  $\mathbf{0.780_{-0.005}^{+0.005}}$ &$0.779_{-0.005}^{+0.005}$ &$0.788_{-0.007}^{+0.006}$ \\
Fe/H & - & $\mathcal{N}_{[-2,+0.5]}$(0.31, 0.05) & $\mathbf{0.240_{-0.021}^{+0.021}}$ &$0.241_{-0.020}^{+0.020}$ &$0.285_{-0.033}^{+0.034}$ \\
ln $f_{\rm var}$ & - & $\mathcal{U}$[-10-0] & $\mathbf{-4.301_{-0.024}^{+0.024}}$ &$-4.300_{-0.024}^{+0.023}$ &$-4.304_{-0.023}^{+0.025}$ \\
$\Delta$BIC & - & - & $\checkmark$ & 12.0 & 25.0 \\ 
AD Stat. & - & - & $\mathbf{2.017}$ & 2.004 & 2.092 \\
\hline
\multicolumn{6}{c}{\textbf{WFC3 G102}} \\ 
$T_1$ & K & $\mathcal{U}$[2300-6500] & $4748_{-13}^{+13}$ &$\mathbf{4966_{-41}^{+33}}$ &$4975_{-50}^{+35}$ \\
$T_2$ & K & $\mathcal{U}$[2300-6500] &   - &$\mathbf{3617_{-149}^{+125}}$ &$3610_{-188}^{+174}$ \\
$T_3$ & K & $\mathcal{U}$[2300-6500] &  - & - &$4407_{-857}^{+549}$ \\
$f_1$ & - & $\mathrm{Dirichlet}^{\dagger}$ &    - &$\mathbf{0.736_{-0.038}^{+0.043}}$ &$0.684_{-0.097}^{+0.055}$ \\
$f_2$ & - & $\mathrm{Dirichlet}^{\dagger}$ &  - &$\mathbf{0.264_{-0.043}^{+0.038}}$ &$0.236_{-0.051}^{+0.048}$ \\
$f_3$ & - & $\mathrm{Dirichlet}^{\dagger}$ &    - & - &$0.091_{-0.059}^{+0.074}$ \\
$R_*$ & R$_\odot$ & $\mathcal{N}$({0.760,0.048}) &  $0.765_{-0.006}^{+0.006}$ &$\mathbf{0.782_{-0.007}^{+0.007}}$ &$0.782_{-0.007}^{+0.007}$ \\
Fe/H & - & $\mathcal{N}_{[-2,+0.5]}$(0.31, 0.05) & $0.324_{-0.044}^{+0.044}$ &$\mathbf{0.369_{-0.040}^{+0.041}}$ &$0.369_{-0.040}^{+0.041}$ \\
ln $f_{\rm var}$ & - & $\mathcal{U}$[-10-0] & $-4.599_{-0.069}^{+0.071}$ &$\mathbf{-4.710_{-0.066}^{+0.069}}$ &$-4.712_{-0.065}^{+0.065}$ \\
$\Delta$BIC & - & - & 16.0 & $\checkmark$ & 10.0 \\
AD Stat. & - & - & 1.273 & $\mathbf{0.793}$ & 0.798 \\
\hline
\multicolumn{6}{c}{\textbf{WFC3 G141}} \\ 
$T_1$ & K & $\mathcal{U}$[2300-6500] & $4888_{-15}^{+14}$ &$\mathbf{4950_{-27}^{+32}}$ &$4961_{-36}^{+41}$ \\
$T_2$ & K & $\mathcal{U}$[2300-6500] &   - &$\mathbf{3326_{-119}^{+110}}$ &$3336_{-130}^{+130}$ \\
$T_3$ & K & $\mathcal{U}$[2300-6500] &  - & - &$4623_{-1291}^{+495}$ \\
$f_1$ & - & $\mathrm{Dirichlet}^{\dagger}$ &    - &$\mathbf{0.718_{-0.045}^{+0.051}}$ &$0.670_{-0.089}^{+0.056}$ \\
$f_2$ & - & $\mathrm{Dirichlet}^{\dagger}$ &  - &$\mathbf{0.282_{-0.051}^{+0.045}}$ &$0.259_{-0.063}^{+0.053}$ \\
$f_3$ & - & $\mathrm{Dirichlet}^{\dagger}$ &    - & - &$0.080_{-0.058}^{+0.071}$ \\
$R_*$ & R$_\odot$ & $\mathcal{N}$({0.760,0.048}) &  $0.730_{-0.006}^{+0.006}$ &$\mathbf{0.802_{-0.016}^{+0.015}}$ &$0.803_{-0.014}^{+0.014}$ \\
Fe/H & - & $\mathcal{N}_{[-2,+0.5]}$(0.31, 0.05) & $0.289_{-0.037}^{+0.037}$ &$\mathbf{0.300_{-0.034}^{+0.037}}$ &$0.303_{-0.034}^{+0.036}$ \\
ln $f_{\rm var}$ & - & $\mathcal{U}$[-10-0] & $-4.650_{-0.075}^{+0.079}$ &$\mathbf{-4.722_{-0.070}^{+0.073}}$ &$-4.749_{-0.070}^{+0.071}$ \\
$\Delta$BIC & - & - & 22.0 & $\checkmark$ & 9.0 \\ 
AD Stat. & - & - & 0.416 & $\mathbf{0.135}$ & 0.118 \\
\hline
\enddata
\tablecomments{The preferred model for each instrument, selected based on the minimum BIC, is shown in bold. \\$^\dagger$A Dirichlet prior was used for filling factor such  that $f_1>f_2>f_3$ and $\Sigma f_i = 1$.}
\end{deluxetable*}

\end{document}